\documentclass[prx,longbibliography,showpacs,twocolumn,superscriptaddress,amsmath,amssymb,verbatim]{revtex4-2}
\usepackage{amsmath,amssymb,amsfonts,stmaryrd,wasysym,graphicx,multirow,textcomp,subfigure}
\usepackage{url}
\usepackage[colorlinks=true,citecolor=blue,urlcolor=blue]{hyperref}
\usepackage{romannum}
\usepackage{mathtools}
\usepackage[utf8]{inputenc}
\usepackage{braket}
\usepackage{amsmath}
\usepackage{xcolor}
\usepackage{color,soul} 
\usepackage{bm,xfrac}
\usepackage[normalem]{ulem}

\usepackage{enumitem}
\usepackage{dcolumn}
\usepackage{bm}
\usepackage[flushleft]{threeparttable}
\usepackage{braket}
\usepackage{array}
\usepackage{booktabs}
\usepackage{float}
\usepackage{setspace}
\hypersetup{colorlinks=true, urlcolor=blue, citecolor=cyan, pdfborder={0 0 0},}

\hypersetup{colorlinks=true, urlcolor=blue, citecolor=cyan, pdfborder={0 0 0}}

\newcommand{\cyan}[1]{{\color{cyan}{\em {#1}}}} 

\newcommand{\vv}[1]{{\boldsymbol #1}}

\def\be{\begin{equation}}
	\def\ee{\end{equation}}
\def\bea{\begin{eqnarray}}
	\def\eea{\end{eqnarray}}

\begin{document}
\title{Topological Phase Transitions of Generalized Brillouin Zone}

	\begin{abstract}
It has been known that the bulk-boundary correspondence (BBC) of the non-Hermitian skin effect is characterized by the topology of the complex eigenvalue spectra, while the topology of the wave function gives rise to Hermitian BBC with conventional boundary modes. In this work, we go beyond the known description of the non-Hermitian topological phase by discovering a new type of BBC that appears in generalized boundary conditions. The generalized Brillouin zone (GBZ) possesses non-trivial topological structures in the intermediate boundary condition between open and periodic boundary conditions. Unlike the conventional BBC, the topological phase transition is characterized by the generalized momentum touching of GBZ, which manifests as exceptional points. As a realization of our proposal, we suggest the non-reciprocal Kuramoto oscillator lattice, where the phase slips accompany the exceptional points  as a signature of such topological phase transition. Our work establishes an understanding of non-Hermitian topological matter by complementing the non-Hermitian BBC as a general foundation of the non-Hermitian topological systems. 
	\end{abstract}

\author{Sonu Verma}
\affiliation{Center for Theoretical Physics of Complex Systems, Institute for Basic Science (IBS) Daejeon 34126, Republic of Korea}
\author{Moon Jip Park}
\email{moonjippark@ibs.re.kr}
\affiliation{Center for Theoretical Physics of Complex Systems, Institute for Basic Science (IBS) Daejeon 34126, Republic of Korea}
\affiliation{Department of Physics, Hanyang University, Seoul, 04763, Republic of Korea}
\maketitle

\pagenumbering{arabic}

The rapid progress of non-Hermitian mechanics has paved a new direction to obtain unconventional phases of matter that have no Hermitian analog \cite{Ueda_review_non_Hermitian_2021, Sato_review_topology_2022, Chen_Fang_review_NHSE_2022}. The representative example is the non-Hermitian skin effect (NHSE), which exhibits the macroscopic localization of bulk states on the boundary. There have been theoretical efforts to understand non-Hermitian BBC using the bulk topology in the periodic boundary condition (PBC). It has been shown that the winding of complex eigenvalue  in one-dimensional systems signifies the presence of the NHSE \cite{Wang_non_bloch_TI_2018,Murakami_non_bloch_TI_2019,Sato_NHSE_topo_2020,Chen_Fang_NHSE_topo_2020}. In a more general sense, the macroscopic spectral collapse of the NHSE can be understood within the theory of generalized Brillouin zone (GBZ) defined in the complex plane.
	
\begin{figure}[t!]
		\centering
		\includegraphics[width=1\linewidth]{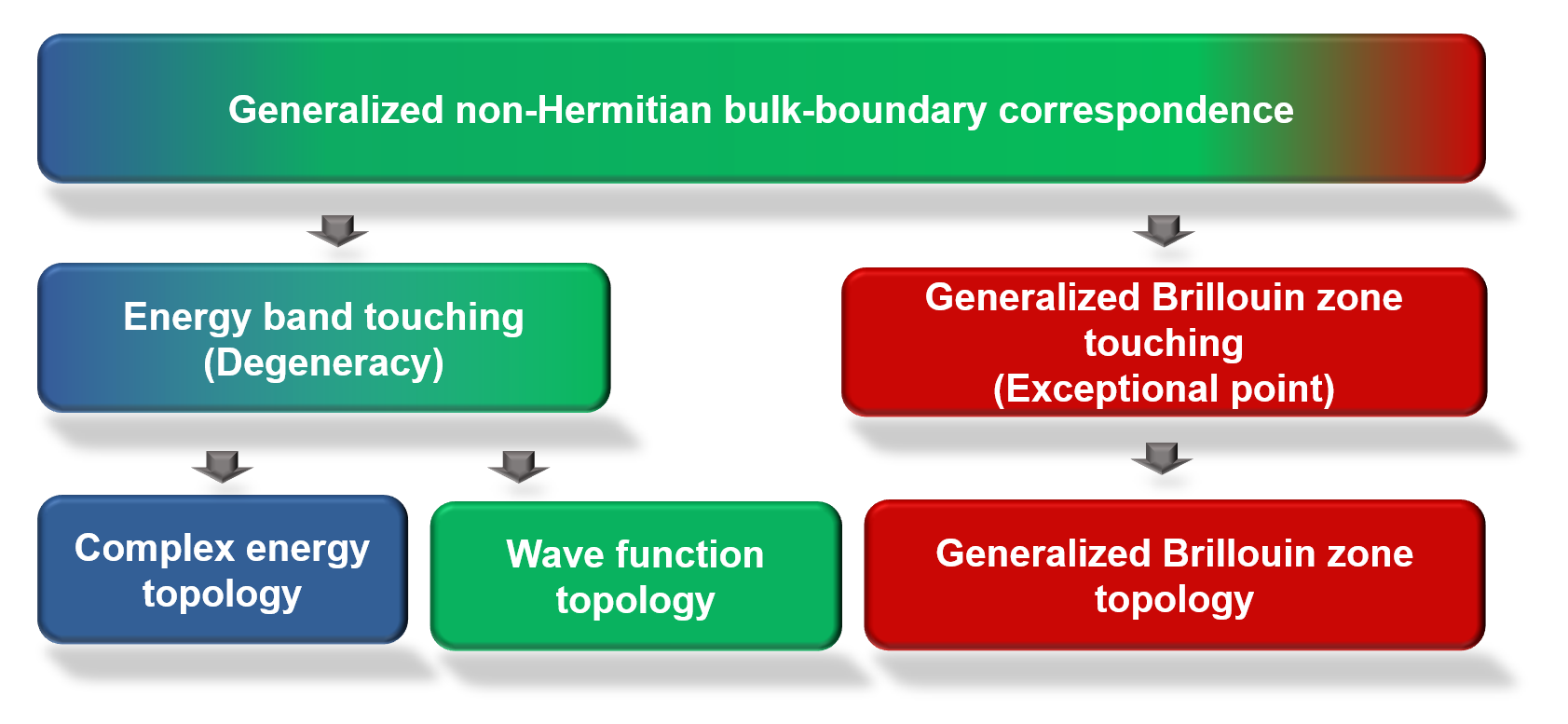}
		\caption{Schematic illustration of non-Hermitian bulk-boundary correspondence in generalized boundary condition(GBC). In the GBC, the GBZ can possess the non-trivial topological invariant, which manifests as the topological boundary mode. These topological boundary modes are not captured in the conventional band topology. The topological phase transition is characterized by the touchings of the GBZs, which manifests as exceptional points.}
		\label{fig:schematic}
		\centering
	\end{figure}

So far, it has been shown that the BBC of non-Hermitian systems is characterized by two distinct types of topology: \textit{wave function topology} in GBZ and \textit{complex eigenvalue spectra topology} in conventional BZ. While the wave function topology gives rise to the Hermitian-like boundary modes, the winding number of the eigenvalue spectra in the complex plane signals the presence of the NHSE in OBC. Here, for the first time, we present the third type of non-Hermitian BBC that emerges in generalized boundary condition (GBC). Unlike the case of PBC or OBC, the single wave functions in GBC possess distinct momenta with multiple GBZs. The existence of multiple GBZ can provoke the intrinsic topological structure of GBZ [See Fig.\ref{fig:schematic}], where the topological phase transitions are accompanied by exceptional points with touchings of GBZs. The information of the complex eigenvalue spectra itself is ignorant of such concomitant topological phase transition. As a result, our result falls in neither of the two previously known categories of non-Hermitian BBC. 
	\begin{figure*}[htbp!]
		\centering
		\includegraphics[width=1.0\linewidth]{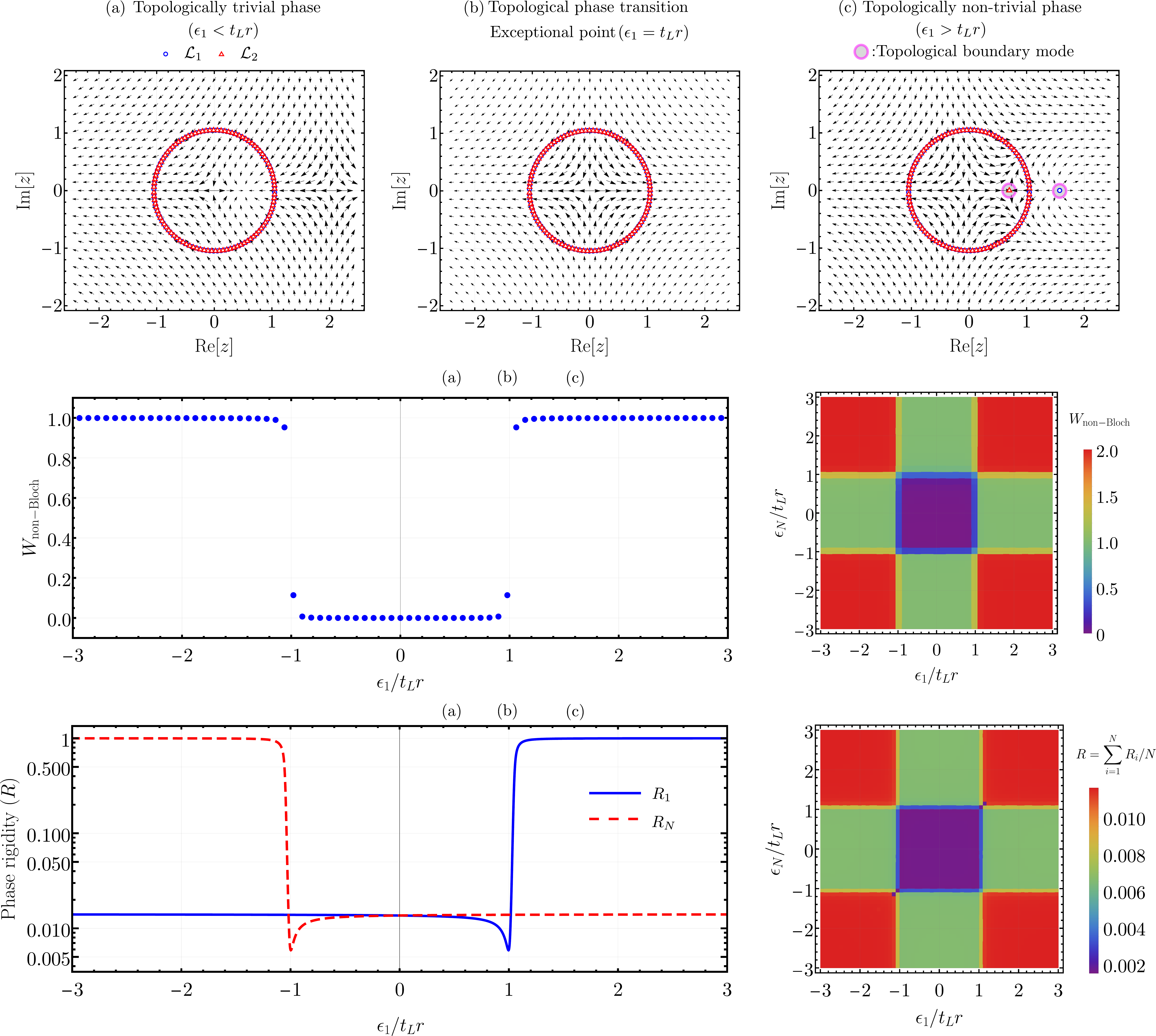}
		\caption{Top: (a), (b), and (c) dictating topological as well as exceptional phase transition of GBZs $\mathcal{L}_1,~\mathcal{L}_2$ together with the complex vector plot of the meromorphic function related to non-Bloch topological invariant in the generalized boundary conditions -(I) (GBC-(I)) (See text).
        Middle left: non-Bloch topological invariant ($W_{\rm non-Bloch}$) vs $\epsilon_1$ which takes quantized value after $\epsilon_1=t_L r$, and defines the topological phase transition of GBZs. Middle right:
        Non-Bloch topological invariant ($W_{\rm non-Bloch}$) in $(\epsilon_1,\epsilon_N)$-plane. Here the value of $W_{\rm non-Bloch}$ also represents the number of boundary states due to GBC-(I). 
	Bottom left: Phase rigidity for eigenstates $|\psi_1\rangle$ and         $|\psi_N\rangle$ vs onsite potential $\epsilon_1$. The phase 
        rigidity vanishes in large $N$-limit at $\epsilon_1=t_L r$ which defines the exceptional point and hence exceptional transition. Bottom right: Total phase rigidity in $(\epsilon_1,\epsilon_N)$-plane also captures the topological as well as exceptional phase transition of GBZs (See text).}
	\label{fig:GOBC}
	\end{figure*}

In this work, we formulate our theory by presenting the example of the Hatano-Nelson model in GBC. After constructing the single-particle theory, we propose the non-reciprocal Kuramoto lattice as an example of many-body systems. The topological phase transitions between different time-dependent phases are accompanied by the touching of two generalized momenta. Unlike the energy band touching, the touchings of GBZs generally manifest as the coalescence of the two eigenstates, known as the exceptional point (EP). The emergence of the exceptional point in the dynamical spectral results in the novel class of non-equilibrium phase transition, known as exceptional transitions (ET) \cite{Fruchart2021}. Our result establishes the important connection between the non-Hermitian BBC and the ET in the non-equilibrium phases of many-body interacting collective behaviors. 

    \cyan{Non-Bloch band theory in GBC}-- We start our discussion by considering the Hatano-Nelson (HN) model \cite{Hatano_Nelson_1996} with the GBCs. The Hamiltonian is given as,
	\bea 
	\label{general_model_ham}
	\hat{H}
	&=& \sum_{i=1}^{N-1}
	\big[ t_R \hat{c}^\dagger_{i+1}\hat{c}_{i}
	+t_L \hat{c}^\dagger_{i}\hat{c}_{i+1}
	\big]
	\\
	&+&
	t_R'\hat{c}^\dagger_{1}\hat{c}_{N}+t_L'\hat{c}^\dagger_{N}\hat{c}_{1}
	+\epsilon_1 \hat{c}^\dagger_{1}\hat{c}_{1}
	+\epsilon_N \hat{c}^\dagger_{N}\hat{c}_{N},
	\nonumber
	\eea
	where $t_{R(L)}$ represents the hopping toward the right (left) nearest neighbor site. The second line represents the general deformation of the boundary terms, which include the cases of OBC ($t'_R=t'_L=\epsilon_1=\epsilon_N=0$), PBC ($t'_R=t_R, t'_L=t_L, \epsilon_1=\epsilon_N=0$) and any intermediate boundary conditions between them (See \ref{sec:def_GBC} for full classifications of the boundary conditions).
	
	We represent the eigenstates as, $|\psi\rangle =1/\sqrt{\mathcal{N}} \sum_{i=1}^{N}\psi_i |i\rangle$, where $\psi_i=\sum_{\alpha} c_\alpha z_\alpha^i$ and $z_\alpha \in \mathbb{C}$ is the generalized complex momenta. The eigenstates in the GBC are required to simultaneously satisfy bulk and boundary equations. In the case of the Hatano-Nelson model having the only nearest neighbor couplings, the bulk equation is given by the second-order complex polynomial as, 
	\bea 
	\label{eq:bulk_eqn}
	\frac{t_R}{z_i}  -E +t_L z_i  = 0,
	\eea 
	which admits the pair of the complex momenta solutions, $(z_1,~z_2)$. We call the two momenta paired if the two momenta $(z_1,z_2)$ satisfy the bulk equation with the same eigenvalue, $E$, and the paired momenta $(z_1,z_2)$ are related by the condition $z_1z_2=r^2$, where $r^2=\frac{t_R}{t_L}$. In addition to the bulk equation, the eigenstates are required to satisfy the boundary equation, $H_B(c_1,~c_2)^\textrm{T}=0$, where
	\begin{align}
	\label{Eq_bmat}
	H_\textrm{B}=
	\begin{pmatrix}
	A(z_1) & A(z_2) \\
	B(z_1) & B(z_2)
	\end{pmatrix}.
	\end{align}
	The coefficients of the boundary matrix $H_\textrm{B}$ are explicitly given as, $A(z)=t_R-\epsilon_1 z - t'_R z^{N}$, $B(z)=t_L' z + \epsilon_N z^N -t_L z^{N+1}$. In practice, we determine the solutions for $(z_1,z_2)$ by solving Eq. \eqref{Eq_bmat}, and subsequently solve for the eigenvalue with Eq.~\eqref{eq:bulk_eqn}. 
	
	Except for PBC, where the translational symmetry is intact, a wave function is generally described by the two different paired momenta  $(z_1,z_2)\in\mathbb{C}^2$, which  map to the same eigenvalue. The contour of different momenta constitutes the generalized Brillouin zones $\mathcal{L}_1\times \mathcal{L}_2$ \cite{chenGBC2021}. In the case of the Hatano-Nelson model, the GBZs form two circles with the radii $|z_1|,~|z_2|$ in the complex plane. For example, in the case of the OBC, the two GBZs overlap each other ($|z_1|=|z_2|=r$), which forms a single GBZ as discussed in the previous studies \cite{Wang_non_bloch_TI_2018, Murakami_non_bloch_TI_2019}. 
 
    In contrast, in the case of the GBCs, the two GBZs can form two disjoint contours, which is the focus of our study. Generalized momenta $z_i$ outside (inside) the unit circle ($|z_i|=1$) or GBZs describes the evanescent wave localized on the left (right) boundary, which corresponds to the boundary states. Touchings of the  unpaired momenta of the two GBZs correspond to the coalescence of the wave functions, which give rise to the EP (See the representative examples of GBZs and eigenvalue spectra for different GBCs in Sec. \ref{sec:examples_GBZ}). 
	
	\begin{figure*}[htbp]
		\centering
		\includegraphics[width=1.0\linewidth]{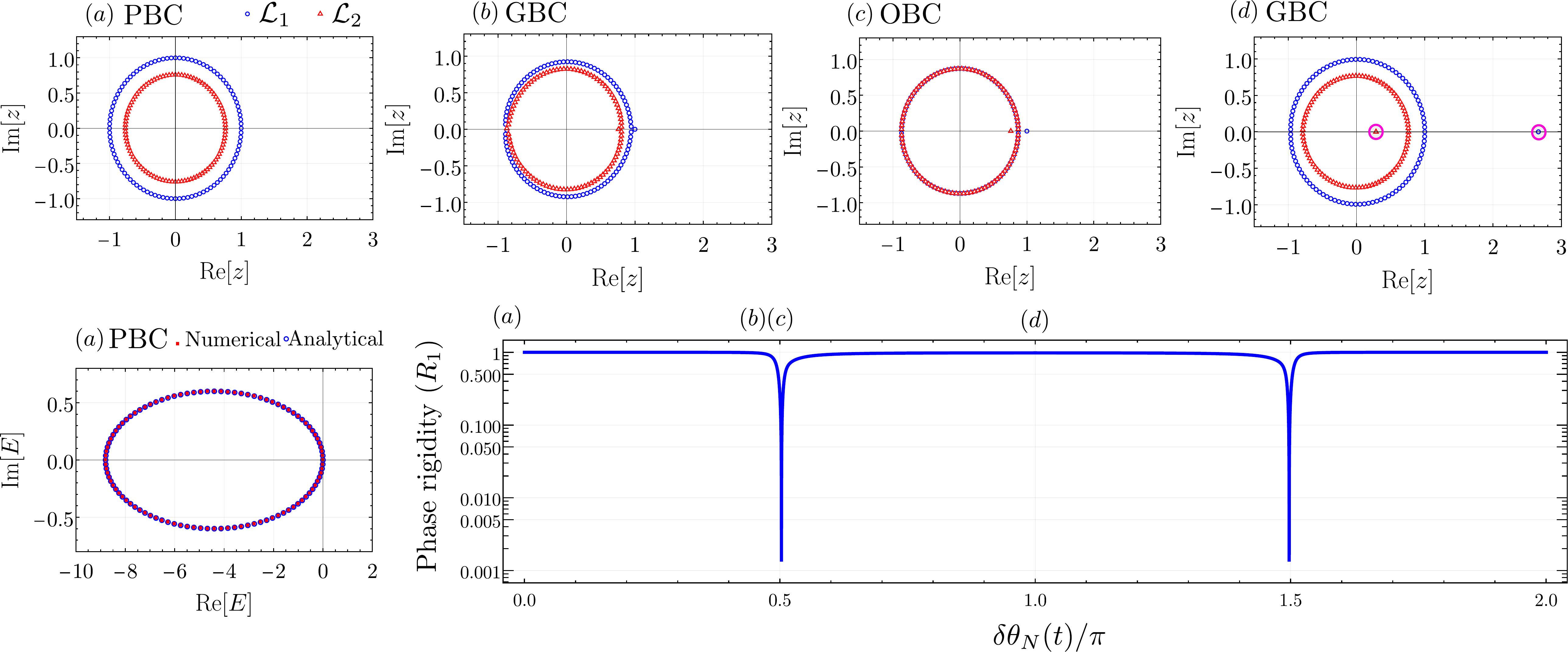}
		\caption{Bottom left: Eigenvalue spectrum of Jacobian for a chiral phase, bottom right: phase rigidity of the first eigenmode vs phase difference $\delta\theta_{N}$ during the interval of the phase slip (See text). Top panel: Generalized Brillouin zones for different $\delta\theta_N$ between $[0,\pi]$ (the behavior of GBZs is similar in the other half). At $\delta\theta_N=\pi/2$, the two GBZs merge with each other defining the exceptional transition. Parameters: $K_R=2.5,~K_L=1.9,~N=100.$}
		\label{fig:exceptional_Kuramoto}
		\centering
	\end{figure*}

\cyan{Topological boundary matrix}-- The presence of the multiple GBZs promotes the spinor representations of the eigenstates. For example, even though the system consists of a single site unit cell, the Bloch wave function can be represented as the expanded spinor $(c_1,c_2)^\textrm{T}$, where each component represents the coefficients of the paired momenta $z_1$ and $z_2$ respectively. Correspondingly, the boundary equation can be transformed as the eigenvalue equation, $\tilde{H}_B(c_1,~c_2)^\textrm{T}=(c_1,~c_2)^\textrm{T}$, with the effective chiral symmetry as, 
	\begin{align}
	\label{Eq:Hb}
	\tilde{H}_B=&\begin{pmatrix}
	0& 	h_B^+(z_1,z_2) \\
	h_B^-(z_1,z_2)& 0
	\end{pmatrix}
	\end{align}
	where $ h_B^+(z_1,z_2)=A(z_2)/A(z_1)$, $ h_B^-(z_1,z_2) = B(z_1)/B(z_2)$. The transformed boundary matrix $\tilde{H}_B$ preserves the chiral symmetry by satisfying the following conditions,  $\{ \tilde{H}_B ,\sigma_z \}=0$. The presence of the chiral symmetry indicates that for any right (left) eigenstates $|\chi_{R(L)}\rangle$ with the complex eigenvalue $E (E^*)$, we can define the right (left) chiral partner states $|\tilde{\chi}_{R(L)}\rangle \equiv\sigma_z|\chi_{R(L)}\rangle$ with the eigenvalue $-E (-E^*)$,  which satisfy the bi-orthogonality relations: $\langle \chi_L|\chi_R\rangle=\langle \tilde{\chi}_L|\tilde{\chi}_R\rangle=1, \langle \tilde{\chi}_L|\chi_R\rangle=\langle \chi_L|\tilde{\chi}_R\rangle=0$.
 
There is a suggestive similarity between the boundary equation [Eq. \eqref{Eq:Hb}] and the Bloch Hamiltonian of the non-Hermitian SSH model \cite{Wang_non_bloch_TI_2018,Murakami_non_bloch_TI_2019}. We can assign the topological winding number $W$ to classify the boundary matrix as,
\bea
\label{eq:non_bloch_TI}
W=\frac{W_+-W_-}{2},~\quad~W_{\pm} = \frac{1}{2\pi}[\textrm{arg}~h_B^{\pm}]_{\mathcal{L}_1\times \mathcal{L}_2},
\eea
where $[\textrm{arg}~h_B^{\pm}]_{\mathcal{L}_1\times \mathcal{L}_2}$ is the change of phase of $h_B^{\pm}$ as $(z_1,z_2)$ goes along the GBZ. In general, another non-zero topological invariant, $W'=(W_++W_-)/2$, can exist. However, for $(z_1,z_2)\in \mathcal{L}_1\times\mathcal{L}_2$, the GBZ ensures that $W_+=-W_-$ due to the boundary equation $\det(H_B)=0$. As a result, $W$ can only have non-trivial topological numbers. The physical manifestation of the non-trivial winding number manifests as the topological bound state localized on the boundary of the chain [See Fig.~\ref{fig:GOBC} (c) and also Sec. \ref{sec:top_GBZ} for more details on the symmetry and topology of the boundary matrix ]. 

In the limit of GBC-(I) ($t'_{L(R)}= 0,~\epsilon_1,\epsilon_{N}\neq 0$), the boundary matrix $H_B$ becomes Hermitian, where the winding number is given by the conventional Berry phase. In other GBCs,  $\mathcal{L}_{1}$ and $\mathcal{L}_{2}$ can form two disjoint contours in the complex plane. At the topological phase transition of the winding number $W_{\rm non-Bloch}$, where $h^+_B(z_1,z_2)$ or $h^-_B(z_1,z_2)$ vanishes, the topological boundary states absorb into the continuum of the bulk bands in the complex momentum plane. The merging of the two generalized momenta manifests as the exceptional point, which is signatured by the coalescence of distinct eigenstates [See Fig.~\ref{fig:GOBC} Bottom left]. We refer to the appearance of the exceptional point during the topological phase transition of the GBZ as the exceptional phase transition. We point out that the observed topological boundary mode defies the standard phenomenology of the conventional topological insulators as our model only contains a single band (See \ref{sec:Exceptionaltransitions} for explicit studies in the general boundary conditions).

	In general non-Hermitian one-dimensional systems, the one-dimensional non-Hermitian Hamiltonian can be written as, 
	\bea
	\hat{H}=\sum_{i=1}^{N}\sum_{\alpha,\beta=1}^{p}\sum_{m=- n_\textrm{L}}^{n_\textrm{R} }t_{m,\alpha\beta}\hat{c}^{\dagger}_{i+m,\alpha}\hat{c}_{i,\beta},
	\eea 
	where $\alpha,\beta$ takes values from $[1,~p]$ with $p$ being total number of internal degrees of freedom. $t_m$ describes the hopping from $i$-th site to $i+m$-th site. The Hamiltonian can be represented as $(n_{\textrm{L}}+n_{\textrm{R}})$-order polynomial, and the wavefunction represented as $\psi_{n\alpha}=\sum_{i=1}^{p(n_{\textrm{L}}+n_{\textrm{R}})}c_{i\alpha} z_{i\alpha}^n$ is required to simultaneously satisfy the boundary equation, $H_{\textrm{B}}(z_{i\alpha})c_{i\alpha}=0$ with $\alpha=1,...,~p$. By requiring the satisfaction of the boundary equation, each point in the GBZs is described by the effective eigenspinor $(c_{11},...,~c_{1p},~c_{21},...,~{c_{2p}},...,~c_{(n_{\textrm{L}}+n_{\textrm{R}})p})^T$, following the symmetry group of the $H_{\textrm{B}}$. The non-trivial homotopy group features the intrinsic topology of the GBZ. 

	\cyan{Generalization to many-body systems}-- We consider the one-dimensional array of the Kuramoto oscillators as a generalization to the many-body systems \cite{Ritort_review_Kuramoto_2005}. 
	\bea
	\label{Eq:Kuramoto_model}
	d\theta_i(t)/dt = 
	K_R \sin(\delta\theta_{i+1,i}) 
	+ K_L \sin(\delta\theta_{i-1,i}) ,
	\eea
Here,  the dynamics of $i$-th oscillator is described by the phase $\theta_i(t)$. $K_{R, L}$ is the coupling with the right and left nearest-neighbor oscillator. $\delta\theta_{ij}=\theta_{i}-\theta_j$ is the phase difference. The model admits the stable limit cycles with the constant phase gradient $\delta\theta$. Due to the PBC, the phase gradient is quantized as,
	$\omega=\sum_{i=1}^N \delta\theta_{i+1,i}\in \mathbb{Z}$. (See supplementary materials for the exact solutions and stability analysis).

	The linearized excitation near the limit-cycle phase is described by the Jacobian of Eq. \eqref{Eq:Kuramoto_model}, $ \mathcal{J}_{i j}[\theta ]=-\partial_{i} (\partial_t \theta_j)$, which is given as,
	\bea
	\mathcal{J}&=&\sum_{i=1}^N K_{R} \cos (\delta\theta_{i+1,i})| i+1\rangle\langle i|+K_{L} \cos (\delta\theta_{i-1,i})| i-1\rangle\langle i|
	\nonumber
	\\
	&-&(K_{R}\cos (\delta\theta_{i+1,i})+K_{L}\cos (\delta\theta_{i-1,i})) | i\rangle\langle i|.
	\label{Eq:jacobian}
	\eea
	 The Jacobian of the limit cycle phases emulates the HN model with the PBC [Fig.~\ref{fig:exceptional_Kuramoto}(a)]. The eigenvalue spectra of the Jacobian form a center-shifted circle in a complex plane comprised of negative real eigenvalues [Fig.~\ref{fig:exceptional_Kuramoto}(e)]. The negative eigenvalue spectrum of the Jacobian characterizes the stability of the limit cycle in non-linear systems \cite{Strogatz_sync_basin_2006, max_no_fixed_points_1_ochab_2010}.

	\begin{figure}[htbp]
		\centering
		\includegraphics[width=1\linewidth]{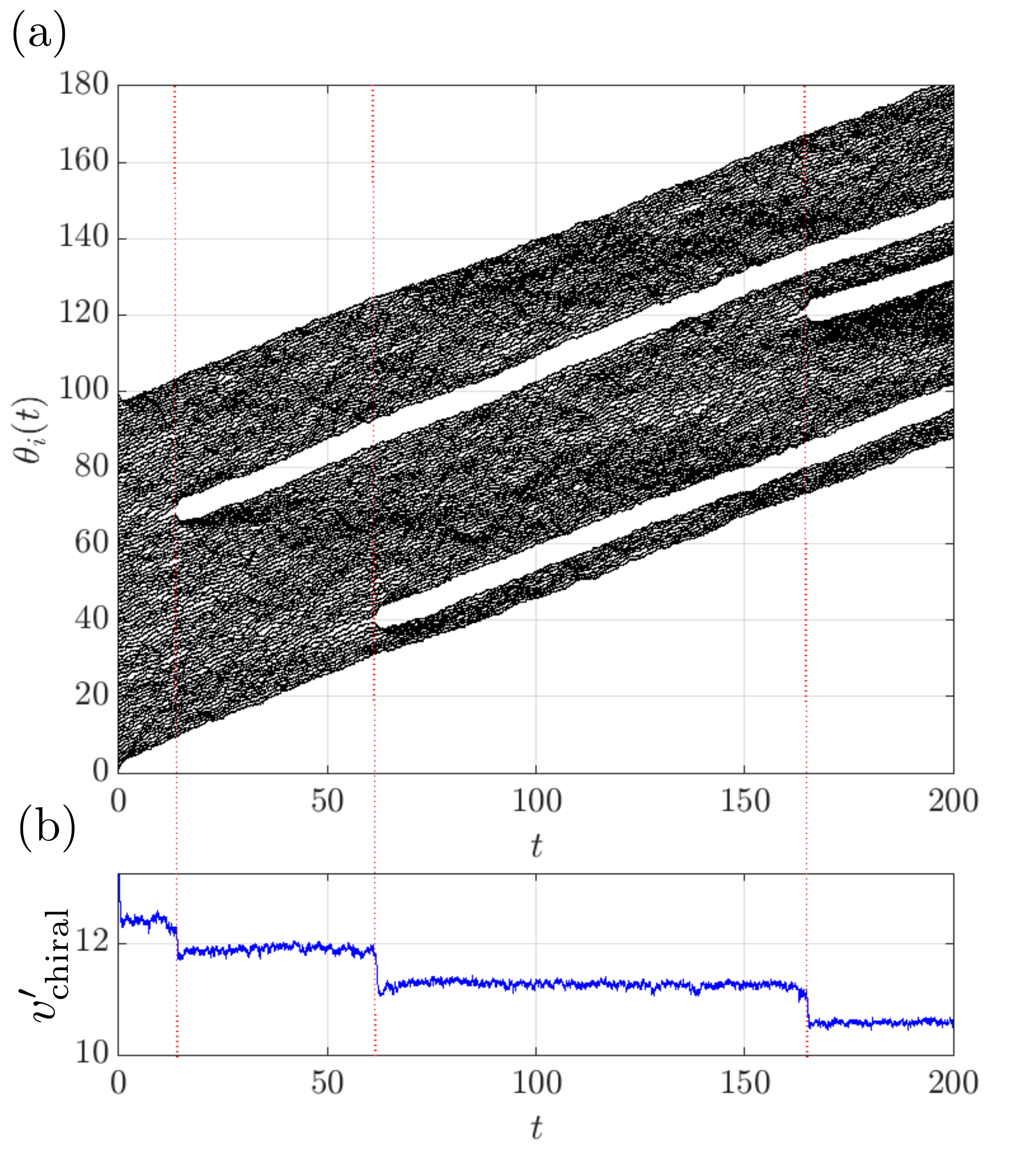}
		\caption{Noise-induced phase transition: (a) time evolutions of phases in the presence of random white noise. The dotted lines mark the positions of phase slips which defines the phase transition in between different chiral phases. (b) Time evolution of the total chiral velocity ($v'_{\rm chiral}=v_{\rm chiral}/(K_R-K_L)$) in the presence of random white noise. The discrete change of the total drift velocity serves as the physical observable of noise-induced non-equilibrium phase transition (See text). Parameters: $K_R=2.5,~K_L=1.9,~D=0.15,~N=100.$}
		\label{fig:noise_phase_transition}
		\centering
	\end{figure}
 
	The phase transitions between the limit cycles with different phase gradients (say $\omega$ and $\omega'$) accompany the phase slip of $2\pi(\omega-\omega')$, where the phase slipping sites can be considered as the boundary.  During $\frac{\pi}{2}<\delta\theta_{N}<\frac{3\pi}{2}$, the Jacobian matrix undergoes the topological phase transition of GBZ and it is characterized by non-trivial winding number. The corresponding topological boundary state is energetically separated from the negative-eigenvalue bulk states and forms the positive real eigenvalue [See Fig.~\ref{fig:exceptional_Kuramoto}(d)], which physically represents the onset of the limit-cycle instability during the phase transitions.
	
	In addition to the negative eigenvalue bulk states, we additionally identify the Goldstone mode associated with the global U(1)  symmetry ($\theta_i \rightarrow \theta_i+\phi$). Regardless of the phases of oscillators, the Jacobian matrix forms a zero-line sum (ZLS) matrix \cite{Boukas_zero_line_sum_matrices}, which has zero-sum of elements in each row and column \textit{i.e.} $\sum\limits_{i=1}^N \mathcal{J}_{i j} = \sum\limits_{j=1}^N\mathcal{J}_{i j}=0$. The physical manifestation is the emergence of Goldstone mode, $|\psi_{\textrm{gs}}\rangle \sim (1,1,1,..1)^T$ with the zero eigenvalue $E=0$. In GBZ, the Goldstone mode touches the singular point of the BZ, $(z_1,z_2)=(1,r^2)$, where $h_B^\pm$ is ill-defined.

	At the topological phase transition, $\delta\theta_{N}=\frac{\pi}{2},\frac{3\pi}{2}$, the Jacobian exhibits the OBC. The GBZ of the bulk and the topological boundary state meet each other [Fig. \ref{fig:exceptional_Kuramoto}(c)]. Correspondingly, we observe the coalescence of the Goldstone mode with the normal modes, observed by the vanishing phase rigidity, 
	$
	R_1= \frac{\langle \psi_{1}^{R}|\psi_{1}^{L}\rangle}{\langle \psi_{1}^{R}|\psi_{1}^{R}\rangle},
	$
	where $|\psi_{1}^{L(R)}\rangle$ are the left (right) eigenmodes corresponding to the first eigenstate. The vanishing phase rigidity can be understood as the consequence of the NHSE where the left eigenstates and the right eigenstates are localized at the opposite boundaries, causing the vanishing wavefunction overlap (See supplementary materials). This coalescence of the Goldstone mode signifies the non-equilibrium phase transition, which has been referred to as ETs \cite{Fruchart2021}. The EP of the Goldstone mode proliferates the temporal fluctuations associated with U(1) symmetry. The physical consequence is the dynamical restoration of  the spontaneously broken U(1) rotation symmetry.
 
    We propose that the overall chiral velocity serves as a macroscopic order parameter that characterizes such topological non-reciprocal phase transitions,
    \bea
    v_{\textrm{chiral}}=\sum_{i=1}^N\frac{d\theta_i (t)}{dt} =             \omega(K_{R}-K_{L}) .
    \eea
    The finite chiral motion occurs for each limit-cycles due to the non-reciprocity of the couplings. In the previous work \cite{Fruchart2021}, a similar chiral motion has been studied in the presence of random noise. Here, we note the crucial difference that the stability of the chiral motion does not require random noise due to the peculiar property coming from periodic boundary conditions. Instead, the presence of random noise destabilizes the chiral motion by inducing phase slips. During the phase slip, the ET proliferates  the Goldstone mode excitations inducing the abrupt change in the chiral velocity [See Fig.~\ref{fig:noise_phase_transition}]. In the long time limit, the system eventually stabilizes into the limit cycle with the zero winding number.

	\cyan{Discussions -} 	The non-trivial topology of GBZ we reveal here can be readily realized in designable material platforms such as active matter \cite{Ramaswamy_active_matter_2022, Vitelli_topological_active_matter_2022}, metamaterials in the fields of optics and photonics \cite{metamaterial_optics_photonics_2019}, acoustics \cite{metamaterial_acoustic_2014}, robotics \cite{Non_reciprocal_robotic_metamaterial_2019}, and electric circuit networks \cite{metamatrial_topoelectric_circuit_2020}. In a more general sense, the excitations living on systems with the translational symmetry-breaking defect can be the subject of our study. We have generalized our result to many-body collective behaviors by proposing non-reciprocally coupled oscillators. Various dynamical systems described by coupled non-linear oscillators such as active matter, coupled Josephson junctions, and XY spin model belong to this class of dynamical systems \cite{ Ramaswamy_active_matter_2022, Vitelli_topological_active_matter_2022, Synchronization_josephson_strogatz_1998, Ramaswamy_driven_XY_2002}. In condensed matter systems, recently proposed studies of spin dynamics with the spin current can be an interesting platform to realize the ET of collective phenomena  \cite{Tokura_condensed_matter_nonreciprocal_2018}.

	
	\section*{Acknowledgments}
	M.J.P. and S.V. acknowledge financial support from the Institute for Basic Science in the Republic of Korea through the project IBS-R024-D1. This work was supported by the research fund of Hanyang University(HY-202300000001149).

    \bibliography{reference}

    \clearpage
    \newpage

    \renewcommand{\thefigure}{S\arabic{figure}}
    \setcounter{figure}{0}
    \renewcommand{\theequation}{S\arabic{equation}}
    \setcounter{equation}{0}

\begin{widetext}
\tableofcontents

\clearpage
\newpage

\section*{Supplementary Materials}

\section{Non-Bloch band theory in generalized boundary conditions}

\subsection{Classifications of generalized boundary conditions}\label{sec:def_GBC}

The non-Bloch band theory \cite{Murakami_non_bloch_TI_2019, chenGBC2021} describes the band structures of non-Hermitian systems in generalized boundary conditions (GBC), where the conventional real Bloch wave vector is replaced by the generalized wave vector lies in the complex plane. Here, we define the generalized Brillouin zone (GBZ) as the set of the complex wave vector that constitutes the eigenstates. As a representative example, we consider the one-dimensional Hantano-Nelson model \cite{Hatano_Nelson_1996} in the GBC. The full Hamiltonian operator can be decomposed as the  bulk term ($\hat{H}_{\rm bulk}$) and the boundary term ($\hat{H}_{\rm boundary}$) as follows,

    \begin{align}\label{supp:general_model_ham}
	\hat{H}&=\hat{H}_{\rm bulk}+\hat{H}_{\rm boundary}\nonumber\\
	&=\Big(\sum_{i=2}^{N-1}t_R| i\rangle\langle i-1|+t_L| i\rangle\langle i+1|+\epsilon_0|i\rangle\langle i|\Big)
	+\Big(t_L| 1\rangle\langle 2|+ t_R^{\prime}| 1\rangle\langle N|+\epsilon_1|1\rangle\langle 1|+t_R| N\rangle\langle N-1|+t_L^{\prime}| N\rangle\langle 1| + \epsilon_N|N\rangle\langle N|\Big),
    \end{align}
	where $|i\rangle \equiv \hat{c}^{\dagger}_i|0\rangle$ is the localized state at $i$-th site. The matrix elements of the boundary term demonstrate various GBCs. For example, $(t_R^{\prime},t_L^{\prime},\epsilon_1,\epsilon_N)=(t_R,t_L,\epsilon_0,\epsilon_0)$ corresponds to the periodic boundary condition (PBC), while $(t_R^{\prime},t_L^{\prime},\epsilon_1,\epsilon_N)=(0,0,\epsilon_0,\epsilon_0)$ corresponds to the open boundary condition (OBC). In a general setting, different combinations of onsite potentials $e_1,~e_N$ and hopping terms $t_R^{\prime}$, $t_L^{\prime}$ fully describe the possible sets of the GBCs.  Table~\ref{tab:def_GBC} summarizes the definitions of the different types of boundary conditions.

 \begin{figure}[htbp]
		\centering
		\includegraphics[width=0.65\linewidth]{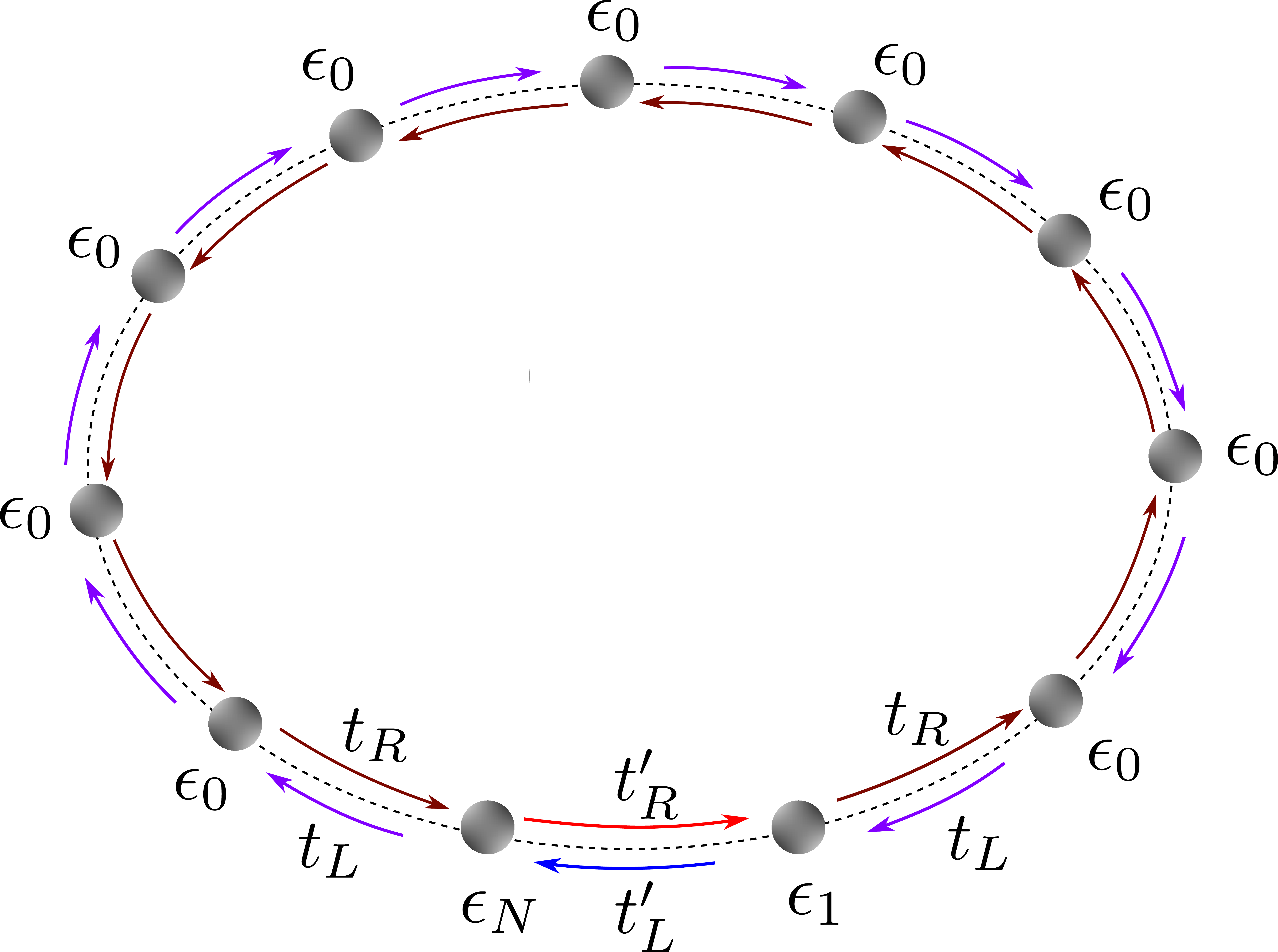}
		\caption{Hatano-Nelson model for 1D chain in generalized boundary conditions.}
		\label{fig:GBC_scheamatic}
\end{figure}
 
    \begin{center}
	\begin{table}[h]
    \def\arraystretch{1.5}
    \begin{tabular}{|l|l|}
    \hline
    Open boundary condition (OBC) & $t'_R=0,~t'_L=0$ and $\epsilon_1=\epsilon_N=\epsilon_0$ \\
     \hline
    Periodic boundary condition (PBC) & $t'_R=t_R,~t'_L=t_L$ and $\epsilon_1=\epsilon_N=\epsilon_0$ \\
    \hline
    Generalized boundary conditions (GBC) & All possible combinations of $t'_R,~t'_L,~\epsilon_1,~\epsilon_N$ \\
    (i) Generalized boundary conditions-(I) (GBC-(I)) &$t'_R=0,~t'_L=0$ and $\epsilon_1,~\epsilon_N\neq\epsilon_0$ \\
    (ii) Generalized boundary conditions-(II) (GBC-(II)) & $t'_R=t_R,~t'_L=t_L$ and $\epsilon_1,~\epsilon_N\neq\epsilon_0$ \\
    (iii) Generalized boundary conditions-(III) (GBC-(III)) & $t'_R<t_R$, $t'_L<t_L$ and $\epsilon_1,~\epsilon_N=\epsilon_0$ \\
    (iv) Generalized boundary conditions-(IV) (GBC-(IV)) & $t'_R>t_R$, $t'_L>t_L$ and $\epsilon_1,~\epsilon_N=\epsilon_0$ \\
    \hline
    \end{tabular}
    \caption{Definitions of different types of generalized boundary conditions.}\label{tab:def_GBC}
    \end{table}
    \end{center}

	\subsection{Detailed methods of non-Bloch band theory}\label{sec:non_Bloch_theory}
	
	In this section, we describe the detailed methods of the non-Bloch band theory \cite{Murakami_non_bloch_TI_2019, chenGBC2021} to find the GBZ. The eigenstate of Eq. \eqref{supp:general_model_ham}, $|\psi\rangle=\sum_i \psi_i | i \rangle$, with the eigenvalue $E$ is required to simultaneously satisfy the bulk and the boundary equations. We analyze the bulk and the boundary equations separately.
	
	\textbf{Bulk equation}: The bulk equations are given as,
	
	\begin{align}\label{bulk_eqn_g}
	t_R \psi_{i-1} + (\epsilon_{0} - E) \psi_i + t_L \psi_{i+1}  = 0 ~~{\rm for}~~i\in[2,N-1],
	\end{align}
	To solve for the bulk equation, we take the single wave vector ansatz of the wave function, $\psi_n \propto z_j^n$, where the bulk equation transforms as,
	\begin{align}\label{supp:bulk_poly_eqn}
	t_{R} z_{j}^{i-1} + (\epsilon_{0} - E) z_{j}^i + t_L z_{j}^{i+1} = 0.
	\end{align}

There are the two solutions of the ansatz for a given eigenvalue $E$, which constitute the paired momenta as,
	\begin{align}
	z_{j} \equiv z_{1,2} = -\frac{(\epsilon_0 -E)}{2t_L} \pm \sqrt{\Bigg(\frac{\epsilon_0 -E}{2 t_R}\Bigg)^2-\Bigg(
		\frac{t_R}{t_L}\Bigg)}, 
	\end{align}
	In addition, regardless of the eigenvalue $E$, the two solutions are related as,
	\begin{align}
	\label{eqs:z1z2}
	z_{1}z_{2} = \frac{t_L}
	{t_R}\equiv r^2, ~~{\rm with}~~ r=\sqrt{\frac{t_R}{t_L}},
	\end{align}
	which provide the relations between the two paired momenta. In general, the linear superposition of the two ansatz forms general solutions of the bulk equations, 
	
	\bea
	\label{supp:bulk_ansatz}
	\psi_n=c_1 z_1^n +c_2 z_2^n.
	\eea

	\textbf{Boundary equation}: The boundary equations are given as,
	\begin{align}\label{boundary_eqn_g_i}
	t_{R}^{\prime} \psi_{N} + (\epsilon_{1} - E) \psi_1 + t_L \psi_{2}  &= 0, \\
	t_{L}^{\prime} \psi_{1} + (\epsilon_{N} - E) \psi_N + t_R \psi_{N-1}  & = 0 .
	\end{align}
With the help of bulk equations, we simplify the boundary equations which take the following form,
	\begin{align}\label{boundary_eqn_g_f}
	  t_R^{\prime} \psi_{N}+(\epsilon_1 - \epsilon_{0}) \psi_1 -t_R \psi_{0}  &= 0,\nonumber\\
	t_L^{\prime} \psi_{1} + (\epsilon_N-\epsilon_{0}) \psi_{N} - t_L \psi_{N+1}  &= 0.
	\end{align}

By plugging in the bulk ansatz in Eq. \eqref{supp:bulk_ansatz}, the boundary equation [Eq. \eqref{boundary_eqn_g_f}] can be compactly rewritten as the form of the eigenvalue equation,
	\begin{align}
	\label{eqs:H_b}
	H_B \begin{pmatrix}
	c_1\\
	c_2\end{pmatrix} =0, ~~{\rm with}~~ H_B = \begin{pmatrix}
	A(z_1)& A(z_2) \\
	B(z_1)& B(z_2)
	\end{pmatrix},
	\end{align}
	where $A(z) = t_R^{\prime} z^{N}+ (\epsilon_1-\epsilon_0)z -t_R$, $B(z) = t_L^{\prime}z +(\epsilon_N-\epsilon_0)z^N - t_L z^{N+1}$.

\textbf{Eigenstate Solutions :}	We derive the non-zero solution of the boundary equation [Eq. \eqref{eqs:H_b}]. The non-zero solutions are obtained by solving for the equation, $\det[H_B] =0$, which takes the explicit form as follows,

	\begin{align}\label{eqn_for_sol_g_z}
	(z_2^{N+1}-z_1^{N+1})-\Bigg(
	\frac{(\epsilon_1-\epsilon_0)+(\epsilon_N-\epsilon_0)}
	{t_L}\Bigg)(z_2^N-z_1^N) &+ \Bigg(
	\frac{(\epsilon_1-\epsilon_0)(\epsilon_N-\epsilon_0)-t_R^{\prime}t_L^{\prime}}
	{t_L^2}\Bigg)(z_2^{N-1}-z_1^{N-1})\nonumber\\
	&-
	\Bigg(\frac{t_L^{\prime}}{t_L}+\frac{t_R^{\prime}}{t_R}r^{2N}\Bigg)(z_2-z_1)=0.
	\end{align}
By solving the above boundary equation with the complementary condition [Eq. \eqref{eqs:z1z2}], we derive the spectrum of the generalized Brillouin zones. 
	
The paired momenta satisfying Eq. \eqref{eqs:z1z2} can be expressed in the polar coordinates as follows,
\begin{align}
\label{eqn_for_r}
z_1 = r e^{i\phi}, ~~z_2 = r e^{-i\phi}, 
\end{align}
where $\phi\in\mathbb{C}$ is a complex number. When the solutions of $\phi$ are real number, the GBZs of the two paired momenta coincides. Accordingly, Eq.~\eqref{eqn_for_sol_g_z} can be rewritten as the following form,
\begin{align}\label{eqn_for_sol_g_phi}
\sin[(N+1)\phi]-\eta_0\sin[N\phi]+\eta_1\sin[(N-1)\phi]-\eta_2\sin[\phi]=0,
\end{align}
	where
	\begin{align}
	\eta_0 = \Bigg(
	\frac{(\epsilon_1-\epsilon_0)+(\epsilon_N-\epsilon_0)}
	{t_L r}\Bigg),~~\eta_1 = \Bigg(
	\frac{(\epsilon_1-\epsilon_0)(\epsilon_N-\epsilon_0)-t_R^{\prime}t_L^{\prime}}
	{t_R t_L}\Bigg),~~\eta_2 = \Bigg(\frac{t_L^{\prime}}{t_L}r^{-N}+\frac{t_R^{\prime}}{t_R}r^{N}\Bigg).
	\end{align}
In practice, we solve Eq.~\eqref{eqn_for_sol_g_phi} to find the solutions of the GBZ.
	
 It is important to point out that, when $\phi=0 $ or $\pi$, the two paired momenta are equal, $z_1=z_2 =r$ or $-r$. Accordingly, Eq. \eqref{eqs:H_b} leads to the condition $c_1=-c_2$, which indicates the vanishing wave function, $\psi_n = 0$ (\textit{i.e.} non-physical solution). However, this conclusion is not correct. For the equal solutions $z_1 = z_2 = z $, the general solution of the bulk equation is given by the form of $\psi_n = (c_1 + c_2 n)z^n$ rather than $\psi_n = c_1 z_1^n + c_2 z_2^n$.
 
Consequently, the boundary equations will be modified which are expressed as follows,
	\begin{align}
	H_B^{\prime} \begin{pmatrix}
	c_1\\
	c_2\end{pmatrix} =0, ~~{\rm with}~~ H_B^{\prime} = \begin{pmatrix}
	t_R - (\epsilon_1-\epsilon_{0})z-t_R^{\prime} z^{N}& 	 - (\epsilon_1-\epsilon_{0})z-N t_R^{\prime} z^{N} \\
	t_L^{\prime} z + (\epsilon_{N}-\epsilon_{0}) z^N -t_L z^{N+1}& t_L^{\prime} z +N (\epsilon_{N}-\epsilon_{0}) z^N -(N+1)t_L z^{N+1}
	\end{pmatrix}.
	\end{align}
 where $\psi_n = (c_1 +c_2 n)z^n$ is the general solution of the eigenstates.

	\begin{figure}[htbp]
		\centering
		\includegraphics[width=1.0\linewidth]{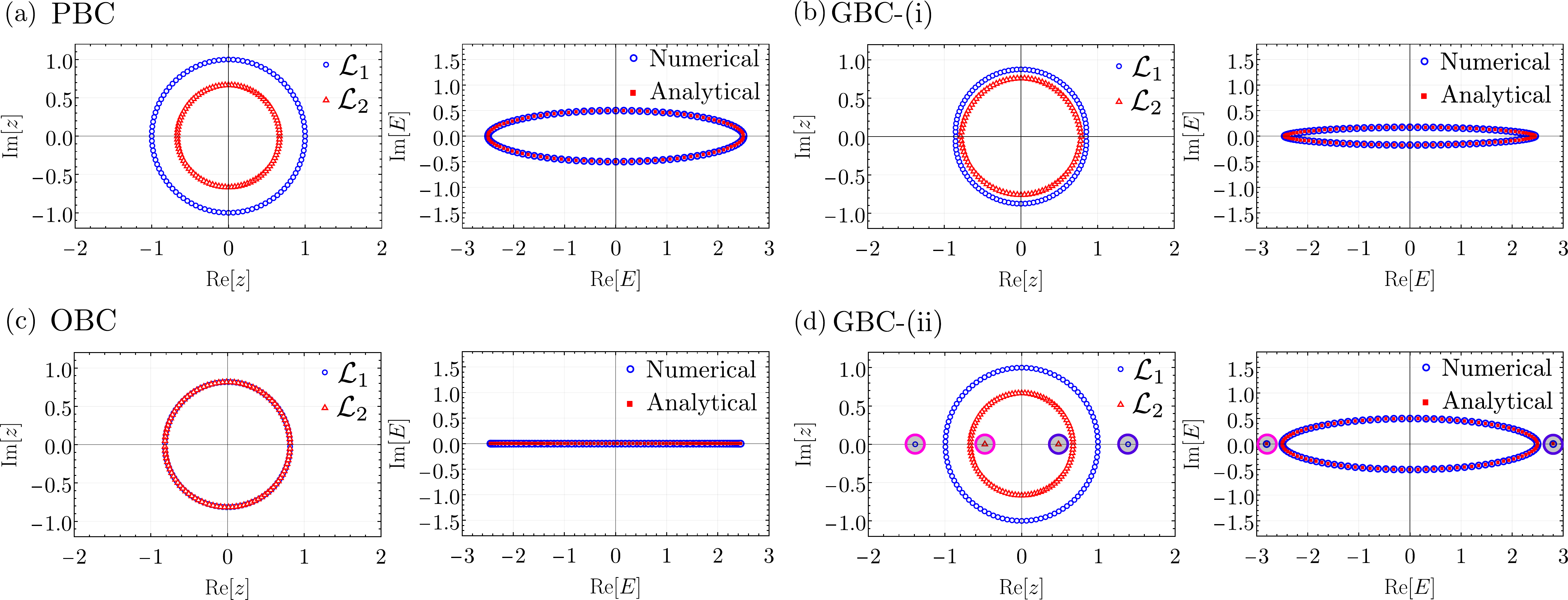}
		\caption{Generalized Brillouin zones $\mathcal{L}_1,~\mathcal{L}_2$ (left) and  eigenvalue spectra (right) for different boundary conditions: (a) PBC, (b) GBC-(III), (c) OBC, (d) GBC-(IV). In the right panel of (d), the real eigenvalues which are marked with circles represent the localized boundary states due to GBC. These boundary states are characterized by the disjoint generalized momenta outside or inside GBZs which are also marked with circles in the left panel of (d). Parameters: $N=100,~t_L=1.5,~t_R=1.0,~\epsilon_1= 0,~\epsilon_N=0.$ GBC-(i): $t'_R<t_R,~t'_L<t_L$. GBC-(ii): $t'_R>t_R,~t'_L>t_L$.}
		\label{fig:GBZ_energy_GBC}
	\end{figure}

\subsection{Representative examples of Generalized Brillouin zones}\label{sec:examples_GBZ}
In this section, we present the representative solutions of the GBZs in GBCs. We determine the eigenvalues and corresponding eigenstates for a finite-size system in the generalized boundary condition. Using the paired momenta described by Eq. \eqref{eqn_for_r}, the eigenvalues $E$ and corresponding wave functions can be expressed in terms of generalized momenta as follows,
\begin{align}
E&=\epsilon_0 + t_L z_i + t_R z_i^{-1}
=\epsilon_0 + 2\sqrt{t_R t_L}\cos\phi,~\text{with} ~\phi\in \mathbb{C},
\end{align}
\begin{align}
\psi_n &= c_1 z_1^n + c_2 z_2^n
=2r^n((c_1+c_2)\cos[n\phi]+(c_1-c_2)i\sin[n\phi]).
\end{align}
	
Fig.~\ref{fig:GBZ_energy_GBC} compares the calculated GBZs with the result of numerical exact diagonalizations of the Hamiltonian. We find that the two methods agree well. We detail the results of some of the boundary conditions as follows:
	
\subsubsection{Periodic boundary condition}
	In this section, we describe the GBZs and eigenvalue spectra for periodic boundary condition (PBC) defined with $t_L^{\prime}=t_L,~t_R^{\prime}=t_R$, and $\epsilon_1=\epsilon_N = \epsilon_0$. In this case, Eq.~\eqref{eqn_for_sol_g_phi}
	\begin{align}\label{eqn_for_sol_g_phi_PBC}
	\sin[(N+1)\phi]-\sin[(N-1)\phi]+(r^N+r^{-N})\sin[\phi]&=0, 
	\end{align}
	 have $N$ physical solutions $\phi_m =\pm i \log r\pm 2 m\pi/N,~ m\in \lbrace 0,1,...,N-1\rbrace$. Accordingly, we have $N$ pairs of distinct generalized momenta $ (z_1,~z_2)=( r e^{i\phi_m},~r e^{-i \phi_m})$, which depict the two distinct GBZs with different radii with $|z_1|\neq|z_2|$ in the complex momenta plane (See Fig.~\ref{fig:GBZ_energy_GBC} (a)). 
  The corresponding eigenvalue spectra are given by,
	\begin{align}
	E_m =2\sqrt{t_L t_R}\cos(\phi_m), ~\phi_m =\pm i \log r\pm 2 m\pi/N, ~m\in \lbrace 0,1,...,N-1\rbrace.
	\end{align}
	which forms a closed curve (an ellipse) in the complex eigenvalue plane and agrees with that of numerical diagonalization (See Fig.~\ref{fig:GBZ_energy_GBC}(a)). The boundary equations together with GBZs determine the wave functions which take the following form,
	\begin{align}
	\psi_n = c_1 r^n e^{i\phi_m}, ~\phi_m = i \log r+ 2 m\pi/N, ~m\in \lbrace 0,1,...,N-1\rbrace.
	\end{align}
	
	\subsubsection{Open boundary condition}
	Now, we consider open boundary condition (OBC) with parameters $t_L^{\prime}=t_R^{\prime}=0$, and $\epsilon_1=\epsilon_N = \epsilon_0$. In this case, Eq.~\eqref{eqn_for_sol_g_phi} can be simplified as
	\begin{align}\label{eqn_for_sol_g_phi_OBC}
	\sin[(N+1)\phi]&=0,
	\end{align}
which possesses $N$ real physical solutions of $\phi_m =m\pi/(N+1)$, where $m\in \lbrace 1,2,...,N\rbrace$. Thus we have $N$ pairs of generalized momenta $(z_1,~z_2)=( r e^{i\phi_m},~r e^{-i \phi_m})$ with $|z_1|=|z_2|=r$, and as result the two GBZs coincide (See Fig.~\ref{fig:GBZ_energy_GBC}). The eigenvalue spectra are given by,
	\begin{align}
	E_m =2\sqrt{t_L t_R}\cos(\phi_m), ~\phi_m&=m\pi/(N+1), ~m\in \lbrace 1,2,...,N\rbrace.
	\end{align}
	which form an open segment in the complex eigenvalue plane and collapse on the real eigenvalue axis for the present model (See Fig.~\ref{fig:GBZ_energy_GBC} (c)). Also from boundary equations, we have $c_1=-c_2 $ and thus the wave functions become,
	\begin{align}
	\psi_n =2 i c_1 r^n\sin(n\phi_m), ~\phi_m&=m\pi/(N+1), m\in \lbrace 1,2,...,N\rbrace.
	\end{align}
	It is clear from the above expressions that all the eigenstates localize at one of the boundaries of the system depending on whether $r<1$ or $ r>1$. This anomalous localization phenomenon is known as the NHSE~\cite{Ueda_review_non_Hermitian_2021, Sato_review_topology_2022, Chen_Fang_review_NHSE_2022}. The eigenvalue spectra and corresponding eigenstates differ from that of PBC, which shows the extreme sensitivity of non-Hermitian systems to the boundary conditions~\cite{Ueda_review_non_Hermitian_2021, Sato_review_topology_2022, Chen_Fang_review_NHSE_2022}.

	\subsubsection{Generalized boundary conditions-(III)}\label{sec:GBC-(III)} 
	 In this section, we discuss GBZs and eigenvalue spectra for one of the generalized boundary conditions defined as $t_L^{\prime}\neq t_L,~t_R^{\prime}\neq t_R$ such that $t_L^{\prime}<t_L~(t'_R<t_R)$, and $\epsilon_1=\epsilon_N = \epsilon_0$. In this case, Eq.~\eqref{eqn_for_sol_g_phi} is written as
	\begin{align}
	\sin[(N+1)\phi]-\eta_1 \sin[(N-1)\phi]+\eta_2\sin[\phi]&=0,
	\end{align}
	which possesses $N$ different physical solutions $\phi_m\in \mathbb{C}$. We note that the solutions of Intermediate boundary condition-(i) have been discussed in Ref.~\cite{chenGBC2021}. Similar to Ref.~\cite{chenGBC2021}, Fig.~\ref{fig:GBZ_energy_GBC} (b) clearly shows that the two GBZs start approaching each other for smaller values of $t'_L/t_L$ and $t'_R/t_R$ which corresponds to the flattening of the eigenvalue spectra (i.e. approaching to the OBC).

 \subsubsection{Generalized boundary conditions-(IV)} 
 
 In this section, we consider the generalized boundary condition which is defined as $t_L^{\prime}\neq t_L,~t_R^{\prime}\neq t_R$ such that $t_L^{\prime}~(t'_R)$ can exceed $t_L~(t_R)$, and $\epsilon_1=\epsilon_N = \epsilon_0$. In this case, Eq.~(\ref{eqn_for_sol_g_phi}) can be written as,
	\begin{align}
	\sin[(N+1)\phi]-\eta_1 \sin[(N-1)\phi]+\eta_2\sin[\phi]&=0,
	\end{align}
 which has $N$ pairs of different physical solutions $\phi_m\in \mathbb{C}$. In contrast to the IBC-(i), we find the localized boundary states in addition to the states of the bulk GBZs (See the left panel of Fig. \ref{fig:GBZ_energy_GBC} (d)). More specifically, among total $N$ paired momenta, $N-2$ pairs of generalized momenta with $|z_1| \neq |z_2|$ form the continuum spectra in the complex plane. The two remaining pairs are isolatedly placed on the real momenta axis, \textit{i.e.} $z_1, ~z_2 \in \mathbb{R},$ (for $r<1$). As we explain later, these states correspond to the localized states characterized by the topological winding number. The eigenvalue spectra are again similar to that of PBC together with isolated real eigenvalues representing the localized boundary states due to GBC (See the right panel of Fig. \ref{fig:GBZ_energy_GBC} (d)).

\subsection{Topological classifications of GBZ}\label{sec:top_GBZ}

For each paired momenta ($z_1,z_2$) in GBZ, the real space wave function is described as $\psi_n (z_1,z_2) = c_1 z_1^n +c_2 z_2^n$ where the adiabatic change of the wave function can be described in terms of the effective spinor $|\chi\rangle =(c_1,~c_2)^T$. In this section, we show that the non-trivial topology of the GBZ manifests as the emergence of the topological boundary states.


	

\subsubsection{Symmetry of boundary equation}\label{sec:sym_HB}
 
Since the effective spinor satisfies the boundary equations $H_B(z_i)|\chi\rangle =0$, the symmetry group of the boundary matrix $H_B$ characterizes the topology of GBZ. The boundary equations can be rewritten as the eigenvalue problem,
	\begin{align}\label{eqn:Hb_effective}
	\tilde{H}_B (c_1,c_2)^{\rm T}= (c_1,c_2)^{\rm T},~~{\rm with}~~\tilde{H}_B=&\begin{pmatrix}
	0& 	h_B^+(z_1,z_2) \\
	h_B^-(z_1,z_2)& 0
	\end{pmatrix},
 \end{align}
	where $ h_B^+(z_1,z_2)=A(z_2)/A(z_1)$, $h_B^-(z_1,z_2) = B(z_1)/B(z_2)$. $A(z)=t_R-\epsilon_1 z - t'_R z^{N}$, $B(z)=t_L' z + \epsilon_N z^N -t_L z^{N+1}$. In general, the non-Hermitian effective boundary matrix satisfies the following chiral symmetry by construction,
 \bea
 \lbrace\sigma_z, \tilde{H}_B\rbrace=0.
 \eea
In the presence of the chiral symmetry, for any right (left) eigenstates $|\chi_{R(L)}\rangle$ with complex eigenvalue $E$ $(E^*)$, there always exists right (left) chiral eigenstates $|\tilde{\chi}_{R(L)}\rangle=\sigma_z |\chi_{R(L)}\rangle$ with eigenvalues $-E$ ($-E^*$) which satisfy the bi-orthogonality relations: $\langle \chi_L|\chi_R\rangle=\langle \tilde{\chi}_L|\tilde{\chi}_R\rangle=1, \langle \tilde{\chi}_L|\chi_R\rangle=\langle \chi_L|\tilde{\chi}_R\rangle=0$.

In addition, for GBC-(I) (See Table~\ref{tab:def_GBC}), $\tilde{H}_B$ becomes a Hermitian matrix since the meromorphic function $h_B^+$ is unimodular ($|h_B^+|=1$) together with $\det[H_B]=0$. Whereas, for other possible sets of generalized boundary conditions (See Table~\ref{tab:def_GBC}), the meromorphic function $h_B^+$ is not unimodular ($|h_B^+|\neq 1$), and hence $\tilde{H}_B $ remains a non-Hermitian matrix.

\subsubsection{Winding number of GBZ}\label{sec:winding}
     
The winding number of the boundary matrix can be defined using Cauchy's argument principle, which relates the number of zeros and poles of a meromorphic function, $h^\pm_B(z_1,z_2)$, to the contour integration of its logarithmic derivative as follows:
    \begin{align}\label{eq:winding_arg}
    W_{\pm}=\frac{1}{2\pi i}\oint_{\mathcal{L}_1\times\mathcal{L}_2}\frac{1}{h_B^{\pm}(z_1,z_2)}\frac{d h_B^{\pm}(z_1,z_2)}{dz_1} dz_1
    = \frac{1}{2\pi}[\textrm{arg}~h_B^{\pm}]_{\mathcal{L}_1\times \mathcal{L}_2}
    = Z - P,
    \end{align}
where the integration contour $\mathcal{L}_1\times\mathcal{L}_2$ indicates the continuous contour of the GBZs satisfying $z_1 z_2 = t_R/t_L$.  $[\textrm{arg}~h_B^{\pm}]_{\mathcal{L}_1\times \mathcal{L}_2}$ is the change of phase of $h_B^{\pm}$ as $(z_1,z_2)$ goes along the GBZ. $Z$($P$) represents the total number of zeros(poles) inside GBZs counting the multiplicities. Fig.~\ref{fig:GOBC_en_TPT}(a)-(c) shows exemplifies the GBZs with different winding numbers for GBC-(I) (See Sec.~\ref{sec:analytical_GOBC1} for more details). In the topologically trivial GBZ [$W=0$, Fig.~\ref{fig:GOBC_en_TPT}(a)], the interior of the GBZ encloses an equal number of poles and zeros of $h^+_B$. As the system undergoes the topological phase transition [Fig.~\ref{fig:GOBC_en_TPT}(b)], the zero of $h^+_B$ touches the GBZ. Finally, the GBZ encircles a pair of poles, which manifest as the non-trivial value of the winding number [Fig.~\ref{fig:GOBC_en_TPT}(c)].
     
In principle, the winding numbers $W_+$ and $W_-$ are unequal. Namely, we can separately define the winding numbers $W$ and $W'$ as,
\bea\label{eq:non_bloch_TI1}
W=\frac{W_+-W_-}{2},\quad W'=\frac{W_++W_-}{2}.
\eea
However, for $(z_1,z_2)\in \mathcal{L}_1\times\mathcal{L}_2$, the boundary equation $\det[H_B]=0$ ensure that $W_+=-W_-$. As a result, $W$ can only have non-zero numbers. In the next section, we discuss that there is an ambiguity in defining the topological invariant using Eq.~\eqref{eq:non_bloch_TI1}, and accordingly, we will re-define the topological invariant to characterize the topological phase transition of GBZ.  

 \begin{figure}[htbp]
		\centering
		\includegraphics[width=1.0\linewidth]{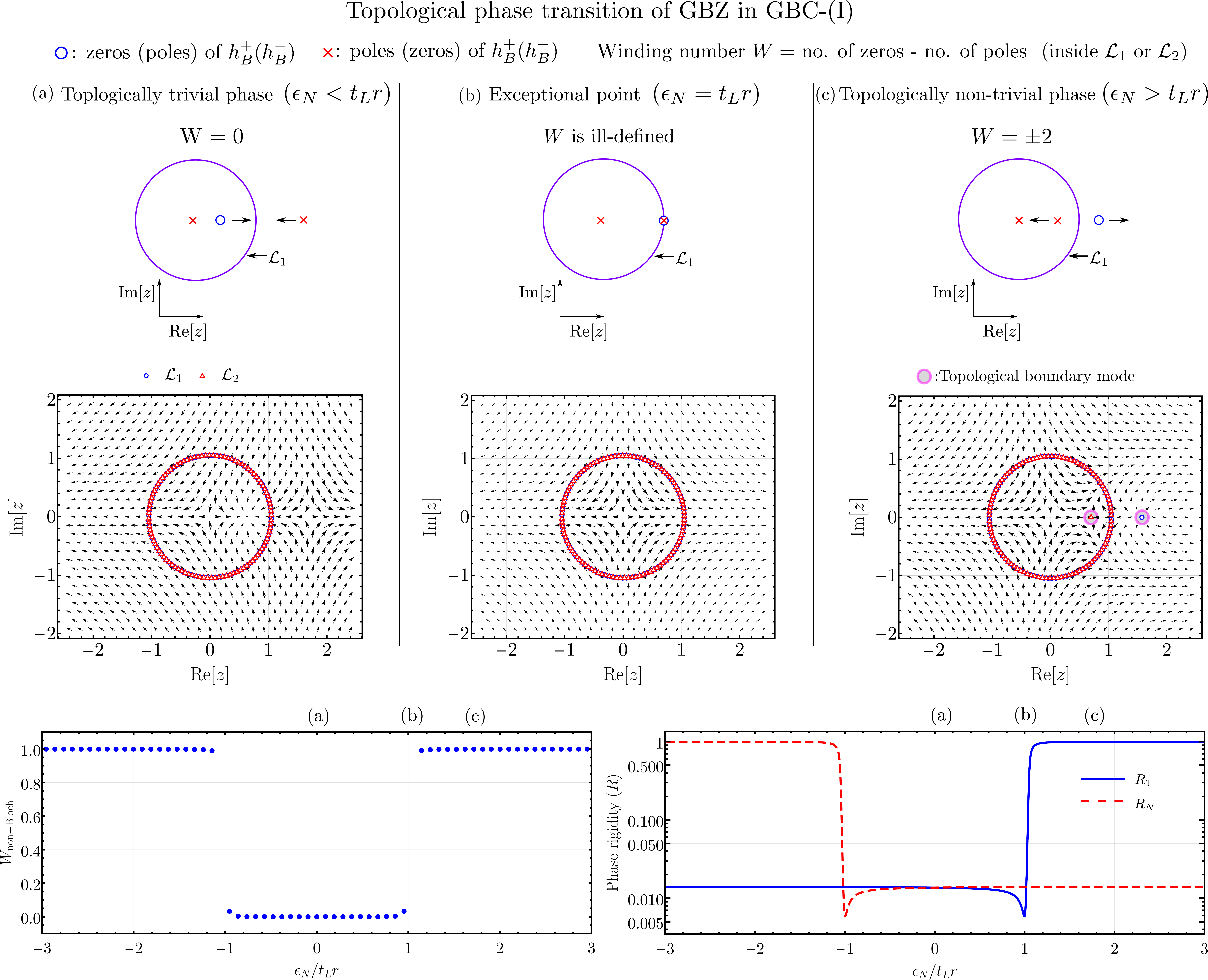}
		\caption{Topological phase transition of GBZ in GBC-(I). Top panel: Vector plot of the meromorphic function $h_B^-$ together with GBZ in the complex plane. We also show the schematics for the positions of poles and zeros of $h_B^+$ together with the contour of GBZ. For $\epsilon_N<t_L r$ (See top panel (a)), there are an equal number of poles and zeros of $h_B^-$ inside the GBZ. Accordingly, the non-topological invariant (defined in Eq.~(\ref{eq:non_bloch_TIf})) vanishes. Hence the parameter regime ($\epsilon_N<t_L r$) defines the topologically trivial phase. In the parameter regime $\epsilon_N>t_L r$ (See top panel (c)), there are a pair of poles inside the GBZ. Accordingly, the non-Bloch topological invariant (defined in Eq.~(\ref{eq:non_bloch_TIf})) takes a non-trivial value, and the parameter regime ($\epsilon_N<t_L r$) defines the topologically non-trivial phase. In this case, the non-zero poles and zeros of $h_B^-$ coincide with the generalized momenta (disjoint from GBZ and marked by circles) characterizing the topological boundary state (See top panel (c)). The topological phase transition of GBZ occurs at $\epsilon_N=t_L r$, where the disjoint generalized momenta, and the contour of the GBZ merge which characterizes the emergence of exceptional points in the eigenvalue spectra (See top panel (b)).
        Bottom panel: Non-Bloch topological invariant (See bottom panel (left)) takes non-trivial values at $\epsilon_N=\pm t_L r$ characterizing the topological phase transitions of GBZs. The emergence of exceptional points at the topological phase transition points is also confirmed by the vanishing of the phase rigidity $R_1$ and $R_N$ (See bottom panel (right)). Parameters: $t_R=1.1,~t_L=0,~t'_L=t'_R=0,~\epsilon_1=\epsilon_0,~N=100, ~\epsilon_N\in[-3 t_L r,~3 t_L r]$.}
	\label{fig:GOBC_en_TPT}
    \end{figure}	
    
\subsubsection{Ambiguity in defining winding numbers and non-Bloch topological invariant}\label{sec:non_bloch}
     Eq.~(\ref{eq:winding_arg}) possesses an ambiguity in defining the winding number. The boundary equations [Eq.~(\ref{eqs:H_b})] obtained with the ansatz $\psi_n=c_1 z_1^n +c_2 z_2^n$ are not unique. Since we can also write the ansatz of the wave function as $\psi_n=c'_1 z_1^{m+n} +c'_2 z_2^{m+n},~m\in\mathbb{Z}$ by re-labeling $(c_1,~c_2)$ to $(c'_1,~c'_2)=(c_1 z_1^{-m},~c_2 z_2^{-m})$. Accordingly, the modified boundary equations (Eq.~(\ref{eqs:H_b})) takes the following form,
    \begin{align}\label{eq:H_b(m)}
	H^{(m)}_B \begin{pmatrix}
	c'_1\\
	c'_2\end{pmatrix} =0, ~~{\rm with}~~ H^{(m)}_B = \begin{pmatrix}
	A^{(m)}(z_1)& A^{(m)}(z_2) \\
	B^{(m)}(z_1)& B^{(m)}(z_2)
	\end{pmatrix},
	\end{align}
	here $A^{(m)}(z) = t_R z^m - (\epsilon_1-\epsilon_0)z^{m+1} -t_R^{\prime} z^{m+N}$, $B^{(m)}(z) = t_L^{\prime}z^{m+1} +(\epsilon_N-\epsilon_0)z^{m+N} - t_L z^{m+N+1}$. In this case, physical solutions $(z_1,~z_2)$ are obtained by solving the equation $\det[H^{(m)}_B]=0$. Now the modified effective boundary equations are written as the following eigenvalue problem
	\begin{align}
	\tilde{H}_B^{(m)} (c'_1,c'_2)^{\rm T}= (c'_1,c'_2)^{\rm T}~~{\rm with}~~\tilde{H}_B^{(m)}=&\begin{pmatrix}
	0& 	h_{B,m}^{+}(z_1,z_2) \\
	h_{B,m}^-(z_1,z_2)& 0
	\end{pmatrix}
 \end{align}
	$h_{B,m}^{+}\equiv h_{B,m}^+(z_1,z_2)=A^{(m)}(z_2)/A^{(m)}(z_1)$, $h_B^{-}\equiv h_{B,m}^-(z_1,z_2) = B^{(m)}(z_1)/B^{(m)}(z_2)$. Hence, the winding numbers (defined in Eq.~(\ref{eq:winding_arg})) are also modified and take the following forms 
	\begin{align}
    W_{\pm}^{(m)}=\frac{1}{2\pi i}\oint_{\mathcal{L}_1\times\mathcal{L}_2}\frac{1}{h_{B,m}^{\pm}(z_1,z_2)}\frac{d h_{B,m}^{\pm}(z_1,z_2)}{dz_1} dz_1, ~z_1 z_2 = t_R/t_L.
    \end{align}
	Accordingly, the topological invariant defined in Eq.~(\ref{eq:non_bloch_TI1}) will also be modified and is given by,
	\bea
    \label{eq:non_bloch_TI2}
    W^{(m)}=-\frac{W_+^{(m)}-W_-^{(m)}}{2},~\quad~W_{\pm}^{(m)}= \frac{1}{2\pi}[\textrm{arg}~h_{B,m}^{\pm}]_{\mathcal{L}_1\times \mathcal{L}_2}.
    \eea
    It is evident from the above equations that the winding numbers and the topological invariant depend on $m$. We elaborate above arguments with the example of the open boundary condition in the following. In the OBC, the modified boundary equations (Eq. \ref{eq:H_b(m)}) are given by,
    \begin{align}
     \frac{c'_2}{c'_1}&=\frac{1}{h_B^+}=-\frac{A^{(m)}(z_1)}{A^{(m)}(z_2)}=-\frac{z_1^m}{z_2^m },\nonumber\\
     \frac{c'_2}{c'_1}&=h_B^-=-\frac{B^{(m)}(z_1)}{B^{(m)}(z_2)}=-\frac{ z_1^{m+N+1}}{z_2^{m+N+1}}.
    \end{align}
   Accordingly, the physical solutions $(z_1,z_2)$ are determined by the solving the following equation
     \begin{align}
        \frac{z_1^m}{z_2^m }&=\frac{z_1^{m+N+1}}{z_2^{m+N+1} }\implies
        \Big(\frac{z_1}{z_2}\Big)^{N+1}=1, 
    \end{align}
    which is independent of $m$. The solutions of the above equations are of the form $(z_1,z_2)=(r e^{i\phi_k},r e^{-i\phi_k})$ with $\phi_k = k\pi/(N+1),~k\in\lbrace 1,2,...,N\rbrace.$ Therefore, the generalized Brillouin zones and eigenvalue spectra do not  depend on $m$, however, $c'_1,~c'_2$ depend on it. Accordingly, the winding numbers associated with the meromorphic functions $h_B^+,~h_B^-$ depend on $m$, which are given by, $W_{\pm}^{(m)}=\mp 2 m$. For instance, we can list the wave functions together with winding numbers for different values of $m$ as follows:
    \\
    
    (i) for $m=0$: $\psi_n = c'_1( z_1^n - z_2^n)$ ($c'_2/c'_1=-1$),  $W_+^{(0)}=0$,
    
    (ii) for $m=1$: $\psi_n = c'_1 z_1 z_2( z_1^n - z_2^n)$ ($c'_2/c'_1=-z_1/z_2$), $W_+^{(1)}=-2$,
    
    (iii) for $m=l$: $\psi_n = c'_1 (z_1z_2)^l( z_1^n - z_2^n)$ ($c'_2/c'_1=-(z_1/z_2)^l$), $W_+^{(l)}=-2l$, and so on.\\
    \\
    Despite of the ambiguity, the topological phase transition always accompanies a change in the total number of zeros and poles of $h_B^+,~h_B^{-}$ inside GBZs. Accordingly, we define our topological invariant as follows:
	\begin{align}\label{eq:non_bloch_TIf}
	    W_{\rm non-Bloch}=\frac{W^{R}\pm W^{L}}{2},~W^{R}\equiv W^{(0)},~W^{L}\equiv W^{(-N-1)},~W^{(m)}=-\frac{W^{(m)}_+-W^{(m)}_-}{2}.
	\end{align}
 We note here that the topological invariant $W^L$ can also be obtained by interchanging the parameters
	$(t_R,t_L,\epsilon_1,\epsilon_N)$ to  $(t_L,t_R,\epsilon_N,\epsilon_1)$ in the eigenvalue equation or vice versa. Hence, we can think of the non-Bloch topological invariant $W^R$ as the topological invariant defined for the right bulk which is defined by the parameters
	$(t_R,t_L,\epsilon_1,\epsilon_N)$, and the non-Bloch topological invariant $W^L$ defined for left bulk by interchanging the parameters
	$(t_R,t_L,\epsilon_1,\epsilon_N)$ to  $(t_L,t_R,\epsilon_N,\epsilon_1)$. We find that the non-Bloch topological invariant $W_{\rm non-Bloch}$ (defined in Eq.~\eqref{eq:non_bloch_TIf}) characterizes the topological phase transition of GBZ for different generalized boundary conditions which we discuss in some detail in the upcoming sections.
	
\subsection{Exceptional and topological phase transitions of GBZ} \label{sec:Exceptionaltransitions}
    
    In contrast to the degeneracy of eigenenergies, the degeneracy of two generalized momenta directly indicates the coalescence of the eigenstates, which manifests as an exceptional point in the spectra. The exceptional point is characterized by vanishing of phase rigidity, (See Fig.~\ref{fig:GOBC_en_TPT})
    \bea\label{def:phase_rigidity}
	R_i=\frac{\langle \Psi_i^R|\Psi_i^L\rangle}{\langle \Psi_i^R|\Psi_i^R\rangle},~~i\in\lbrace 1,2,...,N\rbrace.
    \eea
    here $|\Psi_i^R\rangle$ is the right eigenstates of $\hat{H}$ (Eq.~(\ref{supp:general_model_ham})) with eigenvalue $E_i$ and $|\Psi_j^L\rangle)$ is the left eigenstates of $\hat{H}$ (Eq.~(\ref{supp:general_model_ham})) with eigenvalues $E_i^*$ which satisfy bi-orthogonality relations \textit{i.e.} $\langle \Psi_i^R|\Psi_j^L\rangle=\delta_{i j}$. We calculate the phase rigidity of all the eigenstates by sorting them according to the real parts of the eigenvalues. We elaborate on these findings in upcoming sections when analyzing different types of GBC.

 \subsubsection{Analytical examples (i): Generalized boundary conditions- (I)}
    In this section, we present an analytical study of the topological phase transition of GBZ. First, we consider the GBC-(I) (See Table \ref{tab:def_GBC}) with $t'_R=t'_L=0,~\epsilon_1=\epsilon_0,~\epsilon_N\neq\epsilon_0$. In this case, the boundary equations become,
    \begin{align}
     \frac{c_2}{c_1}&=-\frac{A^{(m)}(z_1)}{A^{(m)}(z_2)}=-\frac{z_1^m}{z_2^m},\nonumber\\
     \frac{c_2}{c_1}&=-\frac{B^{(m)}(z_1)}{B^{(m)}(z_2)}=-\frac{ \epsilon_N z_1^{m+N}-t_L z_1^{m+N+1}}{\epsilon_N z_2^{m+N}-t_L z_2^{m+N+1}}.
    \end{align}
    The paired generalized momenta $(z_1,~z_2)$ are obtained by solving the following equation,
    \begin{align}
    \frac{z_1^m}{z_2^m}&=\frac{ \epsilon_N z_1^{m+N}-t_L z_1^{m+N+1}}{\epsilon_N z_2^{m+N}-t_L z_2^{m+N+1}}\implies
     1=\frac{ \epsilon_N -t_L z_1}{\epsilon_N -t_L z_2}\Big(\frac{z_1}{z_2}\Big)^{N}.
    \end{align}
     We describe the topologically trivial and non-trivial phases in the following manner:

    \textbf{Topologically trivial phase:}
    In the parameter regime $\epsilon_N<t_L r$, two GBZs merge with each other similar to that of the OBC (See top panel (c) of Fig.~\ref{fig:GOBC_en_TPT}). This parameter regime $\epsilon_N<t_L r$ describes topologically trivial GBZ since there is an equal number of zeros and poles of $h_B^-$ inside GBZs, and hence the winding numbers, and non-Bloch topological invariant vanish (See Secs.~\ref{sec:winding} and \ref{sec:non_bloch}). For $\epsilon_N << t_L r$, the bulk states are described by the generalized momenta $(z_1,z_2)=(r e^{i\phi_k},r e^{-i\phi_k})$ with $\phi_k = k\pi/(N+1),~k\in\mathbb{Z}.$ Moreover, the effective spinor, and wave functions are given by,
    \begin{align}
        \frac{c_2}{c_1}&=-\frac{z_1^m}{z_2^m}, \psi_n = c_1 (z_1 z_2)^m( z_1^{n} - z_2^{n}),~n\in\lbrace 1,2,...,N\rbrace.
    \end{align}

    \textbf{Topologically non-trivial phase:}
    For $\epsilon_N > t_L r$, again two GBZs merge with each other similarly to the topologically trivial phase, however, there exists one pair of the real solution $(z_1,z_2) =(t_R/\epsilon_N,\epsilon_N/t_L)$ (See top panel (c) of Fig.~\ref{fig:GOBC_en_TPT}) which describes the topological boundary state ($\psi_n\propto z_2^n$) localized at the boundary of the chain. In this parameter regime (See top panel (c) of Fig.~\ref{fig:GOBC_en_TPT}), the non-zero pole of $h_B^-$ together with the pole at zero lies inside GBZs. Accordingly, the winding numbers and the non-Bloch topological invariant become (See Secs.~\ref{sec:winding} and \ref{sec:non_bloch})
    \begin{align}
        W^{(R)}=0,~W^{(L)}=2 \implies W_{\rm non-Bloch}=1.
    \end{align}
    Therefore, the parameter regime $\epsilon_N >t_L r$ describes the topologically non-trivial phase which supports the emergence of the topological boundary state described by the disjoint generalized momenta $(z_1,z_2) =(t_R/\epsilon_N,\epsilon_N/t_L)$. Hence the parameter regime $\epsilon_N>t_L r$ describes the topologically non-trivial phase.
    Moreover, for $\epsilon_N >> t_L r$, the bulk equations are characterized by the generalized momenta $(z_1,z_2)=(r e^{i\phi_k},r e^{-i\phi_k})$ with $\phi_k = k\pi/(N),~k\in\mathbb{Z}.$ In this case, the expanded spinor, and wave functions are given by,
    \begin{align}
        \frac{c_2}{c_1}&=-\frac{z_1^{m}}{z_2^{m} },~\psi_n =c_1 (z_1 z_2)^{m}( z_1^{n} - z_2^{n}),~n\in\lbrace 1,2,..., N\rbrace.
    \end{align}

    Similar to the previous section (Sec.~(\ref{sec:analytical_GOBC1})), we also find that the non-Bloch topological invariant (the bottom left panel of Fig.~\ref{fig:GOBC_en_TPT}) changes and takes non-trivial value at $\epsilon_N = \pm t_L r$ characterizing the topological phase transition of GBZs. At this transition point the disjoint generalized momenta $(z_1,z_2) =(t_R/\epsilon_N,\epsilon_N/t_L)$ describing the boundary state touches the GBZs (See top panel (b) of Fig.~\ref{fig:GOBC_en_TPT}) which is the exceptional point in the eigenvalue spectra of the $\hat{H}$. The appearance of the exceptional points is confirmed by the vanishing of the phase rigidity $R_1,~R_N$ in the large $N-$limit (see bottom right panel of Fig.~\ref{fig:GOBC_en_TPT}).
    
 \subsubsection{Analytical examples (ii): Generalized boundary conditions- (I)}\label{sec:analytical_GOBC1}
     In this section, we present another analytical study of the topological phase transition of GBZ. Now, we consider the GBC-(I) with $t'_R=t'_L=0,~\epsilon_1\neq\epsilon_0,~\epsilon_N=\epsilon_0$. In this case, the boundary equations become,
    \begin{align}
     \frac{c_2}{c_1}&=-\frac{A^{(m)}(z_1)}{A^{(m)}(z_2)}=-\frac{t_R z_1^m -\epsilon_1 z_1^{m+1}}{t_R z_2^m -\epsilon_1 z_2^{m+1} },\nonumber\\
     \frac{c_2}{c_1}&=-\frac{B^{(m)}(z_1)}{B^{(m)}(z_2)}=-\frac{ z_1^{m+N+1}}{z_2^{m+N+1}}.
    \end{align}
    The paired generalized momenta $(z_1,~z_2)$ are obtained by solving the equation
    \begin{align}
    \frac{t_R z_1^m -\epsilon_1 z_1^{m+1}}{t_R z_2^m -\epsilon_1 z_2^{m+1} }&=\frac{z_1^{m+N+1}}{z_2^{m+N+1} }\implies
     \frac{t_R  -\epsilon_1 z_1}{t_R -\epsilon_1 z_2 }=\frac{z_1^{N+1}}{z_2^{N+1} },
    \end{align}
    which is independent of $m$ and may not be solved analytically. 
    We can understand the topological phase transitions of GBZs by analyzing the boundary equations in the following.
 
    \textbf{Topologically trivial phase:}
    For $\epsilon_1> t_L r$, we find that the two GBZs merge with each other similar to that of OBC. However, in this parameter regime, there is an equal number of poles and zeros of the meromorphic function $h_B^+$ inside GBZs. Accordingly, the winding numbers, and the non-Bloch topological invariant vanish (See Secs.~\ref{sec:winding} and \ref{sec:non_bloch}). Therefore, the parameter regime $\epsilon_1 <t_L r$ describes the topologically trivial phase. In the limiting case ($\epsilon_1 << t_L r$), the approximate solutions of the boundary equations are $(z_1,z_2)=(r e^{i\phi_k},r e^{-i\phi_k})$ with $\phi_k = k\pi/(N+1),~k\in\mathbb{Z}.$ Moreover, the effective spinor and wave functions are given by,
    \begin{align}\label{eqn:GOBC1_trivial}
        \frac{c_2}{c_1}&=-\frac{z_1^m}{z_2^m},~ \psi_n = c_1 (z_1 z_2)^m( z_1^{n} - z_2^{n}),~n\in\lbrace 1,2,..., N\rbrace.
    \end{align}

    \textbf{Topologically non-trivial phase:}
    For $\epsilon_1>t_L r$, we find that the two GBZs merge with
    each other similar to the previous case, and in addition to that, the boundary equations have one pair of the real solution $(z_1, z_2) =
    (t_R/\epsilon_1, r^2\epsilon_1/t_R)$ in the large N limit. This pair of disjoint generalized momenta describes the topological boundary
    state ($\psi_n\propto z_2^n$) localized at the boundary of the chain. In this case, there are a pair of poles of $h_B^+$ inside the GBZs. Accordingly, the winding numbers and the non-Bloch topological invariant become (See Secs.~\ref{sec:winding} and \ref{sec:non_bloch})
    \begin{align}
        W^{(R)}=2,~W^{(L)}=0 \implies W_{\rm non-Bloch}=1.
    \end{align}
     Therefore, the parameter regime $\epsilon_1 >t_L r$ describes the topologically non-trivial phase which supports the emergence of the topological boundary state described by the disjoint generalized momenta $(z_1,z_2)=(t_R/\epsilon_1,r^2 \epsilon_1/t_R)$. For $\epsilon_1 \gg t_L r$, the approximate solutions of the boundary equations are given by $(z_1,z_2)=(r e^{i\phi_k},r e^{-i\phi_k}$) with $\phi_k = k\pi/(N),~k\in\mathbb{Z}.$ In this case, the effective spinor, and wave functions are given by,
    \begin{align}\label{eqn:GOBC1_non_trivial}
        \frac{c_2}{c_1}&=-\frac{z_1^{m+1}}{z_2^{m+1} },~\psi_n =c_1 (z_1 z_2)^{m+1}( z_1^{n-1} - z_2^{n-1}),~n\in \lbrace 1,2,..., N\rbrace.
    \end{align}

    The topological phase transitions of GBZs occur at the exceptional points $\epsilon_1 = \pm t_L r$ where the disjoint generalized momenta $(z_1,z_2)=(t_R/\epsilon_1,r^2 \epsilon_1/t_R)$ describing the topological boundary state touch the GBZs (See Fig.~\ref{fig:GOBC}). The appearance of the exceptional points is also confirmed by the vanishing of the phase rigidity $R_1,~R_N$ in the large $N-$limit (See Fig.~\ref{fig:GOBC}). Moreover, the non-zero zero and pole of $h_B^{+}$ coincide with the disjoint generalized momenta in the topologically non-trivial phase and merge with the GBZs at the topological phase transition point.

    \begin{figure}[htbp]
	\centering
	\includegraphics[width=0.9\linewidth]{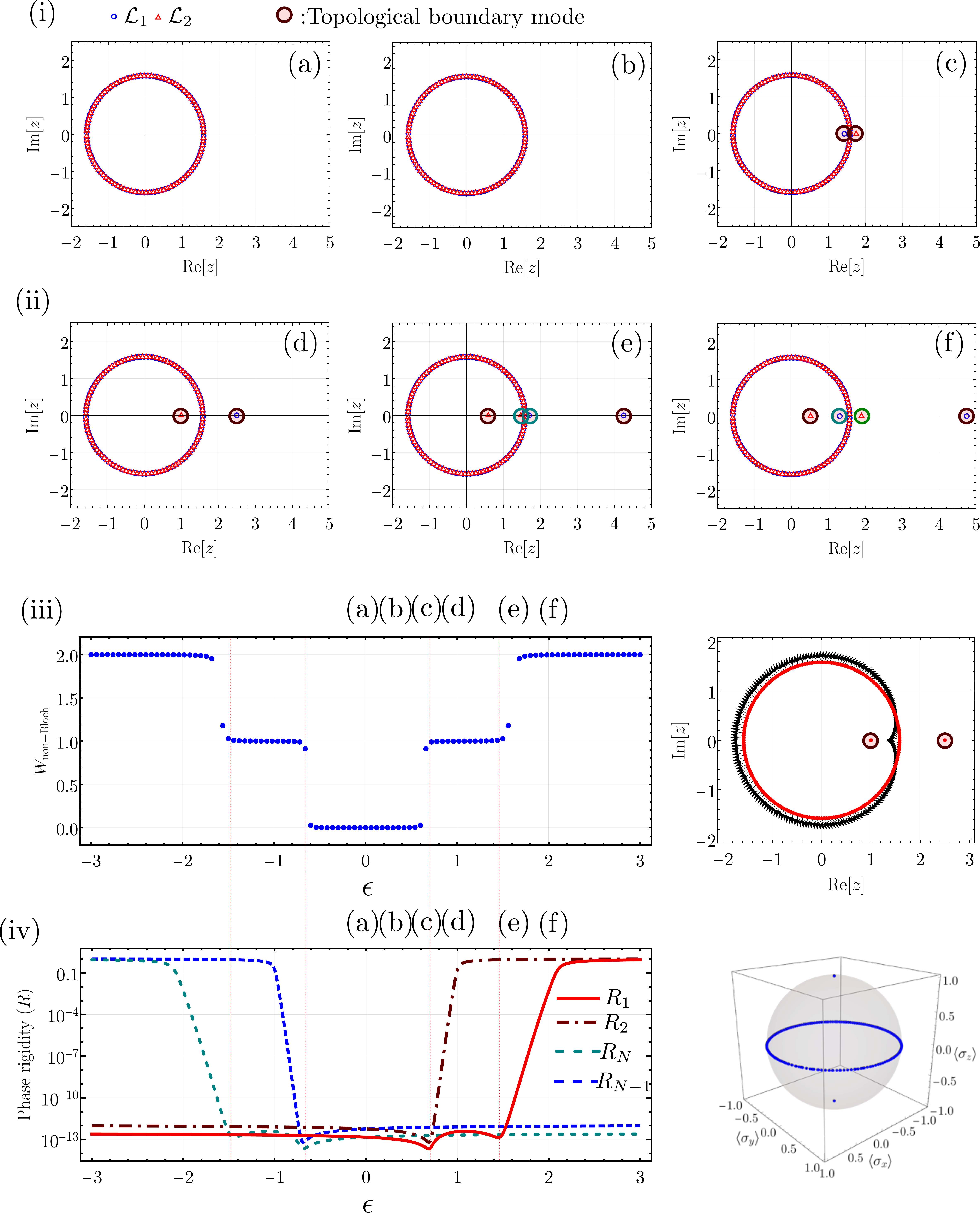}
	\caption{Exceptional and topological phase transition of GBZ in the GBC-(I). Panels (i) and (ii) describe the evolution of GBZs and the emergence of the disjoint generalized momenta describing bound states as we tune the onsite potentials at systems boundaries. In the left panel of (iii), the emergence of boundary states is characterized by the quantized change of non-trivial values of the non-Bloch topological invariant ($W_{\rm non-Bloch}$). The topological phase transition occurs when disjoint generalized momenta merge with GBZs (exceptional points) which is well captured by the change in $W_{\rm non-Bloch}$. The figure (right) of the panel-(iv) shows the vanishing of the phase rigidity (defined in Eq.~(\ref{def:phase_rigidity})) $R_1,~R_2,~R_{N-1},~R_{N}$ at the topological phase transition points. The right panels (iii) and (iv) show the projection of the effective spinor on the GBZ and Bloch sphere in the topologically non-trivial phase (figure (ii)(e)) which clearly show the non-trivial winding and topology of GBZs (See Sec.~\ref{sec:ET_TPT_GBC_I} for more details). Parameters: $N=100,~t'_L=0,~t_L=1.0,~t'_R=0,~t_R=2.1,~\epsilon_1= t_R\epsilon,~\epsilon_N=t_L\epsilon,~\epsilon\in[-3,3].$}
    \label{fig:ET_TPT_GBC_1}
    \end{figure}

   \subsubsection{Example (iii): GBC-(I)}\label{sec:ET_TPT_GBC_I}
    In this section, we discuss the exceptional and topological phase transitions of GBZ for the GBC-(I) which is defined with $t_L^{\prime}=t_R^{\prime}=0$, and $\epsilon_1 (\epsilon_N)$ is not equal to $\epsilon_0$ at the same time but can have any value in general (See Table~\ref{tab:def_GBC}). This type of boundary condition can be understood as an open boundary condition with different onsite potentials at the boundaries of the chain. In this case, the boundary equations are,
	\begin{align}
	t_R \psi_0 - (\epsilon_1 -\epsilon_0)\psi_1 &= 0,\nonumber\\
	(\epsilon_N -\epsilon_0)\psi_N-t_L\psi_{N+1}  &= 0.
	\end{align}

    \textbf{Generalized Brillouin zones and boundary states:}
    Solving the above boundary equations, we get $N$, $N-1$, or $N-2$ (depending on the strengths of $\epsilon_1$ and $\epsilon_N$) pairs of generalized momenta with $|z_1| =|z_2|=r$. Hence, two generalized Brillouin zones merge similarly to OBC (See Figs.~\ref{fig:ET_TPT_GBC_1}(i)-(ii)). The wave functions corresponding to the bulk states can be written as $\psi_n =c_1 z_1^n +c_2 z_2^n, ~(c_1\neq c_2)$, where $|z_1| =|z_2| =r$. These states describe localized states at one of the boundaries of the system (skin states). 
 
    The remaining one pair (or two pairs) of disjoint generalized momenta (See the evolution of GBZs and the emergence of disjoint generalized momenta in Figs.~\ref{fig:ET_TPT_GBC_1}(i)-(ii)) lie on the real momenta axis, \textit{i.e.} $z_1, ~z_2 \in \mathbb{R},$ (for $r<1$). Accordingly, the wave functions corresponding to these disjoint momenta become $\psi_n\propto z_2^n$ and describe the bound states localized at one of the system's boundaries.

    \textbf{Exceptional phase transitions of GBZ:}
    We find exceptional points in the eigenvalue spectra of the Hamiltonian by varying $\epsilon_1=t_R\epsilon$, $\epsilon_N=t_L\epsilon$ simultaneously which are dictated by the vanishing phase rigidity (defined in Eq.~(\ref{def:phase_rigidity})) $R_1$, $R_2$ for positive values of $\epsilon$, and $R_N$, $R_{N-1}$ for negative values of $\epsilon$, respectively (See Fig.~\ref{fig:ET_TPT_GBC_1}(iv)). The dips in the phase rigidity depend on the degree of non-reciprocity ($r^2=t_R/t_L$) and system size ($N$) which become sharper with an increase in the system size. Importantly, we observe that at exceptional points, bound states described by disjoint generalized momenta in the complex plane emerge out of the bulk continuum (See Figs.~\ref{fig:ET_TPT_GBC_1}(i)(c)-(e)). The numerical finding of exceptional points is confirmed by touching of disjoint generalized momenta to the contour of the generalized Brillouin zone (See Fig.~\ref{fig:ET_TPT_GBC_1}(i)(c)-(e) and Fig.~\ref{fig:ET_TPT_GBC_1}(ii)(e)). Therefore, exceptional points characterizing the emergence of bound states from the bulk continuum define the exceptional transitions of GBZ. 

    \textbf{Topological phase transitions of GBZ:}
    We find that boundary states in response to the GBC are characterized by the non-trivial values of the non-Bloch topological invariant $W_{\rm non-Bloch}$. The Fig.~\ref{fig:ET_TPT_GBC_1}(iii) (left) dictates the change of the non-Bloch topological invariant at the exceptional points defined by the touching of disjoint generalized momenta to the contour of GBZ (See vanishing of phase rigidity $R_1$, $R_{2}$, $R_{N-1}$, and $R_N$ in Fig.~\ref{fig:ET_TPT_GBC_1}(iv) (left)) which marks topological phase transitions of the GBZ. Therefore, topological phase transitions of GBZ accompany the exceptional points.

    We also find that in GBC-(I), the effective boundary matrix $\tilde{H}_B$ (See Sec.~\ref{sec:sym_HB}) becomes Hermitian together with the chiral symmetry. Accordingly, in the topologically non-trivial phase, the effective spinor $(c_1,~c_2)^T$ projected on the Bloch sphere always lies on the equator for all generalized momenta of the bulk (See figures (right) of panels (iii) and (iv) of Fig.~\ref{fig:ET_TPT_GBC_1}). Therefore, the associated Berry phase is quantized and serves the purpose of characterizing topological phase transitions. In this case, the non-Bloch topological invariant coincides with the winding number related to the conventional Berry phase associated with the effective spinor $(c_1,~c_2)^T$. 

    \begin{figure}[htbp]
    \centering
    \includegraphics[width=1.0\linewidth]{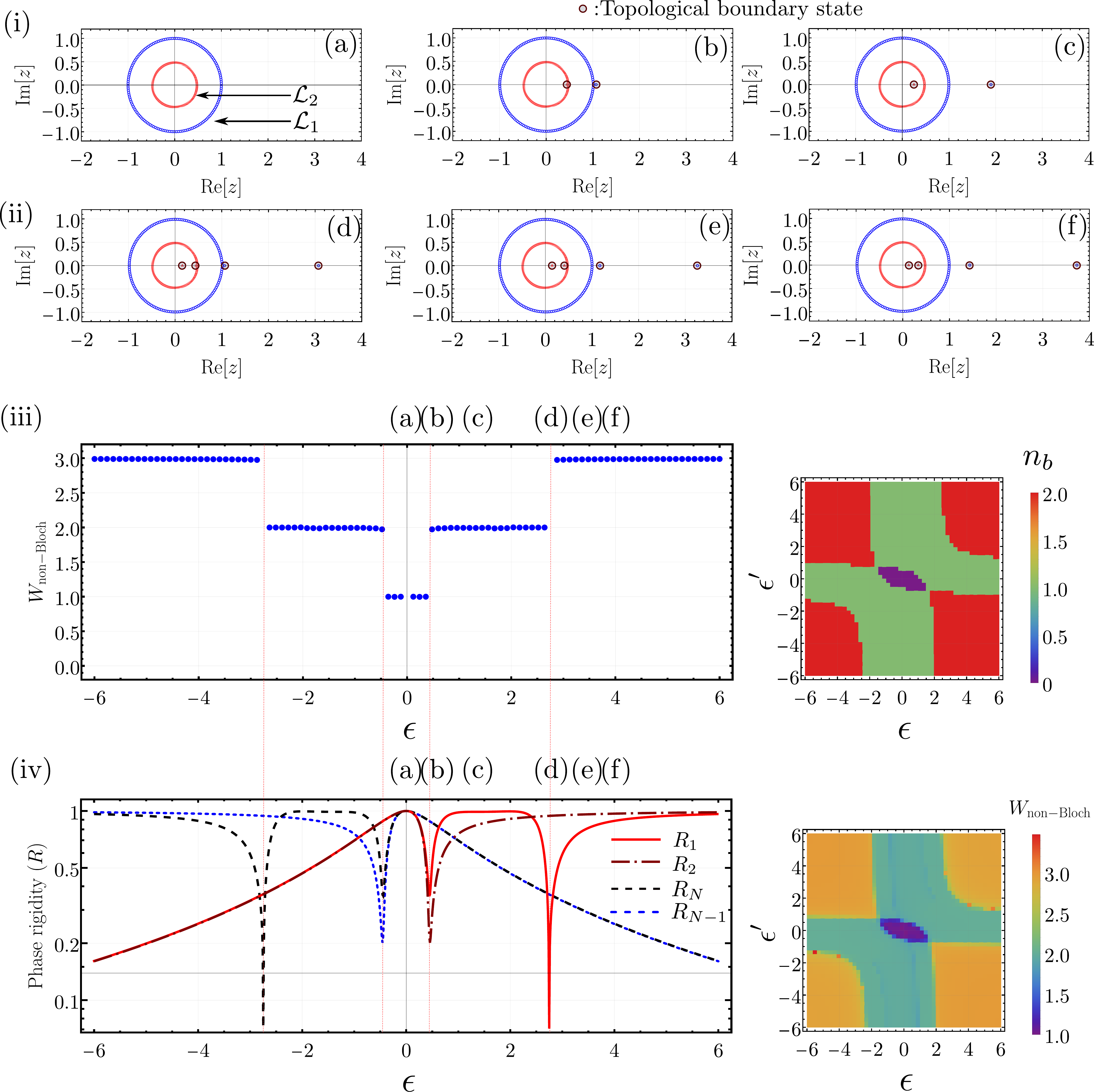}
    \caption{Exceptional and topological phase transition of GBZ in GBC-(II). Panels (i) and (ii) show GBZs and the disjoint generalized momenta for different values of the onsite potentials at the boundaries of the system. Left panels of (iii) and (iv) show the behavior of non-Bloch topological invariant and phase rigidity for the change in the onsite potentials(See Sec.~\ref{sec:ET_TPT_GBC_II} for more details). Parameters: $N=160,~t'_L=t_L=2.1,~t'_R=t_R=1.0,~\epsilon_1=t_R        \epsilon,~\epsilon_N=t_L\epsilon,~\epsilon\in[-6,6].$ \\
    The right panel of (iii) shows the number of bound states ($n_b$) in the $(\epsilon,~\epsilon'$)-plane. In the right panel of (iv), the change in the non-Bloch topological invariant ($W_{\rm non-Bloch}$) characterizes the topological phase transition points in the $(\epsilon,~\epsilon'$)-plane, where topological boundary states emerge out of the bulk continuum (See Sec.~\ref{sec:ET_TPT_GBC_II} for more details).
    Parameters for right panels of (iii) and (iv): $N=80,~t'_L=t_L=2.1,~t'_R=t_R=1.0,~\epsilon_1=t_R \epsilon,~\epsilon_N=t_L\epsilon^{\prime},$ with $\epsilon,~\epsilon^{\prime}\in[-6,6].$}
    \label{fig:ET_TPT_GBC_2}
    \end{figure}
     
     \begin{figure}[htbp]
		\centering
		\includegraphics[width=1.0\linewidth]{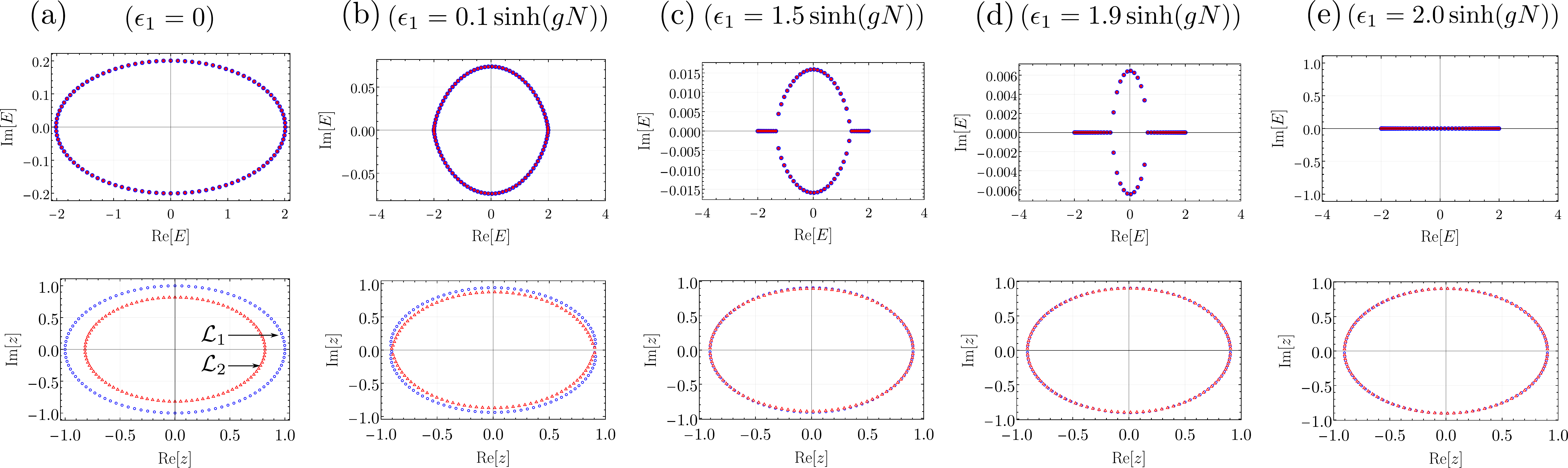}
		\caption{ GBC-(II) with very large onsite potentials at the boundary. The top and bottom panels show the plots of eigenvalue spectra and GBZs for different values of onsite potential $\epsilon_1$. For better visualization of eigenvalue spectra and GBZs, we have omitted the eigenvalue and generalized momenta corresponding to the bound states in (b), (c), (d), and (e). Parameters: $N=100,~t'_L=t_L,~t_L=e^{g},~t'_R=t_R,~t_R=e^{-g},,~\epsilon_N=0,~g=0.1.$}
		\label{fig:GPBC_large_e1}
    \end{figure}
    \subsubsection{Example (iv): GBC-(II)}\label{sec:ET_TPT_GBC_II}
    In this section, we discuss the exceptional and topological phase transitions of GBZ in GBC-(II) which is defined as $t_L^{\prime}= t_L,~t_R^{\prime}= t_R$, and $\epsilon_1\neq\epsilon_N \neq\epsilon_0$. This type of boundary condition can be understood as the periodic boundary conditions with different onsite potentials at two neighboring sites anywhere in the chain.  In this case, the boundary equations are,
	\begin{align}
	t_R\psi_{0} - (\epsilon_1 - \epsilon_{0}) \psi_1 - t_R \psi_{N}  &= 0,\nonumber\\
	t_L \psi_{1} + (\epsilon_N-\epsilon_{0}) \psi_{N} - t_L \psi_{N+1}  &= 0,
	\end{align}
       
    \textbf{Generalized Brillouin zones and boundary states:}
    After solving the above boundary equations, we get $N$, $N-1$ or $N-2$ (depending on the strengths of $\epsilon_1$ and $\epsilon_N$) pairs of generalized momenta with $|z_1| \neq |z_2|$ which form two GBZs. We find that similar to the case of PBC, in this case also, the two generalized Brillouin zones differ from each other. The remaining one pair (or two pairs) of disjoint generalized momenta lie on the real momenta axis, \textit{i.e.} $z_1, ~z_2 \in \mathbb{R},$ (for $r>0$). In this case, also, the wave functions corresponding to these disjoint generalized momenta become $\psi_n \propto z_2^n $ which describe the bound states localized at the boundaries of the system. The evolution of GBZs and the emergence of disjoint generalized momenta by changing the onsite potentials are shown in Figs.~\ref{fig:ET_TPT_GBC_2}(i)-(ii). We also find that at very large values of the onsite potentials ($\epsilon_1$ or $\epsilon_N$ or both), the two GBZs merge, and the bulk eigenvalue spectra collapse on the real eigenvalue axis (See Fig.~\ref{fig:GPBC_large_e1}).
	
    \textbf{Exceptional phase transition of GBZ:}
    We find two exceptional points in the eigenvalue spectra of the Hamiltonian by varying $\epsilon_1$, $\epsilon_N$ simultaneously which are dictated by the vanishing phase rigidity $R_1$, $R_2$ for positive values of $\epsilon_1$, and $R_N$, $R_{N-1}$ for negative values of $\epsilon_1$ (See the left panel of Fig.~\ref{fig:ET_TPT_GBC_2}(iv)). Importantly, we observe that after these exceptional transition points, there appear bound states localized at the boundaries of the system. The numerical finding of exceptional points due to GBC-(II) is confirmed by touching of disjoint generalized momenta to the generalized Brillouin zones (See Figs.~\ref{fig:ET_TPT_GBC_2}(i)-(ii), and left panel of Fig.~\ref{fig:ET_TPT_GBC_2}(iv)). Therefore, the exceptional points characterizing the emergence of bound states outside the bulk eigenvalue spectra define the exceptional transitions of the GBZ. 

    \textbf{Topological phase transition of GBZ:}
    Similar to the GBC-(I), in this case also, the boundary states arising at the exceptional points in the eigenvalue spectra are characterized by non-trivial values of non-Bloch topological invariant ($W_{\rm non-Bloch}$). Moreover, the topological phase transition of the GBZ occurs at these exceptional points which are marked by the quantized change in the non-Bloch topological invariant (See Fig.~\ref{fig:ET_TPT_GBC_2}) (iii) (left)). The emergence of exceptional points at the topological phase transition points is also confirmed by the vanishing of phase rigidity $R_1$, $R_2$, $R_{N-1}$ and $R_{N}$ (See left panels of Fig.~\ref{fig:ET_TPT_GBC_2}(iii)-(iv)). 
    
    We also find that the conventional Berry phase associated with the effective spinor will not be quantized in this case because the effective boundary matrix $\tilde{H}_B$ is non-Hermitian (See Sec.~\ref{sec:sym_HB}). We determine the number of bound states ($n_b$) in the ($\epsilon_1$, $\epsilon_N$)-parameter space ($\epsilon_1$, $\epsilon_N$)-plane (See right panel of Fig.~\ref{fig:ET_TPT_GBC_2} (iii)). The non-Bloch topological invariant show a similar phase diagram in the ($\epsilon_1$, $\epsilon_N$)-plane (See right panels of Figs.~\ref{fig:ET_TPT_GBC_2}(iii)-(iv)) which characterizes the emergence of the boundary states.

    \subsubsection{Example (v): GBC-(III) and (IV)}\label{sec:ET_TPT_GBC_III_IV}
    In this section, we consider the generalized boundary condition by combining the ones described by (iii) and (iv) type of GBC mentioned in Table~\ref{tab:def_GBC}: $t_L^{\prime}\neq t_L,~t_R^{\prime}\neq t_R$, and $\epsilon_1=\epsilon_N =\epsilon_0$. In this case, the boundary equations are,
	\begin{align}
	t_R \psi_{0} - t_R^{\prime} \psi_{N}  &= 0,\nonumber\\
	t_L^{\prime} \psi_{1}  - t_L \psi_{N+1}  &= 0,
	\end{align}
	 \textbf{Generalized Brillouin zones and boundary states:}
    Similar to GBC-(II), solving the above boundary equations, we get $N$, $N-1$, or $N-2$ (depending on the strengths of $t'_R$ and $t'_L$) pairs of generalized momenta with $|z_1| \neq |z_2|$ which form two GBZs. The two generalized Brillouin zones differ from each other except for the OBC ($t'_R=t'_L=0$) where they merge with each other. The remaining one pair (or two pairs) of disjoint generalized momenta lie on the real momenta axis, \textit{i.e.} $z_1, ~z_2 \in \mathbb{R},$ (for $r>0$). In this case, also, the wave functions corresponding to these disjoint generalized momenta become $\psi_n \propto z_2^n $ which describe the bound states localized at the boundaries of the system. The evolution of GBZs ($\mathcal{L}_1,~\mathcal{L}_2$) and the emergence of disjoint generalized momenta by changing the boundary hopping terms are shown in Fig.~\ref{fig:ET_TPT_GBC_3}(i). Also, at very large values of the boundary hopping terms (both $t'_R$ and $t'_L$), the two GBZs merge, and the bulk eigenvalue spectra collapse on the real eigenvalue axis (See  Fig.~\ref{fig:GBC_large_hopp}).
   
    \begin{figure}[htbp]
		\centering
		\includegraphics[width=1.0\linewidth]{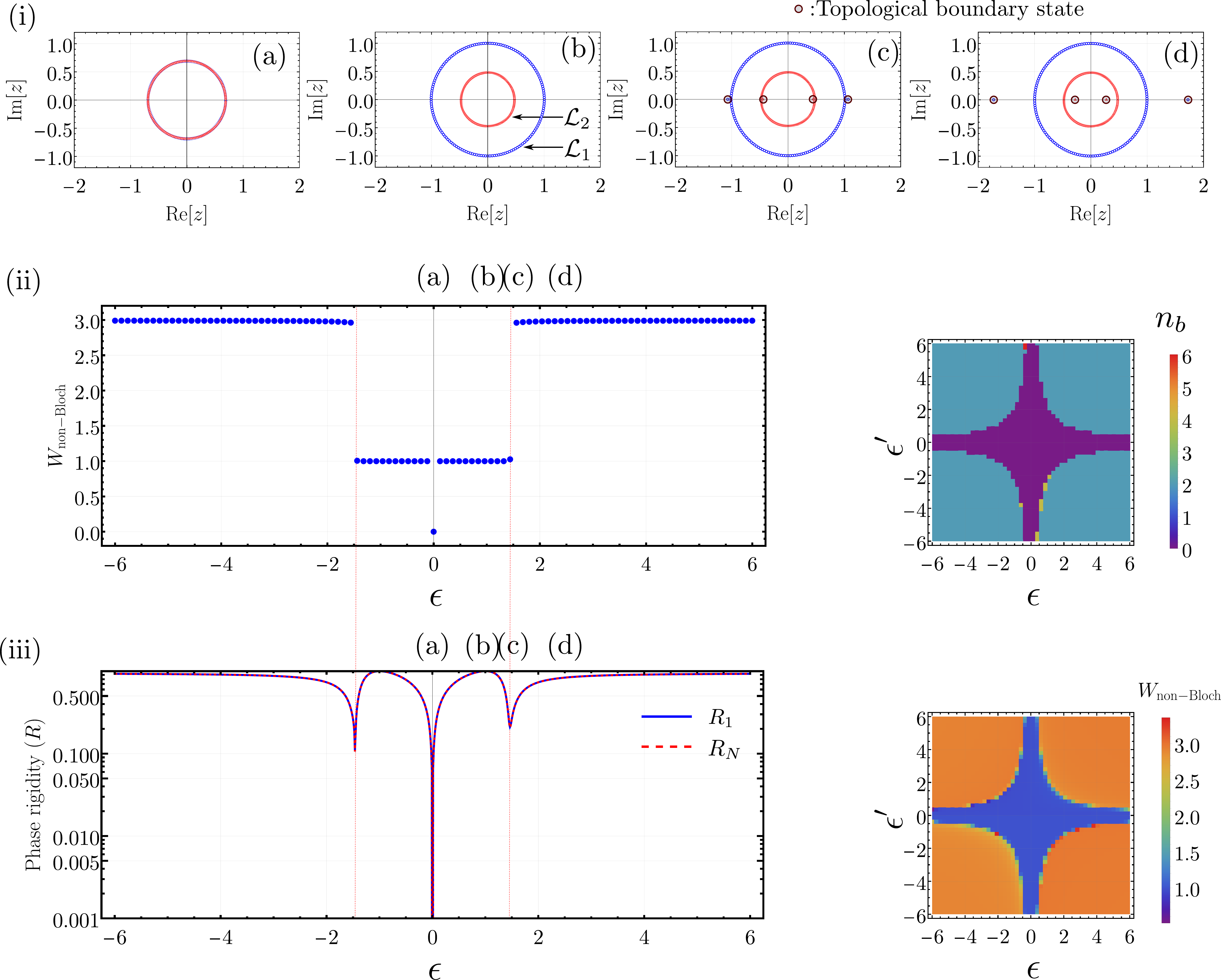}
		\caption{Exceptional and topological phase transition of GBZ in GBC-(III) and (IV). Panel (i) shows GBZs ($\mathcal{L}_1,~\mathcal{L}_2$) together with disjoint generalized momenta for different values of boundary hopping terms. The left panels of (ii) and (iii) show the behavior of non-Bloch topological invariant and phase rigidity with respect to changes in the boundary hopping terms ($t'_R=t_R\epsilon,~t'_L=t_L\epsilon$) (See Sec.~\ref{sec:ET_TPT_GBC_III_IV} for more details). Parameters: $N=160,~t_L=1.0,~t_R=2.1,~\epsilon_1=0,~\epsilon_N=0,~\epsilon\in[-6,6].$ The right panels of (iii) and (iv) show the number of bound states ($n_b$) and non-Bloch topological invariant ($W_{\rm non-Bloch}$) in the $(\epsilon,~\epsilon'$)-plane with $t'_R=t_R\epsilon,~t'_L=t_L\epsilon'$. In the right panel of (iv), the change in the non-Bloch topological invariant  characterizes the topological phase transition points in the $(\epsilon,~\epsilon'$)-plane, where topological boundary states emerge out of the bulk continuum (See Sec.~\ref{sec:ET_TPT_GBC_III_IV} for more details).
        Parameters for right panels of (iii) and (iv): $N=80,~t'_L=t_L\epsilon,~t_L=1.0,~t'_R=t_R\epsilon^{\prime},~t_R=2.1,~\epsilon_1=0,~\epsilon_N=0,$ with $\epsilon,~\epsilon^{\prime}\in[-6,6].$}
		\label{fig:ET_TPT_GBC_3}
    \end{figure}

    \begin{figure}[htbp]
		\centering
		\includegraphics[width=1.0\linewidth]{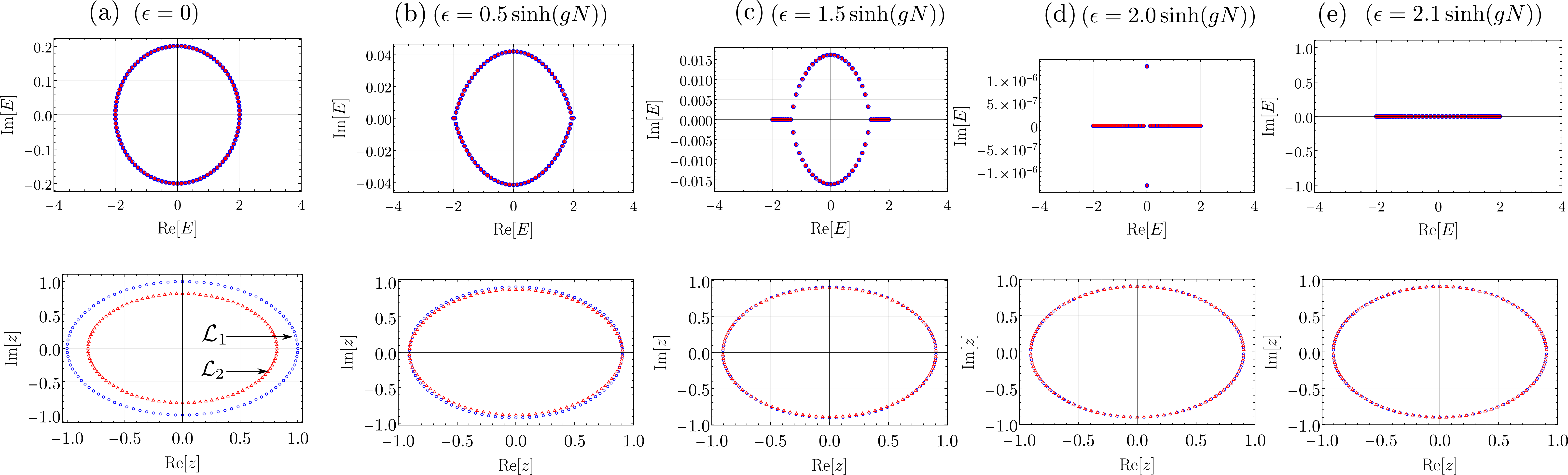}
		\caption{ GBC-(IV) with very large values of hopping terms at the boundaries. The top and bottom panels show the plots of eigenvalue spectra and GBZs for different values of boundary hopping terms $t'_R = t_R \epsilon,~t'_L=t_L\epsilon$. For better visualization of eigenvalue spectra and GBZs, we have omitted the eigenvalue and generalized momenta corresponding to the bound states in (b), (c), (d), and (e). Parameters: $N=100,~t_L=e^{g},~t_R=e^{-g},~\epsilon_1=\epsilon_N=0,~g=0.1.$}
		\label{fig:GBC_large_hopp}
    \end{figure}

    \textbf{Exceptional transition of GBZ:}
    We observe exceptional points by varying $t'_R=t_R\epsilon$, $t'_L=t_L\epsilon$ simultaneously which are dictated by the vanishing phase rigidity $R_1$ and $R_N$ for positive and negative values of $t'_R$, respectively (See Fig.~\ref{fig:ET_TPT_GBC_3}(iii)). Importantly, we find that after the appearance of exceptional points (for non-zero $t'_R, t'_L$), there emerge bound states localized at the boundaries of the system. The boundary states do not arise for arbitrary small values of $t'_L-t_L$ or $ t'_R - t_R$, however, arise for $|t'_R-r t_R|>0$. These boundary states are characterized by the disjoint generalized momenta which touch the contours of GBZs at $t'_R= r t_R$ (See Fig.~\ref{fig:ET_TPT_GBC_3} (i)(c)-(d)). The touching of generalized momenta manifests as the exceptional points in the eigenvalue spectra (See Fig.~\ref{fig:ET_TPT_GBC_3}(i)(c), (iii)). Thus, the numerical finding of exceptional points due to the GBC is confirmed by touching of disjoint generalized momenta to the GBZ (See Fig.~\ref{fig:ET_TPT_GBC_3} (i)(c), (iii)). Therefore, the emergence of the exceptional points in the eigenvalue spectra defines the exceptional transition of the GBZ. 

    We note here that at $t'_R=t'_L=0$, there are additional exceptional points corresponding to the coalescence of all the eigenvalues and eigenstates in the thermodynamic limit which are dictated by the vanishing phase rigidity for all the eigenstates and are confirmed by the merging of two GBZs (See Fig.~\ref{fig:ET_TPT_GBC_3} (i)(a)). At this point, the HN model realizes the OBC and we observe NHSE.

    \textbf{Topological phase transition of GBZ}:
    Similar to the GBC-(I) and GBC-(II), in this case also, we find that the appearance of the boundary states after the exceptional points are marked by the non-trivial values of the non-Bloch topological invariant ($W_{\rm non-Bloch}$) (See Fig.~\ref{fig:ET_TPT_GBC_3}(ii)). Fig.~\ref{fig:ET_TPT_GBC_3}(ii) dictates the quantized change in the non-Bloch topological invariant at the exceptional points which are marked by the vanishing of phase rigidity $R_1$ and $R_{N}$ (See Fig.~\ref{fig:ET_TPT_GBC_3}(iii)). Therefore, the non-Bloch topological invariant characterizes the topological phase transition of GBZ. 
    
    In this case, the effective boundary matrix $\tilde{H}_B$ is non-Hermitian (See Sec.~\ref{sec:sym_HB}), and accordingly, the Berry phase associated with the effective spinor will not be quantized. We also determine the number of bound states ($n_b$) in the ($t'_R,t'_L$)-plane (See Fig.~\ref{fig:ET_TPT_GBC_2}(iii) (right)). The non-Bloch topological invariant ($W_{\rm non-Bloch}$) shows a similar phase diagram in the ($\epsilon$, $\epsilon'$)-plane (See Figs.~\ref{fig:ET_TPT_GBC_2}(iii)-(iv) (right)) which characterizes the topological phase transition of GBZ where boundary states emerge out of bulk continuum.
    
    It is interesting to compare these results with 1D Hermitian systems ($t_R=t_L\implies r=1$) where the bound states appear for arbitrary small values of $t'_L-t_L$. There is no exceptional point in this case, and the bound states arise at certain points in the parameter space when only eigenvalues (at least two) touch each other which is also known as the band-touching point or diabolic point.

\clearpage
\section{Generalization to many body dynamical systems}
To generalize our results to many body systems, we consider the Kuramoto model of coupled oscillators~\cite{Ritort_review_Kuramoto_2005, Gupta_Kuramoto_review_2014, Kurths_Kuramoto_complex_network_2016}, which describes the synchronization phenomenon~\cite{Kurths_book_synchronization_2001} in various physical systems including Josephson junction arrays\cite{Synchronization_josephson_strogatz_1998}, laser arrays~\cite{Exp_twisted_states_Kuramoto_laser_2017, Kuramoto_chimera_laser_2017, Kuramoto_laser_array_Takemura_2021}.
We consider the Kuramoto model of locally coupled oscillators (LCKM: locally coupled Kuramoto model) on a 1D lattice of a finite number of oscillators $N$ with periodic boundary conditions. The Kuramoto model of locally coupled oscillators is given by~\cite{Ritort_review_Kuramoto_2005, Gupta_Kuramoto_review_2014, Kurths_Kuramoto_complex_network_2016},
\begin{align}\label{Kuramoto_model}
\frac{d\theta_i(t)}{dt} &= \omega_i  + K_{R} \sin(\delta\theta_{i+1,i}(t))+ K_{L} \sin(\delta\theta_{i-1,i}(t))+ D\eta_{i}(t),
\end{align}
which describes the time evolution of the phase $\theta_i(t)$ of oscillator $i$ with natural frequency $\omega_i$ for $i\in\lbrace 1,2,...,N \rbrace$ with $\delta\theta_{j, i}(t) = \theta_j(t) -\theta_i(t)$ being the phase difference between the oscillators $i,j \in \lbrace 1,2,...,N\rbrace$. $K_{R}$ and $K_{L}$ denote the strengths of forward and backward nearest neighbor couplings, and $\eta_i(t)$ denotes the random white noise of strength $D$ with $\langle \eta_i(t)\rangle =0 $ and $\langle \eta_i(t) \eta_j(t^{\prime})\rangle=\delta_{i j}\delta(t-t^{\prime})$. Here $\langle...\rangle$ denotes the average over independent noise realizations. In the above model, the non-reciprocity arises with $K_R\neq K_L$. The non-reciprocal interactions can occur in the non-equilibrium environment~\cite{non_equi_Newtons_law_broken_2015}. Examples of systems having non-reciprocal interactions include meta-materials~\cite{metamaterial_acoustic_2014, Non_reciprocal_robotic_metamaterial_2019, metamaterial_optics_photonics_2019, metamatrial_topoelectric_circuit_2020}, active matter~\cite{Ramaswamy_active_matter_2022, Vitelli_topological_active_matter_2022}, living matter~\cite{nonreciprocal_living_matter_2023}.

\subsection{LCKM for two oscillators}
Before discussing the LCKM on a 1D lattice with a finite number of oscillators, we review the LCKM of two oscillators first. We consider the following dynamical equations for two oscillators $\theta_1$ and $\theta_2$,
\begin{align}
    \frac{d\theta_1(t)}{dt} = K_R \sin(\theta_2-\theta_1),\nonumber\\
    \frac{d\theta_2(t)}{dt} = K_L \sin(\theta_1-\theta_2).
\end{align}
\textbf{Stationary synchronized solutions:}
For reciprocal coupling $K_R=K_L=K$, stationary synchronized solutions exist $\delta\theta_{1,2} \equiv \delta\theta=0$ for aligned phase (in phase synchronized solutions) and $\delta\theta =\pi$ for anti-aligned phase (out of phase synchronized solutions)\cite{Strogatz_1988}. 
Similarly, for non-reciprocal couplings $K_R\neq K_L$, the oscillators align with each other for $(K_R+K_L)>0$ or anti-align with each other for $(K_R+K_L)<0$. Therefore, we have two possible synchronized phases which can be understood as follows:

(i) for $(K_R+K_L)>0$, one of the oscillators is going to chase down the second one and they will eventually align (in phase synchronized solutions) (See Fig.~\ref{fig:two_stable_phases}).

(ii) for $(K_R+K_L)<0$, again one of the oscillators will win and they will eventually anti-align (out of phase synchronized solutions) (See Fig.~\ref{fig:two_stable_phases}). 
\begin{figure}[htbp]
	\centering
	\includegraphics[width=1.0\linewidth]{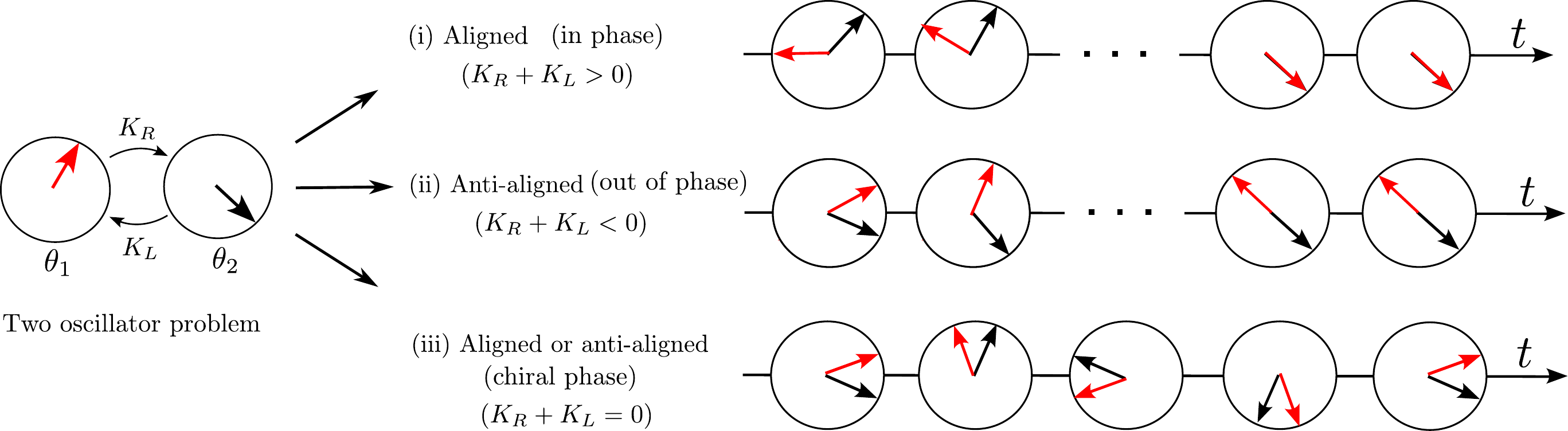}
	\caption{Synchronized solutions for LCKM for two oscillators (See text).}
	\label{fig:two_stable_phases}
\end{figure}

\textbf{Time-dependent synchronized solutions:}
Interestingly, for exact anti-symmetric coupling \textit{i.e} $ K_R=-K_L$, the dynamical equations simplifies in the following manner,
\begin{align}
    \frac{d\delta\theta(t)}{dt} &= 0, ~ \delta\theta(t)=\theta_2(t)-\theta_{1}(t),\nonumber\\
    \frac{d\theta_{\rm avg}(t)}{dt} &= (K_R-K_L)\sin(\delta\theta(t)),~\theta_{\rm avg} = \theta_1(t)+\theta_{2}(t).
\end{align}
As a result, two oscillators move with finite and common drift velocity $ (K_R-K_L)\sin(\delta\theta(0))$ keeping constant phase difference $\delta\theta(0)$ which depends on the initial conditions. This type of synchronized stable phase has been called the \textit{chiral phase}~\cite{Fruchart2021}. Indeed, the system of two oscillators dynamically restores the U(1) ($\theta_i\rightarrow \theta_i + \theta_0$) global rotation symmetry. For $K_R\neq -K_L$, the system collapses to one of the two stable phases (in phase and out of phase).
\\ 

\subsection{LCKM for $N$ oscillators with reciprocal couplings}\label{sec:LCKM_reciprocal}
In this section, we discuss the Kuramoto model of locally coupled oscillators. We start with the stationary solutions of LCKM with reciprocal couplings \cite{Strogatz_1988, Strogatz_sync_basin_2006, max_no_fixed_points_1_ochab_2010, max_no_fixed_points_2_2016}.

\subsubsection{Stationary solutions}\label{sec:LCKM_syn_sol}
Consider the LCKM with reciprocal couplings $K_{R} = K_{L} = K$. The stationary solutions $\boldsymbol{\tilde{\theta}} = (\tilde{\theta}_1,...,~\tilde{\theta}_N)^T$ for the dynamical equations (\ref{Kuramoto_model}) are given by,
\begin{align}\label{fixed_point_1}
\frac{d\theta_i (t)}{dt}|_{\theta_{t} = \tilde{\theta}_{i}}= 0~\forall~i\in\lbrace 1,2,...,N\rbrace
&\implies\sin(\delta\tilde{\theta}_{i+1,i})+\sin(-\delta \tilde{\theta}_{i,i-1})=0,~\forall~i\in\lbrace 1,2,...,N\rbrace,\nonumber\\
&\implies\delta\tilde{\theta}_{i+1,i}= m_{i} \pi + (-1)^{m_i} \delta\tilde{\theta}_{i,i-1}, ~\forall ~i \in\lbrace 1,2,...,N\rbrace,~\text{with}~ m_{i}\in\mathbb{Z},
\end{align} 
which immediately implies that LCKM with reciprocal couplings has two types of stationary solutions $\boldsymbol{\tilde{\theta}} = (\tilde{\theta}_1,...,~\tilde{\theta}_N)^T$:
\begin{align}\label{fixed_point_2}
(i)~~\delta\tilde{\theta}_{i+1,i} - \delta\tilde{\theta}_{i,i-1} = 0~({\rm mod}~2\pi), ~\forall~i\in\lbrace 1,2,...,N\rbrace ,\nonumber\\
{\rm and}~(ii)~~ \delta\tilde{\theta}_{i+1,i} + \delta\tilde{\theta}_{i,i-1} = \pm \pi~({\rm mod}~2\pi), ~\forall~i\in\lbrace 1,2,...,N\rbrace.
\end{align}
It is important to point out that the above stationary solutions described by Eqs.~(\ref{fixed_point_1}) and (\ref{fixed_point_2}) will only hold with reference to the periodic boundary condition (\textit{i.e.} $\theta_{N+i} =\theta_i$: the first oscillator is coupled to the last one in the same fashion as that of any $\theta_i,~i\in \lbrace 2,3,..., N-1\rbrace$). In simple words, the stationary solutions of LCKM are those where the phase differences between all the pairs of oscillators are the same. This leads to two different synchronized states where the neighboring oscillators are aligned with $|\delta\tilde{\theta}_{i+1, i}|<\pi/2$ and anti-aligned with $|\delta\tilde{\theta}_{i+1, i}|>\pi/2$ depending on the coupling $K$. These stationary solutions also hold if we include the next nearest neighbor coupling (or any finite range coupling) in Eq.~(\ref{Kuramoto_model}).

\begin{figure}[htbp]
	\centering
	\includegraphics[width=1.0\linewidth]{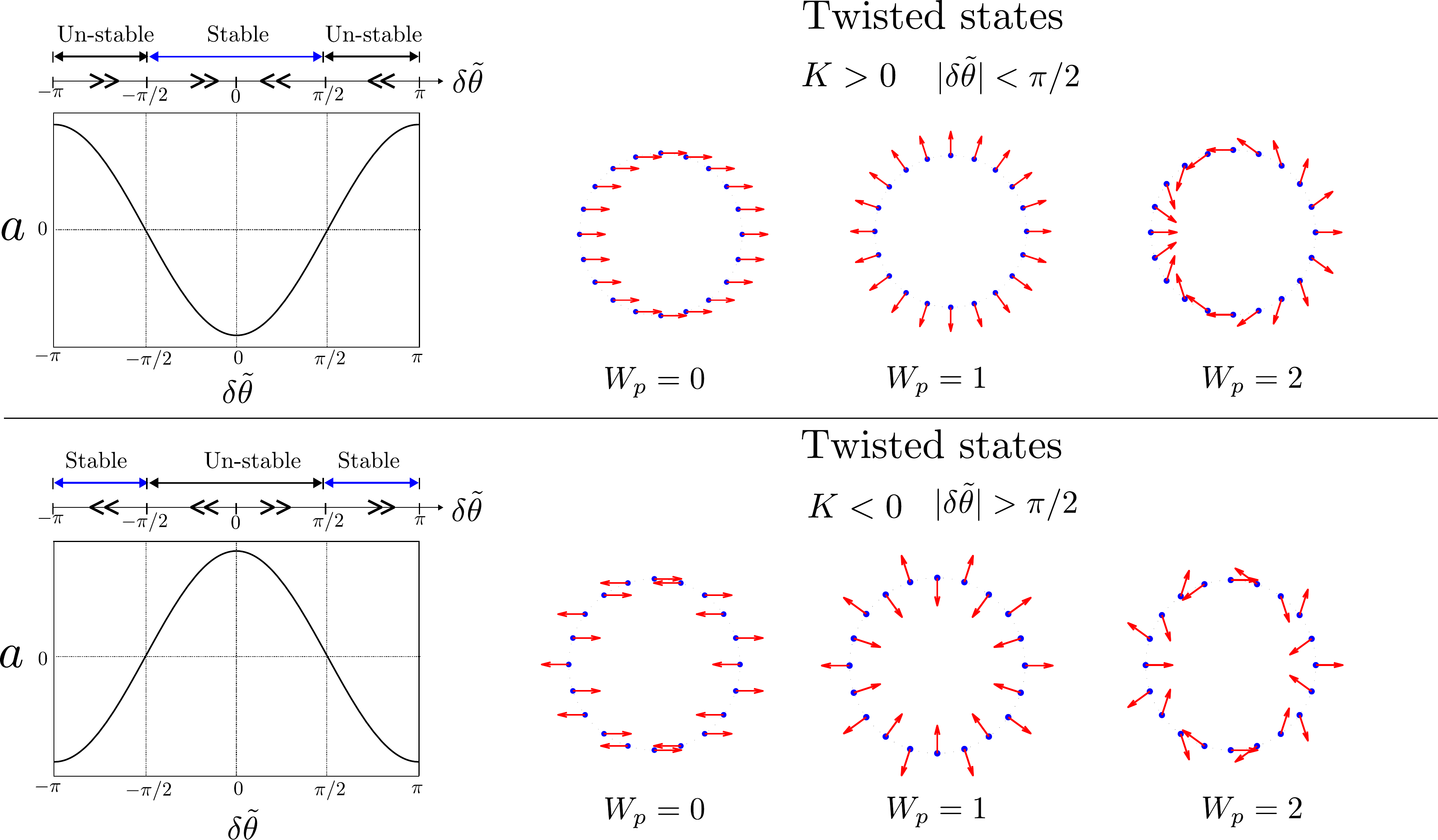}
	\caption{Figures in the left of the top and bottom panels show the variation of center of Gershgorin discs $a_{ii}\equiv a=-2K\cos\delta\tilde{\theta}$ with respect to constant nearest neighbor phase differences $\delta\tilde{\theta}$ for twisted states with $K>0$ and $K<0$, respectively (See Sec.~\ref{sec:local_stability_reciprocal} for more details on the local stability of twisted states). Figures in the right of the top and bottom panels show the vector plots of stable twisted states ($a<0$) with different winding numbers ($W_p$) for $K>0$ and for $K<0$, respectively. }
	\label{fig:N_twisted_phases}
\end{figure}

\textbf{Exact form of synchronized solutions:}
We can determine the exact form of the synchronized solutions analytically. For that, we write the synchronized solutions in  the following general form~\cite{max_no_fixed_points_1_ochab_2010, max_no_fixed_points_2_2016}:
\begin{align}
\theta_{i}(t) &= \omega_{s} t +\tilde{\theta}_{i}, ~\forall ~i\in \lbrace 1,2,...,N\rbrace ~ (\tilde{\theta}_i~\text{may be independent of $i$})
\end{align}
here $\omega_s$ is the common frequency and $\boldsymbol{\tilde{\theta}}=(\tilde{\theta}_1,...,~\tilde{\theta}_N)^T$ denote phases for the oscillators corresponding to the given synchronized solution.
It has been shown in \cite{max_no_fixed_points_1_ochab_2010} that the synchronized solutions of LCKM with reciprocal couplings take the following form:
\begin{align}\label{syn_solution}
\tilde{\theta}_{i,i-1} = \arcsin{\Big(f + \frac{1}{K} \sum_{j=i}^{N-1} (\omega_j -\omega_s)\Big)} + 2p_i \pi,~p_i\in \mathbb{Z},~\forall ~i\in \lbrace 1,2,...,N\rbrace,
\end{align}
here $f=\sin{\delta\tilde{\theta}_{N,N-1}} \in [-1,1]$ as a free parameter and $\omega_s = \frac{1}{N}\sum\limits_{i=1}^N \omega_i$ denote the average or common frequency. After summation, the above equations immediately end up to 
\begin{align}
&\sum_{i=1}^{N}\arcsin{\Big(f + \frac{1}{K} \sum_{j=i}^{N-1} (\omega_j -\omega_s)\Big)} + 2p \pi = 0 ,~\text{with}~\sum_{i=1}^{N} p_i =p \in \mathbb{Z}.
\end{align}
Hence, the synchronized solutions are characterized by the integer `$p$' which for $\delta\tilde{\theta}_{i+1, i}<<1$ is the same as the \textit{topological winding number} $W_p\in\mathbb{Z}$. In other words, a unique synchronized solution $\boldsymbol{\tilde{\theta}}^{(p)}$ is defined via topological winding number:
\begin{align}
W_{p} = \frac{1}{2\pi}\sum_{i=1}^{N}\delta\tilde{\theta}_{i+1,i}.
\end{align}
Therefore, the synchronized solutions have topological protection. Moreover, for small and equal phase differences, $\delta\tilde{\theta}_{i+1, i}=\delta\tilde{\theta} = 2\pi W_p/N,~\forall~i$, phases of individual oscillators are given by $\tilde{\theta}_{i}^{(p)} = 2\pi W_p i/N + \tilde{\theta}_{0}^{(p)}$, $\forall~i\in\lbrace 1,2,..., N\rbrace$. Here $\tilde{\theta}_{0}^{(p)}$ is an arbitrary uniform phase shift. These synchronized solutions take the form of uniformly twisted waves with the number of complete phase twists equal to the value of the topological winding number. Accordingly, these synchronized solutions are also called $W_p$-\textit{twisted states} with topological charge or the number of complete phase twist $W_p\in \mathbb{Z}$ \cite{Strogatz_sync_basin_2006, max_no_fixed_points_1_ochab_2010} (See vector plots of stable twisted states in the top and bottom panels (right) of Fig.~\ref{fig:N_twisted_phases}). We further divide different synchronized phases into the following two parts:

(i) \textbf{State of complete phase synchronization:}
In this case, starting from random initial conditions, after a sufficiently long time the phase differences among the oscillators vanishes \textit{i.e.} $|\theta_{i}(t\rightarrow \infty) - \theta_{j}(t\rightarrow \infty)|\rightarrow 0$ or $\theta_{i}(t\rightarrow \infty) = {\rm const.},~\forall~i\neq j$. Therefore, this synchronized state of oscillators is called \textit{state of complete phase synchronization} which is also known as the \textit{sync}. In this case, the synchronized solutions are characterized by $\boldsymbol{\tilde{\theta}}^{(p)}$ with the topological winding number $W_p=0$ \cite{Strogatz_sync_basin_2006, max_no_fixed_points_1_ochab_2010}. 

(ii) \textbf{Phase-locked state:}
In this case, depending on different random initial conditions the phase differences among the oscillators approach to constant values \textit{i.e.} $|\theta_{i}(t\rightarrow \infty) - \theta_{j}(t\rightarrow \infty)|\rightarrow {\rm const.},~\forall~i\neq j$. Therefore, all the oscillators approach their different steady state value such that $\frac{d\theta_i}{dt} = 0~\forall~i$, and the oscillators are said to be in the \textit{phase-locked state}. Moreover, for small and equal constant phase differences, the phase-locked states are known as $W_p$-\textit{twisted states} with the number of complete phase twist $W_p\in \mathbb{Z}$ \cite{Strogatz_sync_basin_2006, max_no_fixed_points_1_ochab_2010}. This essentially means that stationary synchronized solutions $\boldsymbol{\tilde{\theta}}^{(p)}$ have topological protection identified by topological winding number $W_p\in \mathbb{Z}-\lbrace 0\rbrace$.

\subsubsection{Stability analysis of the synchronized solutions}\label{sec:local_stability_reciprocal}
\paragraph{Jacobian of linear stability:}
One can analyze the linear stability of the above-synchronized solutions by perturbing the system by a small amount such that $\theta_{i}(t)\rightarrow \tilde{\theta}_i + \phi_{i}(t)$, with $|\phi_{i}(t)| \ll 1$. Then the evolution equation for $\phi_i(t)$ is given by,
\begin{align}
\frac{d \phi_i (t)}{d t} \approx (-J_{i,i})\phi_i + J_{i+1, i} \phi_{i+1} + J_{i-1} \phi_{i-1,i},~\forall~i\in\lbrace 1,2,...,N\rbrace.
\end{align}
here $J_{i,i} = -K(\cos{\delta\tilde{\theta}_{i+1,i}} + \cos{\delta\tilde{\theta}_{i-1,i}}),~J_{i+1,i} = K\cos{\delta\tilde{\theta}_{i+1,i}},  J_{i-1,i} = K\cos{\delta\tilde{\theta}_{i-1,i}}$. We can also write the above equations in vector form as $\frac{d \boldsymbol{\phi}}{d t} \approx \hat{\mathcal{J}} \boldsymbol{\phi}$, with $\hat{\mathcal{J}}$ being a $N\times N$ matrix (also known as the Jacobian or perturbation matrix) and $\boldsymbol{\phi} =(\phi_1,...,\phi_N)^T$.
\begin{eqnarray}\label{Jacobian_reciprocal}
\frac{d}{dt}\begin{bmatrix}
		\phi_1\\
		\vdots\\
		\phi_N 
		\end{bmatrix}= \hat{\mathcal{J}}
		\cdot \begin{bmatrix}
             \phi_1\\
		\vdots\\
		\phi_N 
		\end{bmatrix},~~~\hat{\mathcal{J}}=\begin{bmatrix}
				J_{1,1} & K\cos{\delta\tilde{\theta}_{2,1}}&\cdots& K\cos\delta\tilde{\theta}_{N,1}&\\
				K\cos\delta\tilde{\theta}_{1,2}&J_{2,2} & K\cos\delta\tilde{\theta}_{3,2}&\cdots\\
				\vdots  & \vdots  & \ddots & \vdots  \\
				K \cos\delta\tilde{\theta}_{1,N}& 0& \cdots& J_{N,N}\\
			\end{bmatrix}
			\end{eqnarray}
			\\
   We note here the remarkable similarity between the Jacobian $\hat{\mathcal{J}}$ and the one-dimensional tight-binding Hamiltonian with the generalized boundary conditions for 1D lattice systems with reciprocal hopping. We will come to this point later. We summarize the relationship between the Jacobian and local stability of synchronized solutions in the following manner~\cite{Strogatz_sync_basin_2006, max_no_fixed_points_1_ochab_2010}:

   We know that for the first-order ordinary differential equations, any initial conditions non-identical to a stable solution are perturbations for that stable solution. Moreover, the negative eigenvalues of the Jacobian or the perturbation matrix indicate the speeds of return to that stable solution for small perturbations. Therefore the initial conditions very close to the stable solution with negative eigenvalues of the Jacobian will approach the stable solution eventually. We also know that the basins of attraction are connected sets containing stable solutions. Based on these arguments, we assume the correspondence between the volume of the basin of attraction and the stability of the solution, which means that the solutions whose all eigenvalues are strongly negative are more stable and have a larger basin of attraction. Similarly, the solutions with  maximal negative eigenvalues closer to zero are less stable and have smaller basins of attraction ~\cite{Strogatz_sync_basin_2006, max_no_fixed_points_1_ochab_2010}. We analyze the eigenvalue spectra of the Jacobian corresponding to different synchronized solutions in the following sections.

\paragraph{Local stability analyses:}
The stability of the synchronized solutions of LCKM with reciprocal couplings has been studied analytically with the help of \textit{Gershgorin circle theorem}~\cite{Gershgorin_1931} which states that every eigenvalue of a general $N\times N$ complex matrix $ A = [a_{ij}]_{i,j\in\lbrace 1,2,..., N\rbrace}$ with complex entries $a_{ij}$ lies within at least one of the closed \textit{Gershgorin discs} $D(a_{ii}, R_i) \subseteq \mathbb{C},~\forall~i\in\lbrace 1,2,..., N \rbrace$ centered at $a_{ii}$ having radius $R_i = \sum_{j\neq i} |a_{ij}| ~\forall~ i\in\lbrace 1,2,..., N \rbrace$. For the Jacobian matrix (See Eq.~\eqref{Jacobian_reciprocal}), the center and radius of Gershgorin discs are as follows:
\begin{align}
    a_{ii} = -K(\cos{\delta\tilde{\theta}_{i+1,i}} + \cos{\delta\tilde{\theta}_{i-1,i}}),~\text{and}~R_i= |K\cos{\delta\tilde{\theta}_{i+1,i}}+ K\cos{\delta\tilde{\theta}_{i-1,i}}|,~\forall~i\in \lbrace 1,2,..., N \rbrace. 
\end{align}
Utilizing Gershgorin theorem together with numerical analyses for $\delta\tilde{\theta}_{i+1, i} \in [-\pi,\pi]$, there are following findings depending on the non-zero eigenvalues of the Jacobian~\cite{Strogatz_sync_basin_2006, max_no_fixed_points_1_ochab_2010, max_no_fixed_points_2_2016,fixed_point_unstable_1_2017,fixed_point_unstable_2_Taylor_2012, Mihara_stability_R_2019}:
\begin{align*}
&(1a)~\text{If}~ |\delta\tilde{\theta}_{i+1,i}|>\pi/2,~\forall~i,~\text{and}~K>0,~\text{then the eigenvalues}~ \lambda_i~\text{satisfy}:~{\rm Re~ \lambda_i~> 0}\implies \text{unstable solutions},\nonumber\\
&(1b)~\text{If}~ |\delta\tilde{\theta}_{i+1,i}|>\pi/2,~\forall~i,~\text{and}~K<0,~\text{then the eigenvalues}~ \lambda_i~\text{satisfy}:~{\rm Re~ \lambda_i~< 0}\implies \text{stable solutions},\nonumber\\
&(2a)~\text{If}~ |\delta\tilde{\theta}_{i+1,i}|<\pi/2,~\forall~i,~\text{and}~K>0,~\text{then the eigenvalues}~ \lambda_i~\text{satisfy}:~{\rm Re~ \lambda_i~< 0}\implies \text{stable solutions},\nonumber\\
&(2b)~\text{If}~ |\delta\tilde{\theta}_{i+1,i}|<\pi/2,~\forall~i,~\text{and}~K<0,~\text{then the eigenvalues}~ \lambda_i~\text{satisfy}:~{\rm Re~ \lambda_i~> 0}\implies \text{unstable solutions},\nonumber\\
&(3)~\text{If}~\exists ~i: |\delta\tilde{\theta}_{i+1,i}|>\pi/2~\land~ \exists~ j: |\delta\tilde{\theta}_{j+1,j}|<\pi/2,~\text{then  stability of the solutions may not be determined}.
\end{align*}

The Jacobian (Eq.~\eqref{Jacobian_reciprocal}) also has a zero eigenvalue due to the U(1) global translation symmetry of Eq.~\eqref{Kuramoto_model}, $\theta_i \rightarrow \theta_i + \theta_0, ~\forall ~ i\in\lbrace 1,2,...,N \rbrace$. The U(1) symmetry implies that the matrix $\hat{\mathcal{J}}$ has cyclic structure \cite{Strogatz_sync_basin_2006, max_no_fixed_points_1_ochab_2010} \textit{i.e.} 
\begin{align}
\sum\limits_{i=1}^{N}J_{i,j}=0~\forall j\in\lbrace 1,2,...,N\rbrace~\text{or}~ \sum\limits_{j=1}^{N}J_{i,j}=0~\forall i\in\lbrace 1,2,...,N\rbrace.
\end{align} 
The matrices having the property of vanishing of the sum of individual columns/rows are called zero-line sum (ZLS) matrices \cite{Boukas_zero_line_sum_matrices} which possess at least one zero eigenvalue (say $\lambda_1=0$) with constant eigenvector $(1,1,...,1)^T$. Therefore, one of the eigenvalues of $\hat{\mathcal{J}}$ is always zero, which implies that all the solutions are neutrally stable under the rigid rotations (homogeneous/global translations). The normal mode corresponding to this zero eigenvalue is known as the \textbf{\textit{Goldstone mode}}.

Now we are ready to discuss the stability of the twisted states which are the synchronized solutions with small and equal phase differences $\delta \tilde{\theta}_{i+1, i}\equiv \delta\tilde{\theta} =2\pi W_p/N,~\forall~i$, and are characterized by the \textit{topological winding number} $W_p$~\cite{Strogatz_sync_basin_2006, max_no_fixed_points_1_ochab_2010, max_no_fixed_points_2_2016,fixed_point_unstable_1_2017,fixed_point_unstable_2_Taylor_2012, Mihara_stability_R_2019}. In the top and bottom panels of Fig.~\ref{fig:N_twisted_phases}, we show the variation of the center of the Gershgorin disc $a = -2 K \cos\delta\tilde{\theta}$ for the Jacobian matrix corresponding to twisted states as a function of nearest neighbor phase difference $\delta\tilde{\theta}$.  It is easy to observe that for $K>0$, centers of all the Gershgorin discs are negative ($a<0$) and positive ($a>0$) for twisted states with $|\delta\tilde{\theta}|<\pi/2$ and $|\delta\tilde{\theta}|>\pi/2$, respectively. Accordingly, for $K>0$, all non-zero eigenvalues will have a negative (positive) real part for twisted states with $|\delta\tilde{\theta}|<\pi/2$ ($|\delta\tilde{\theta}|>\pi/2$). Therefore, for $K>0$, twisted states with $|\delta\tilde{\theta}|<\pi/2$ ($|\delta\tilde{\theta}|>\pi/2$) will be stable (unstable). Moreover, the stable $W_p-$twisted states for $K>0$ become unstable for $K<0$ and vice versa. We tabulate the exact number of stable and unstable twisted states for a finite-size system in Table~\ref{tab:twisted_stability}.

    \begin{center}
    \begin{table}[htbp]
    \def\arraystretch{2.5}
    \begin{tabular}{|c|c|c|c|c|c|c|c|c|c|c|c|c|c|c|}
    \hline
    $W_p$ & ~~0~~ & ~~1~~& ~...~ &$\frac{N}{4}$-1& ~$\frac{N}{4}$~ & $\frac{N}{4}$+1 & $\frac{N}{4}$+2 & ~...~ &$\frac{3N}{4}$-1& ~$\frac{3N}{4}$~ & $\frac{3N}{4}$+1& ~...~ & $N$-2 &$N$-1 \\
     \hline
    $K>0$  & s & s & s & s & -& u & u & u &u &- &s &s &s& s \\
    \hline
    $K<0$  & u & u & u & u & -& s & s & s &s &- &u &u &u& u \\
    \hline
    \end{tabular}
    \caption{Table for the stability of twisted states for different winding numbers ranging from 0 to $N-1$ for even number ($N$) of  oscillators. Stable, unstable, and neutral phases are dictated by symbols (s), (u), and (-), respectively. For odd $N$, one can accordingly obtain the table using Gershgorin theorem~\cite{Strogatz_sync_basin_2006, max_no_fixed_points_1_ochab_2010, max_no_fixed_points_2_2016,fixed_point_unstable_1_2017,fixed_point_unstable_2_Taylor_2012, Mihara_stability_R_2019}.}\label{tab:twisted_stability}
    \end{table}
    \end{center}

    \begin{figure}[htbp]
	\centering
	\includegraphics[width=1\linewidth]{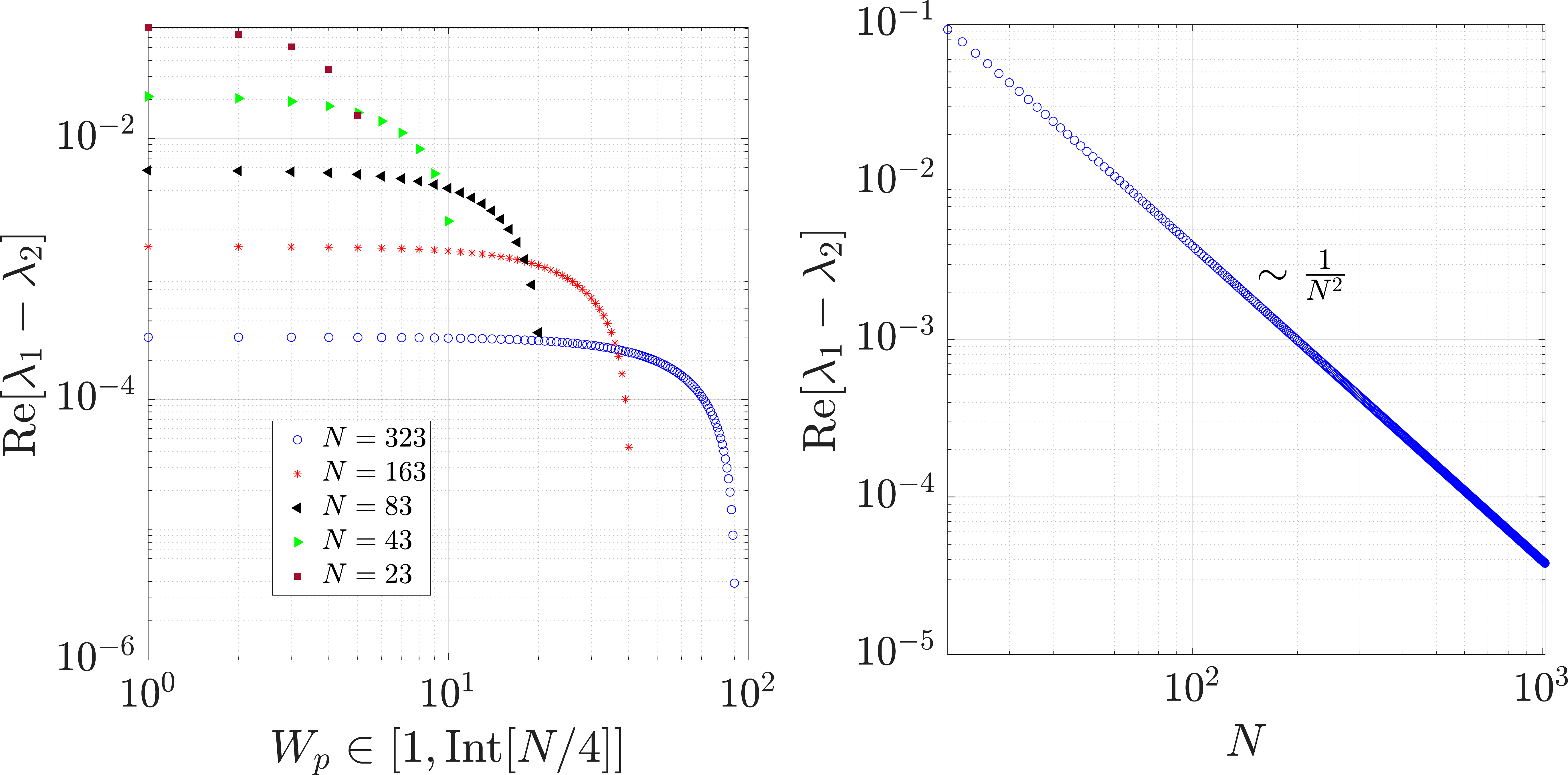}
	\caption{Left panel: behavior of maximal negative eigenvalue close to zero (Re$[\lambda_1-\lambda_2]$)of the Jacobian as a function of winding number ($W_p$) for different system sizes ($N$). Right panel: Re$[\lambda_1-\lambda_2]$ vs $N$ for a given winding number. Parameters: $K_R=1.0,~K_L=1.0$.}
	\label{fig:basin_of_attraction_reciprocal}
    \end{figure}

\paragraph{Basins of attraction and stability of twisted states:}
We discussed in the previous section that for stable twisted states, all the eigenvalues of the Jacobian have negative real parts except $\lambda_1=0$ which corresponds to the Goldstone mode. The eigenvalue spectra are open segments on the negative Re$[\lambda]$ axis in the (Re[$\lambda$], Im[$\lambda$])-plane. We find that the difference between the eigenvalue corresponding to the Goldstone mode ($\lambda_1 =0$) and the maximal negative eigenvalue ($\lambda_2$) decreases with an increase in the value of topological winding number $W_p$ and system size $N$ (See the left panel of Fig.~\ref{fig:basin_of_attraction_reciprocal}). It essentially means that the stable twisted states with higher values of winding numbers have smaller basins of attraction and thus are less stable in comparison to the ones with a smaller winding number which is again confirmed by our numerical simulations. Therefore, the volume of the basins of attraction and the value of the winding numbers are highly correlated and the solutions with large winding numbers are marginalized. Although the decrease in the maximal eigenvalue with the increase in the system size indicates that arbitrary small unbounded noise will destabilize the stable solutions with finite winding numbers in the thermodynamic limit.

 Through our numerical simulations, we observe that starting with the random initial conditions, the probability of finding stable synchronized solutions with higher finding numbers is low as compared to that of smaller winding numbers. We find that the stationary solution with the topological winding number $W_p=0$ is globally stable. Whereas other stable synchronized solutions with $W_p\neq 0$ are only locally stable \textit{i.e.} they have finite basins of attraction which again support the linear stability analyses.  Hence the LCKM with reciprocal coupling hosts stable synchronized solutions for finite sizes of the system. These stable solutions can be found explicitly depending on the initial conditions and their number is proportional to the finite size of the system.

\subsubsection{Stability of synchronized solutions in the presence of random white noise and disorder in frequency}\label{sec:phase_slip_reciprocal}
In this section, we study the stability of the synchronized solutions in the presence of random white noise and also disorder in frequency. Firstly, we describe the numerical method which we have utilized in our numerical simulations.
\paragraph{Details about numerical simulations:}
 In order to incorporate the white noise term correctly, we implement Euler-Maruyama's method and also Heun's method in our numerical simulation to integrate the stochastic differential equation \cite{Stochastic_numerical_diff_2012}. These algorithms are important to use because the standard methods like Euler, Runge-Kutta, \textit{etc.} assume that the function governing the dynamics are well behaved and differentiable. However the white noise is not a differential function even for a single realization, hence its correct incorporation is necessary.  The basic idea to implement these algorithms is to integrate the dynamical equation: although the values of the white noise and its derivatives are not well defined, the first integral of the white noise is, and it is the Wiener process which is a continuous function \cite{Stochastic_numerical_diff_2012, Notes_spin_dynamics_wysin}. In our case, using the Euler-Maruyama scheme, the iterative solutions for the dynamical equation are given by,
\begin{align}
   \vv{\theta}(t+d t) = \vv{\theta}(t) + f(\vv{\theta}(t),~t) d t + D\vv{\eta}\sqrt{d t},
\end{align}
here $d t$ is the time interval, $\boldsymbol{\theta}(t)=(\theta_1, \theta_2,..., \theta_N)^{\rm T}$ is a column vector describing the phases of all oscillators, $f(\vv{\theta}(t),~t) = \vv{\omega} + \vv{S}^{\rm T}(t)\vv{\times} (\mathcal{K}\cdot \vv{S}(t))$  with $\vv{S}(t) = (\cos(\vv{\theta}(t)),\sin(\vv{\theta}(t)))$. $\mathcal{K}$ represents the coupling matrix describing the nearest neighbor couplings of oscillators on a lattice. $\vv{\omega}=(\omega_1, \omega_2,..., \omega_N)^{\rm T}$ describes the natural frequencies for all oscillators and can take random values in the presence of frequency disorder. $\vv{\eta}$ is a vector of random numbers corresponding to the Gaussian white noise. It is important to note here that if the random number generated has a different variance than unity then the above formula should modify accordingly \cite{Notes_spin_dynamics_wysin}. The Euler-Maruyama method has a strong order of convergence of 0.5. We use a better method called Heun's method which has strong order of convergence is 1.0 and will produce better trajectories for our stochastic differential equation. Again, using Heun's scheme, the iterative solution for our stochastic equation becomes,
\begin{align}
   \vv{\theta}(t+d t) = \vv{\theta}(t) + \frac{1}{2}(f_1 + f_2) d t + D\vv{\eta}\sqrt{d t},
\end{align}
here $f_1=f(\vv{\theta}(t),~t), ~f_2=f(\vv{\theta}(t)+d t f_1 + D\vv{\eta}\sqrt{d t},~t+d t)$. The higher schemes for numerical integration become much more complicated and for comparison of numerical results for different schemes for Langevin-type stochastic differential equations has been provided in Ref.~\cite{Stochastic_numerical_diff_2012}. Here we consider Heun's scheme for our numerical simulation to implement the Gaussian white noise. Also, we consider the fourth-order Runge-Kutta method in the absence of the random white noise together with Heun's scheme. For small enough time intervals, both schemes give similar results in the absence of noise. 
\begin{figure}[htbp]
	\centering
	\includegraphics[width=1\linewidth]{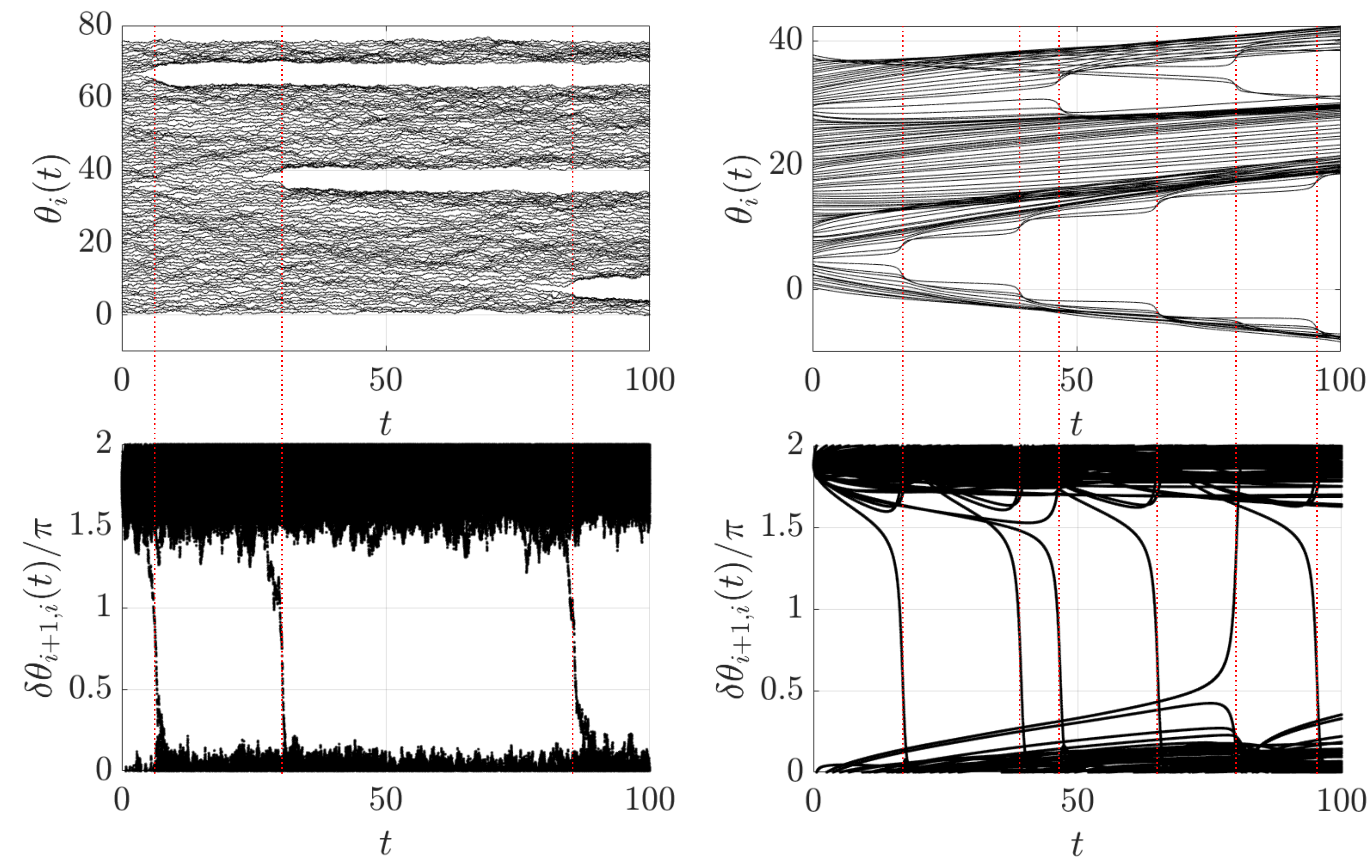}
	\caption{Stability of twisted phases: noise and disorder-induced phase transitions as a direct consequence of phase slips. Time evolution of the individual phases $\theta_i(t)$ (upper panel) and nearest neighbor phase differences $\delta\theta_{i+1, i}(t)$ (lower panel). The red dotted lines mark the noise (left panel) and disorder (right panel)-induced non-equilibrium phase transitions due to phase slips. It is clearly shown that in the time interval of the phase slips, one of the phase differences undergoes a change of $2\pi$ (See related discussion in the Sec.~\ref{sec:phase_slip_reciprocal}). Parameters: Left panel: $K_R=1.0,~K_L=1.0,~D=0.15,~N=100.$ Right panel: $K_R=1.0,~K_L=1.0,~\Delta\omega=0.25,~N=100.$.}
	\label{fig:phase_slip_noise_disorder_reciprocal}
\end{figure}
\paragraph{Noise induced non-equilibrium phase transitions:}

We have discussed in the previous section that the twisted states are characterized by distinct winding numbers $W_p\in \mathbb{Z}$. Therefore, the transition between two different winding number sectors (say $W_p$ and $W_p+1$) is only possible when one of the phases (say $\theta_N$) changes from $2\pi W_p$ to $2\pi (W_p+1)$ which implies that one of the phase differences (say $\delta\theta_{1, N}=\theta_1-\theta_N$) undergoes a change from $\delta\theta_{1, N}$ to $\delta\theta_{1, N}+2\pi$. Therefore, the transition between different winding number sectors representing the twisted states can occur through the abrupt change of one of the phase differences by $2\pi$.  This relatively rapid change of the phase difference by $2\pi$ is called \textit{phase slip} (See Ref.~\cite{Kurths_book_synchronization_2001} for more details).

In the presence of random white noise, the phase slips occur resulting in noise-induced phase transitions to the different twisted states (See left panel of Fig.~\ref{fig:phase_slip_noise_disorder_reciprocal}). After a long enough time, the oscillators finally end up in the complete phase synchronized state with $W_p=0$ and the phase slips occur around this state $W_p=0$. We note here that for a given noise realization, both phase transitions from $W_p$-twisted state to the $(W_p+1)$-twisted state and also $W_p$-state  to the $(W_p-1)$-twisted state can occur. These phase transitions are known as \textit{noise-induced non-equilibrium phase transitions}. In addition, for twisted states with larger winding numbers, the phase slips occur rather quickly for small strengths of white noise due to the larger size of the basins of attraction. Whereas for twisted states with small winding numbers, the phase slips occur after longer times (See Fig.~\ref{fig:phase_slip_noise_disorder_reciprocal} (left)). We also observe similar non-equilibrium phase transitions as a consequence of phase slips due to random disorder in the frequency of Kuramoto oscillators (See Fig.~\ref{fig:phase_slip_noise_disorder_reciprocal} (right)). 

\paragraph{Analytical modeling of noise-induced phase transitions:}
In order to understand the nature of the noise-induced phase transitions from one twisted state to the other, we model the adiabatic evolution of the phases as follows,
\begin{align}
    \theta_{i}(t)= \frac{2\pi W_p i}{N} (1-t) + \frac{2\pi W_{p^{\prime} }i}{N} t,~~(0\leq t\leq 1),~~i\in\lbrace 1,2,...,N\rbrace, 
\end{align}
here $W_p$ and $W_{p^{\prime}}$ are the winding numbers corresponding to different twisted states. At $t=0$, the oscillators are in the $W_p$-twisted state and evolve in an adiabatic manner to $W_{p^{\prime}}$-twisted state at $t=1$ as a consequence of the phase slip in the presence of noise. In the interval of the phase slip, one of the phase differences $\delta\theta_{1, N}$ undergoes a change from $\delta\theta_{1, N}$ to $\delta\theta_{1, N} + 2\pi$, and it passes through the points $\delta\theta_{1, N}=\pi/2,3\pi/2$. 

At the non-equilibrium phase transition, the Jacobian corresponding to a given twisted state ($W_p$) starts with periodic boundary condition (PBC), then $\delta\theta_{1, N}=\pi/2,3\pi/2$ it realizes the open boundary condition (OBC) and finally, end up into the Jacobian with PBC corresponding to different twisted state ($W_{p^{\prime}}$). Interestingly, in the interval of the phase slip, the Jacobian is similar to the one-dimensional (1D) tight bonding Hamiltonian with generalized boundary conditions (See Sec.~\ref{sec:def_GBC}).  During $\pi/2<\delta\theta_{1,N}<3\pi/2$, the Jacobian has an eigenvalue with the positive real part which physically represents the appearance of the phase instability during the non-equilibrium phase transition from one stable twisted state to the other. From the perspective of the tight-binding model for the 1D chain, this eigenmode with positive eigenvalue is similar to the bound state localized at the edge of the system due to the generalized boundary conditions. The non-Bloch band theory (See Sec.~\ref{sec:non_Bloch_theory}) is also applicable here, however, we will not discuss this in detail (See for more detail in Secs.~\ref{sec:noise_chiral_phase_transition} and \ref{sec:non_Bloch_theory_Jaco}).

\subsection{LCKM for $N$ oscillators with non-reciprocal couplings}\label{sec:non_reciprocal_chiral} 
    We consider the LCKM on a lattice with the periodic boundary condition consisting of $N$ oscillators which are coupled through non-reciprocal couplings defined as $K_{R} \neq K_{L}$ and denote $\Delta K = K_R-K_L$. We first analyze the situation with $\omega_i = 0,~D=0$. In this case, the dynamical equations will modify to the following,
    \begin{align}
    \frac{d\delta\theta_{i+1,i}(t)}{dt} &= 0, ~ \delta\theta_{i+1,i}(t)=\theta_{i+1}(t)-\theta_{i}(t),~i\in\lbrace 1,2,...,N\rbrace,\nonumber\\
    \frac{d\theta_{\rm avg}(t)}{dt} &= \sum_{i=1}^{N}(K_R-K_L)\sin(\delta\theta_{i+1,i}(t)),~\theta_{\rm avg} = \sum_{i=1}^{N}\theta_i(t).
    \end{align}
    looking at the above equations, it is straightforward to find out that the present problem is similar to the two oscillators problem. In this case, each oscillator moves with a finite common drift velocity having a constant phase difference with their neighbors. We also call these synchronized phases as \textit{chiral phases}. These chiral phases are nothing but twisted states (See Sec.~\ref{sec:LCKM_syn_sol}) with finite drift velocities which are characterized by distinct topological winding numbers. Hence the total number of the chiral phases is proportional to the number of oscillators similar to the LCKM with reciprocal coupling.
    
    It is important to note here that in contrast to the two oscillator problems, the stable chiral phases can appear for arbitrary values of non-reciprocal couplings. In this case of LCKM with a non-reciprocal coupling on a lattice with periodic boundary condition, the stable chiral phases appear as a consequence of chase and runaway motion and do not require random noise in contrast to globally coupled Kuramoto model with non-reciprocal couplings \cite{Fruchart2021}. We note here that for our case the periodic boundary condition is necessary to realize chiral phases which is also the case for realizing the twisted states of LCKM with reciprocal coupling (See Sec.~\ref{sec:LCKM_syn_sol}). 

    \subsubsection{Synchronized solutions}
    In order to find out the synchronized solutions for Eq.~\eqref{Kuramoto_model}, we take the following ansatz for individual phases
    \begin{align}\label{syn_sol}
    \theta_{i}(t) = v_{s} t +\tilde{\theta}_{i}, ~\forall ~i\in\lbrace 1,2,...,N\rbrace~~ (\tilde{\theta}_i~\text{may be independent of $i$})
    \end{align}
    here $v_s$ is the common frequency or drift velocity for each oscillator and $\boldsymbol{\tilde{\theta}}$ denote constant phases for the oscillators corresponding to the given synchronized solution. Suppose $f_i = \sin\delta\tilde{\theta}_{i,i-1}\in [-1,1]$, then Eq.~\eqref{Kuramoto_model} with the help of above solution
\begin{align}
K_R \sin(\tilde{\theta}_i -\tilde{\theta}_{i+1})+ K_L \sin(\tilde{\theta}_i - \tilde{\theta}_{i-1}) = \omega_{i}-v_s \equiv \Omega_i
\end{align}
simplifies to
\begin{align}
f_i =\frac{\Omega_i}{K_L} + \alpha f_{i+1},~~{\rm for}~~ i\in \lbrace 1,2,...,N\rbrace.
\end{align}
with $\alpha = K_R/K_L$. With the help of the above equation, we can express, $f_i$ in terms of $f_j$ with $j\in \lbrace i+1,i+2,...,i+N\rbrace$ in the following manner,
\begin{align}
f_i &=\frac{\Omega_i}{K_L} + \alpha f_{i+1},\nonumber\\
f_i &=\frac{\Omega_i}{K_L} +  \alpha \frac{\Omega_{i+1}}{K_L} + \alpha^2 f_{i+2},\nonumber\\
f_i &=\frac{\Omega_i}{K_L} +  \alpha \frac{\Omega_{i+1}}{K_L} + \alpha^2 \frac{\Omega_{i+2}}{K_L} + \alpha^3 f_{i+3},\nonumber\\
.\nonumber\\
.\nonumber\\
.\nonumber\\
f_i &=\frac{\Omega_i}{K_L} +  \alpha \frac{\Omega_{i+1}}{K_L} + \alpha^2 \frac{\Omega_{i+2}}{K_L} +...+\alpha^{N-1} \frac{\Omega_{i+N-1}}{K_L} + \alpha^N f_{i+N},
\end{align}
which again can be re-written as follows,
\begin{align}
f_i &=\sum_{j=0}^{N-1}\alpha^j \frac{\Omega_{i+j}}{K_L} + \alpha^N f_{i+N},~~{\rm for}~~ i\in \lbrace 1,2,...,N\rbrace.
\end{align}
We utilize periodic boundary conditions $\tilde{\theta}_i = \tilde{\theta}_{i+N}$ in above equation which guarantees that $f_{i+N} = \sin(\tilde{\theta}_{i+N} -\tilde{\theta}_{i-1+N}) = \sin(\tilde{\theta}_i-\tilde{\theta}_{i-1}) = f_i$. Thus above equation reduces to,
\begin{align}\label{fixed_non_reci}
f_i &=\sum_{j=0}^{N-1}\alpha^j \frac{\Omega_{i+j}}{K_L} + \alpha^N f_{i},~~{\rm for}~~ i\in \lbrace 1,2,...,N\rbrace,\nonumber\\
\textit{i.e.}~~(1-\alpha^N) f_i &=\sum_{j=0}^{N-1}\alpha^j \frac{\Omega_{i+j}}{K_L} ,~~{\rm for}~~ i\in \lbrace 1,2,...,N\rbrace.
\end{align}

\begin{figure}[htbp]
	\centering
	\includegraphics[width=1.0\linewidth]{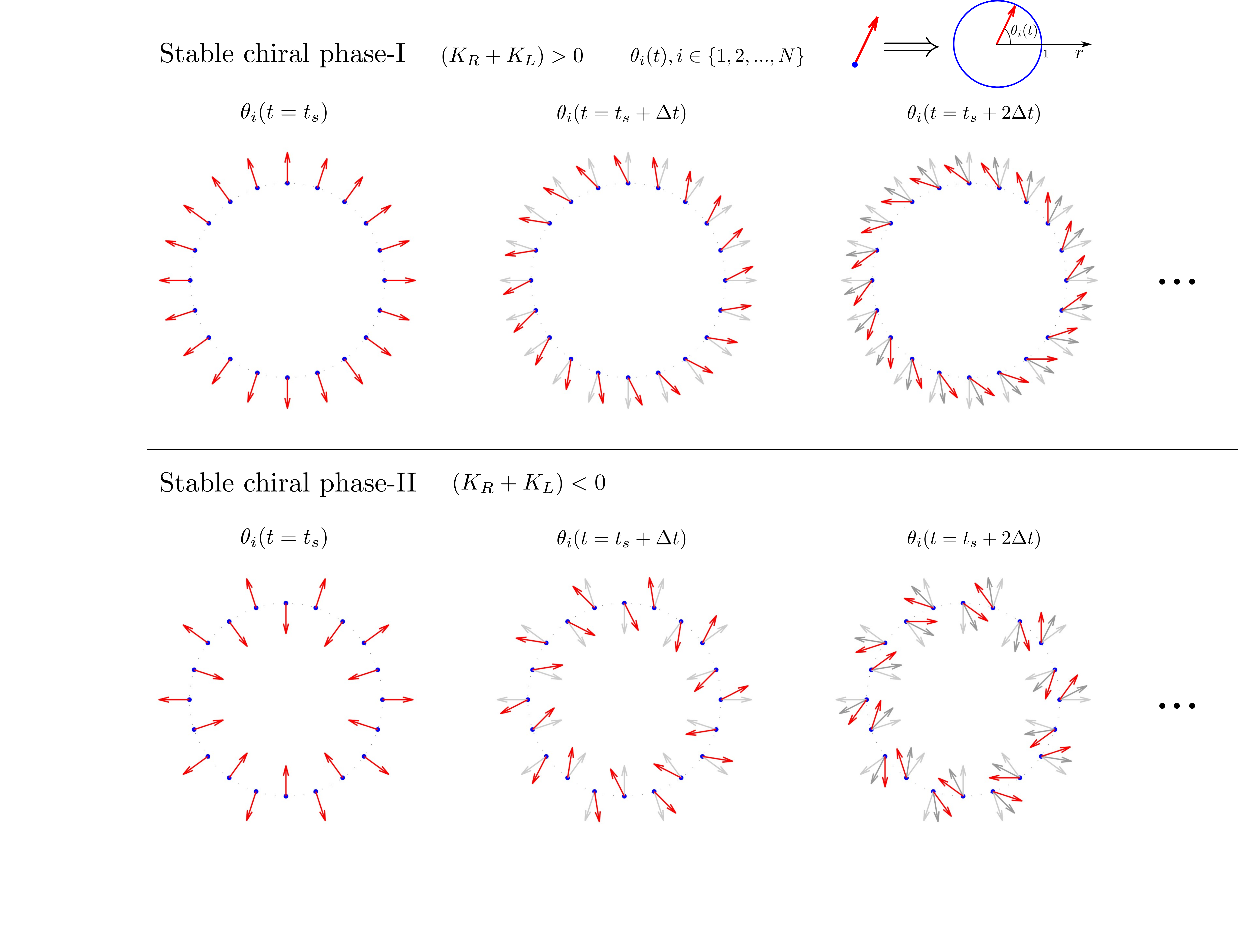}
	\caption{Vector plots of the time-evolution of phases $\theta_{i}(t)$ in stable chiral phase with winding number ($W_p=1$) for $(K_R+K_L)>0$ (top panel) and for $(K_R+K_L)<0$ (bottom panel) (See Sec.~\ref{sec:non_reciprocal_chiral}).}
	\label{fig:N_chiral_phase}
\end{figure}
\textbf{Quantized common frequency (drift velocity):} It is straightforward to see from Eqs.~\eqref{Kuramoto_model} and \eqref{syn_sol} that the common frequency is modified in comparison to the LCKM with reciprocal couplings. In this case, the common frequency has an extra term proportional to the difference in the non-reciprocal couplings  and takes the following form,
\begin{align}\label{vel_chiral}
v_s = \frac{1}{N}\sum_{i=1}^{N}\omega_i +\frac{\Delta K}{N}\sum_{i=1}^{N} \sin{\delta\tilde{\theta}_{i+1,i}},
\end{align}
which reduces to 
\begin{align}
v_s = \frac{\Delta K}{N}\sum_{i=1}^{N} \delta\tilde{\theta}_{i+1,i} = 2\pi W_p \frac{\Delta K}{N},~W_p\in \mathbb{Z}
\end{align}
for $\omega_i=0$, and $\delta\tilde{\theta}_{i+1,i}\ll 1,~\forall~i\in \lbrace 1,2,..., N \rbrace$. Here $W_p$ is the topological winding number. This is a remarkable result that indicates that synchronized solutions of this kind have quantized finite total velocity (in units of $2\pi\Delta K$) equal to the topological winding numbers $W_p\in \mathbb{Z}$. 

\textbf{Chiral phases:}
Now we derive the expression of phases of the oscillators for the synchronized solutions which are characterized by the topological winding number. Considering $\omega_i = 0$, and $f_i = f_0~({\rm const.}) ~\forall~ i\in \lbrace 1,2,..., N \rbrace $, Eqs.~\eqref{fixed_non_reci} and \eqref{vel_chiral} implies,
\begin{align}
(1-\alpha^N)f_0&=-\frac{v_s}{K_L}\sum_{j=0}^{N-1}\alpha^j=-\frac{V_s}{K_L} \frac{(1-\alpha^N)}{1-\alpha},\nonumber\\	
\implies f_0&=\frac{v_s}{\Delta K}=\frac{1}{N}\sum_{j=1}^{N}\sin(\tilde{\theta}_{j+1}-\tilde{\theta}_{j}),\nonumber\\
\implies \delta\tilde{\theta}_{i,i-1}&\approx \frac{2\pi W_p}{N}, ~~{\rm for}~~ \delta\tilde{\theta}_{i,i-1}\ll 1.
\end{align}
Hence we have proven for small phase differences, the synchronized solutions with finite total velocity (in units of $2\pi\Delta K$) for LCKM with non-reciprocal couplings are the same as that of twisted states for LCKM with reciprocal coupling and are characterized by the topological winding number $W_p\in\mathbb{Z}$. We note the difference from the LCKM with reciprocal coupling where the total velocity vanishes for $\omega_i=0$, and does not have quantized values for $\omega_i\neq 0$. In addition, the individual phases of the oscillators for a synchronized solution with the topological winding number $W_p$ is given by
\begin{align}
    \tilde{\theta}_{i} = \frac{2\pi W_p i}{N},~~{\rm with}~~\delta\tilde{\theta}_{i+1, i}\equiv \delta\tilde{\theta}=\frac{2\pi W_p}{N}~~{\rm for}~~ i\in \lbrace 1,2,...,N\rbrace.
\end{align}

For all $\omega_i =0$, the oscillators oscillate with the non-vanishing quantized total velocity (in units of $2\pi\Delta K$) while keeping the constant phase differences between their neighbors (See Fig.~\ref{fig:N_chiral_phase}). We call these time-dependent synchronized solutions as \textit{chiral phases} whose physical origin is discussed in the previous section.

Considering $\omega_i =0$, and different $f_i$'s, one can find the phase differences $\delta\tilde{\theta}_{i,i-1}$ corresponding to the synchronized solutions by solving Eqs.~\eqref{fixed_non_reci} and \eqref{vel_chiral}. Through our numerical analysis, we find that even in this case for small phase differences, the synchronized solutions are characterized by $\boldsymbol{\tilde{\theta}^{(p)}}$ with the winding number $W_p$. 

Similar to the reciprocal case, we also found through our numerical simulations that the realization of different chiral phases depends on the initial conditions. We mention some of the cases in the following:

(i) If the initial conditions are such that that the $\boldsymbol{\theta}$ belongs to the basin of attraction of the solution characterized by winding number $W_p = 0 $, then the oscillators will evolve to a state of \textit{complete phase synchronization} with vanishing common frequency without finite $\omega_i$.

(ii) If the initial conditions are such that the $\boldsymbol{\theta}$ belongs to the basin of attraction of the solution characterized by a given winding number $W_p \in \mathbb{Z}-\lbrace 0 \rbrace $, then each oscillator will evolve in time to the \textit{chiral phase} characterized by the winding number $W_p$ with a finite common frequency proportional to the winding number and difference in the non-reciprocal couplings $v_s=2\pi W_p \Delta K /N$. This essentially means that the chiral phases have topological protection.

\begin{figure}[htbp]
	\centering
	\includegraphics[width=1.0\linewidth]{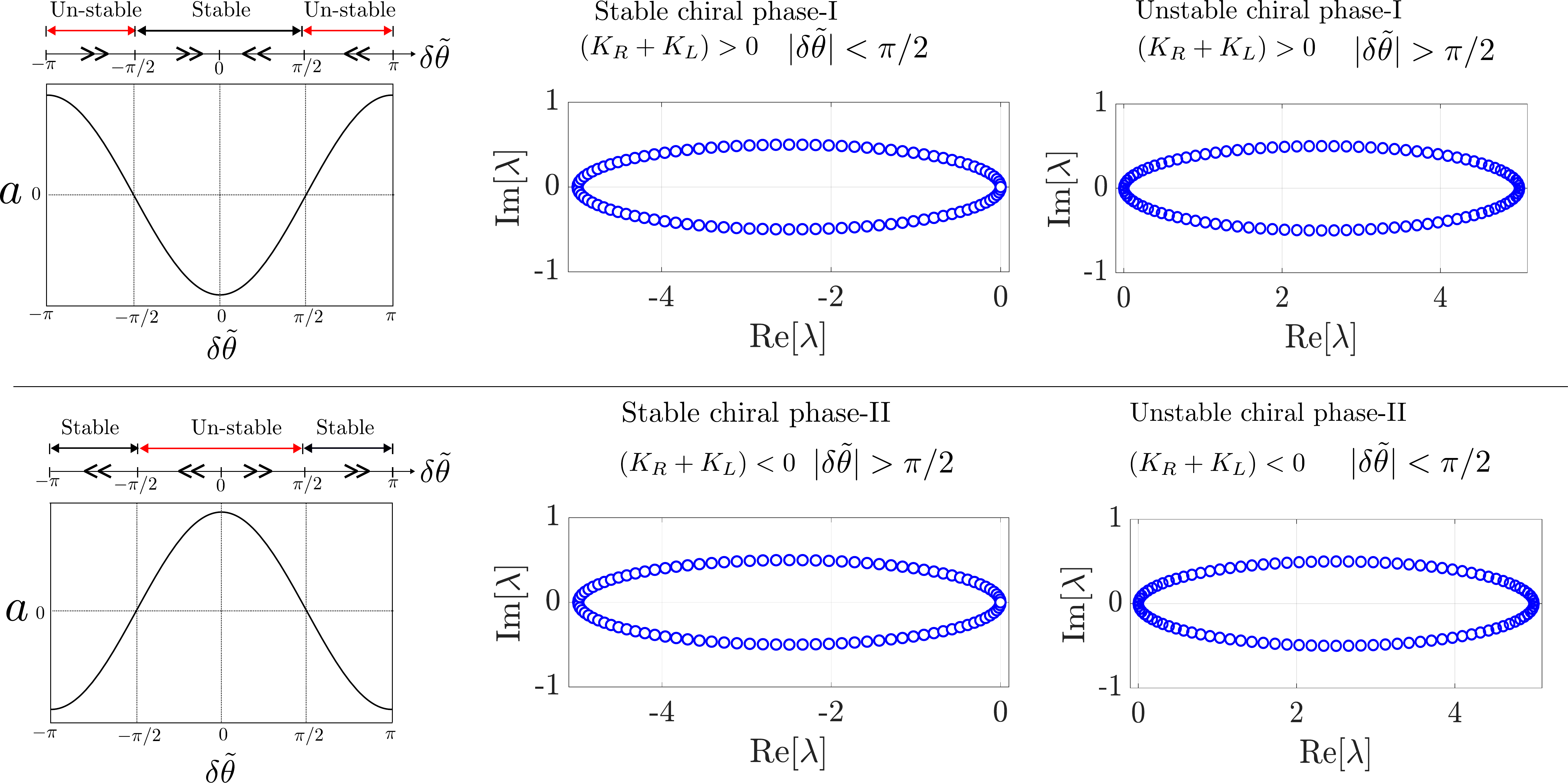}
	\caption{Figures in the left of the top and bottom panels show the variation of center of Gershgorin discs $a_{ii}\equiv a=-(K_R+K_L)\cos\delta\tilde{\theta}$ with respect to constant nearest neighbor phase differences $\delta\tilde{\theta}$ for chiral phases with $(K_R+K_L)>0$ and $(K_R+K_L)<0$, respectively. Figures on the left of the top (bottom) panel show the eigenvalue spectra of the Jacobian matrix for stable and unstable chiral phases for $(K_R+K_L)>0$ ($(K_R+K_L)<0$) (See Sec.~\ref{sec:local_stability_non_reciprocal} for more details on the local stability of twisted states). }
    \label{fig:N_stability_chiral_phases}
\end{figure}

\subsubsection{Stability of the chiral phases: linear stability analysis}\label{sec:local_stability_non_reciprocal}
\paragraph{Jacobian of linear stability:}
We analyze the linear stability of the chiral phases via perturbing the system around them by a small amount such that $\theta_{i}(t)\rightarrow \tilde{\theta}_i + \phi_{i}(t)$, with $|\phi_{i}(t)| \ll 1$. Then the time evolution of $\phi_i(t)$ is given by,
\begin{align}
\frac{d \phi_i (t)}{d t} \approx (-J_{i,i})\phi_i + J_{i, i+1} \phi_{i+1} + J_{i-1, i} \phi_{i-1},~\forall~i\in \lbrace 1,2,...,N\rbrace.
\end{align}
here $J_{i, i} = -(K_R\cos{\delta\tilde{\theta}_{i+1,i}} + K_L\cos{\delta\tilde{\theta}_{i-1,i}}),~J_{i,i+1} = K_R\cos{\delta\tilde{\theta}_{i+1,i}},  J_{i-1,i} = K_L\cos{\delta\tilde{\theta}_{i-1,i}}$. In other words, we can write the above equations in vector form as $\frac{d \boldsymbol{\phi}}{d t} \approx \mathcal{J} \boldsymbol{\phi}$, with $\mathcal{J}$ being a $N\times N$ matrix (also known as the Jacobian) and $\boldsymbol{\phi} = [\phi_1,...,\phi_N]^T$. 
\begin{eqnarray*}
\frac{d}{dt}\begin{bmatrix}
		\phi_1\\
		\vdots\\
		\phi_N 
		\end{bmatrix}= \mathcal{J}
		\cdot \begin{bmatrix}
             \phi_1\\
		\vdots\\
		\phi_N 
		\end{bmatrix},~~~\mathcal{J}=\begin{bmatrix}
				J_{1,1} & K_R\cos{\delta\tilde{\theta}_{2,1}}&\cdots& K_L\cos\delta\tilde{\theta}_{N,1}&\\
				K_L\cos\delta\tilde{\theta}_{1,2}&J_{2,2} & K_R\cos\delta\tilde{\theta}_{3,2}&\cdots\\
				\vdots  & \vdots  & \ddots & \vdots  \\
				K_R \cos\delta\tilde{\theta}_{1,N}& 0& \cdots& J_{N,N}\\
			\end{bmatrix}
			\end{eqnarray*}
			\\
The above Jacobian matrix is similar to the HN model Hamiltonian for 1D lattice with non-reciprocal nearest neighbor hopping (See Secs.~\ref{sec:noise_chiral_phase_transition} and~\ref{sec:non_Bloch_theory_Jaco} for more discussion).

\paragraph{Local stability analyses:}
We found through analytical and numerical analysis that the stability of the synchronized chiral phases depends not only on the phase differences between the neighboring oscillators (which was the case for the LCKM with reciprocal couplings) but also on the coupling strengths $K_R$ and $K_L$. In order to analyze the stability of the chiral phase, we again utilize the \textit{Gershgorin theorem} for the Jacobian $\hat{\mathcal{J}}$ corresponding to Eq.~\eqref{Kuramoto_model} with non-reciprocal couplings. In this case, the center $a_{ii}$ and radius $R_i$ of Gershgorin discs $ D(a_{ii}, R_i)\subseteq \mathbb{C} $ are as follows:
\begin{align}
a_{ii} & = -K_R \cos{\delta\tilde{\theta}_{i+1,i}} - K_L \cos{\delta\tilde{\theta}_{i-1,i}},
~{\rm and} ~ R_i  = |K_R\cos{\delta\tilde{\theta}_{i+1,i}}+ K_L\cos{\delta\tilde{\theta}_{i-1,i}}|,~ \forall~ i\in \lbrace 1,2,...,N\rbrace. 
\end{align}
    
The Jacobian for LCKM with non-reciprocal coupling always possesses a vanishing eigenvalue $\lambda_1=0$ (the Goldstone mode) due to the U(1) global rotation symmetry of Eq.~(\ref{Kuramoto_model}). Utilizing Gershgorin theorem together with numerical analyses for $\delta\tilde{\theta}_{i+1, i} \in [-\pi,\pi]$, the results for the local stability analyses are summarized as follows:
\begin{align*}
&(1 a)~\text{If}~ |\delta\tilde{\theta}_{i+1,i}|>\pi/2,~\forall~i, ~\text{and}~ (K_R + K_L)>0,~\text{then the eigenvalues $\lambda_i$ satisfy}:~{\rm Re~\lambda_i > 0}\implies \text{unstable solutions},\nonumber\\
&(1 b)~\text{If}~ |\delta\tilde{\theta}_{i+1,i}|>\pi/2,~\forall~i, ~\text{and}~ (K_R + K_L)<0,~\text{then the eigenvalues $\lambda_i$ satisfy}:~{\rm Re~\lambda_i > 0}\implies \text{stable solutions},\nonumber\\
&(2 a)~\text{If}~ |\delta\tilde{\theta}_{i+1,i}|<\pi/2,~\forall~i,  ~\text{and} ~(K_R + K_L)>0,~\text{then the eigenvalues $\lambda_i$ satisfy}:~{\rm Re~\lambda_i < 0}\implies \text{stable solutions},\nonumber\\
&(2 b)~\text{If}~ |\delta\tilde{\theta}_{i+1,i}|<\pi/2,~\forall~i,  ~\text{and} ~(K_R + K_L)<0,~\text{then the eigenvalues $\lambda_i$ satisfy}:~{\rm Re~\lambda_i < 0}\implies \text{unstable solutions},\nonumber\\
&(3)~\text{If}~\exists ~i: |\delta\tilde{\theta}_{i+1,i}|>\pi/2~\land~ \exists~ j:|\delta\tilde{\theta}_{j+1,j}|<\pi/2~\text{then  stability of the solutions may not be determined}.
\end{align*}

We have discussed that the chiral phases are the synchronized solutions with small and equal phase differences $\delta \tilde{\theta}_{i+1, i}\equiv \delta\tilde{\theta} =2\pi W_p/N,~\forall~i \in \lbrace 1,2,..., N \rbrace$ which are characterized by the \textit{topological winding number}. We show the behavior of the center of the Gershgorin disc $a = -R_i=-(K_R + K_L) \cos\delta\tilde{\theta}$ of the Jacobian matrix for chiral phases as a function of nearest neighbor phase difference $\delta\tilde{\theta}$ in Fig.~\ref{fig:N_stability_chiral_phases}. It is easy to observe that for $(K_R + K_L)>0$, all centers of the Gershgorin discs $a<0$ ($a>0$) for chiral phases with $|\delta\tilde{\theta}|<\pi/2 ~(>\pi/2)$. Accordingly, for $(K_R + K_L)>0$, all the non-zero eigenvalues of the Jacobian matrix for chiral phases with $|\delta\tilde{\theta}|<\pi/2 ~(>\pi/2)$ have negative (positive) real parts (See the top and bottom panels (right) of Fig.~\ref{fig:N_stability_chiral_phases}). Therefore, for $(K_R + K_L)>0$, chiral phases with $|\delta\tilde{\theta}|<\pi/2 ~(>\pi/2)$ will be stable (unstable). Moreover, the stable chiral states for $(K_R + K_L)>0$ become unstable for $(K_R + K_L)<0$ and vice versa. We tabulate the exact number of stable and unstable chiral phases for a finite-size system in Table~\ref{tab:chiral_stability}.

\begin{center}
    \begin{table}[htbp]
    \def\arraystretch{2.5}
    \begin{tabular}{|c|c|c|c|c|c|c|c|c|c|c|c|c|c|c|}
    \hline
    $W_p$ & ~~0~~ & ~~1~~& ~...~ &$\frac{N}{4}$-1& ~$\frac{N}{4}$~ & $\frac{N}{4}$+1 & $\frac{N}{4}$+2 & ~...~ &$\frac{3N}{4}$-1& ~$\frac{3N}{4}$~ & $\frac{3N}{4}$+1& ~...~ & $N$-2 &$N$-1 \\
     \hline
    $(K_R+K_L)>0$  & s & s & s & s & -& u & u & u &u &- &s &s &s& s \\
    \hline
    $(K_R+K_L)<0$  & u & u & u & u & -& s & s & s &s &- &u &u &u& u \\
    \hline
    \end{tabular}
    \caption{Table for the stability of chiral phases for different winding numbers ranging from 0 to $N-1$ for even number ($N$) of  oscillators. Stable, unstable, and neutral phases are dictated by symbols (s), (u), and (-), respectively. For odd $N$, one can accordingly obtain the table using the Gershgorin theorem.}\label{tab:chiral_stability}
    \end{table}
\end{center}

\begin{figure}[htbp]
	\centering
	\includegraphics[width=1\linewidth]{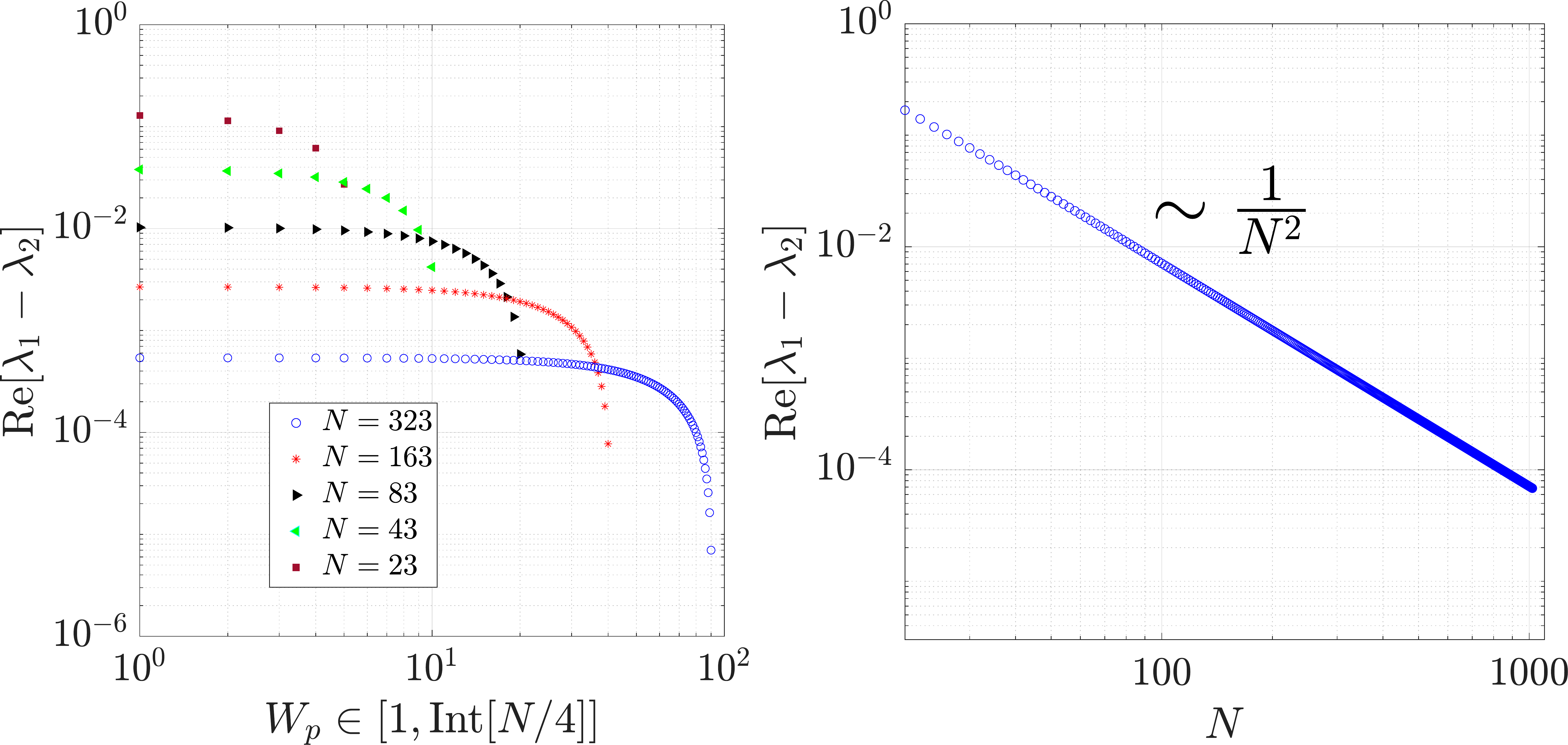}
	\caption{Left panel: behavior of maximal negative eigenvalue close to zero (Re$[\lambda_1-\lambda_2]$)of the Jacobian as a function of winding number ($W_p$) for different system sizes ($N$). Inset: Re$[\lambda_1-\lambda_2]$ vs $N$ for a given winding number. Right panel: behavior of global stability measure $\Gamma(W_p)$ as a function of winding number for different system sizes ($N$). Parameters: $K_R=1.9,~K_L=1.7.$}
    \label{fig:basin_of_attraction_non_reciprocal}
\end{figure}

\paragraph{Basins of attraction and stability of chiral phases:}
We discussed in the previous section that for stable chiral phases, all the eigenvalues of the Jacobian matrix have negative real parts except $\lambda_1=0$ which corresponds to the Goldstone mode. In this case, the eigenvalue spectra are center-shifted ellipse on the negative Re$[\lambda]$ axis in the (Re[$\lambda$], Im[$\lambda$])-plane (See Fig.~\ref{fig:N_chiral_phase}). Similar to the LCKM with reciprocal coupling, the difference between the eigenvalue corresponding to the Goldstone mode ($\lambda_1 =0$) and the maximal negative eigenvalue ($\lambda_2$) decreases with an increase in the value of topological winding number $W_p$ and system size $N$ (See Fig.~\ref{fig:basin_of_attraction_non_reciprocal}). It essentially means that the stable chiral phases with higher values of winding numbers have smaller basins of attraction and thus are less stable in comparison to the ones with a smaller winding number which is again confirmed by our numerical simulations. Therefore the volume of the basins of attraction and the value of the winding numbers are highly correlated and the phases with large winding numbers are marginalized. Also, the decrease of the maximal negative eigenvalue with the system size indicates that arbitrary small unbounded noise will destabilize the stable chiral phases with finite winding numbers in the thermodynamic limit and also in the long-enough time limit. It is important to mention here that the eigenvalue spectra and hence the local stability also depend on the degree of non-reciprocity which is defined by $\Delta K = K_R-K_L$. For example, in the case of fully non-reciprocal coupling ($K_R=-K_L$), the eigenvalue spectra collapse on the Im$[\lambda]$ axis consisting of degenerate eigenvalues which correspond to the marginal orbits.


Therefore, the stability analysis of the chiral phases is similar to that of the twisted states for LCKM with reciprocal coupling as both are described by the topological winding numbers. In nutshell, the LCKM with non-reciprocal coupling hosts stable time-dependent synchronized phases for finite sizes of the system. These stable solutions can be found explicitly depending on the initial conditions and their number is proportional to the finite size of the system. The synchronized solution with the topological winding number $W_p=0$ is globally stable. Whereas other stable synchronized solutions (chiral phases) with $W_p\neq 0$ are only locally stable. We also observe the correlation between the winding number and stability of the chiral phases similar to the reciprocal case. 

\begin{figure}[htbp]
	\centering
	\includegraphics[width=1\linewidth]{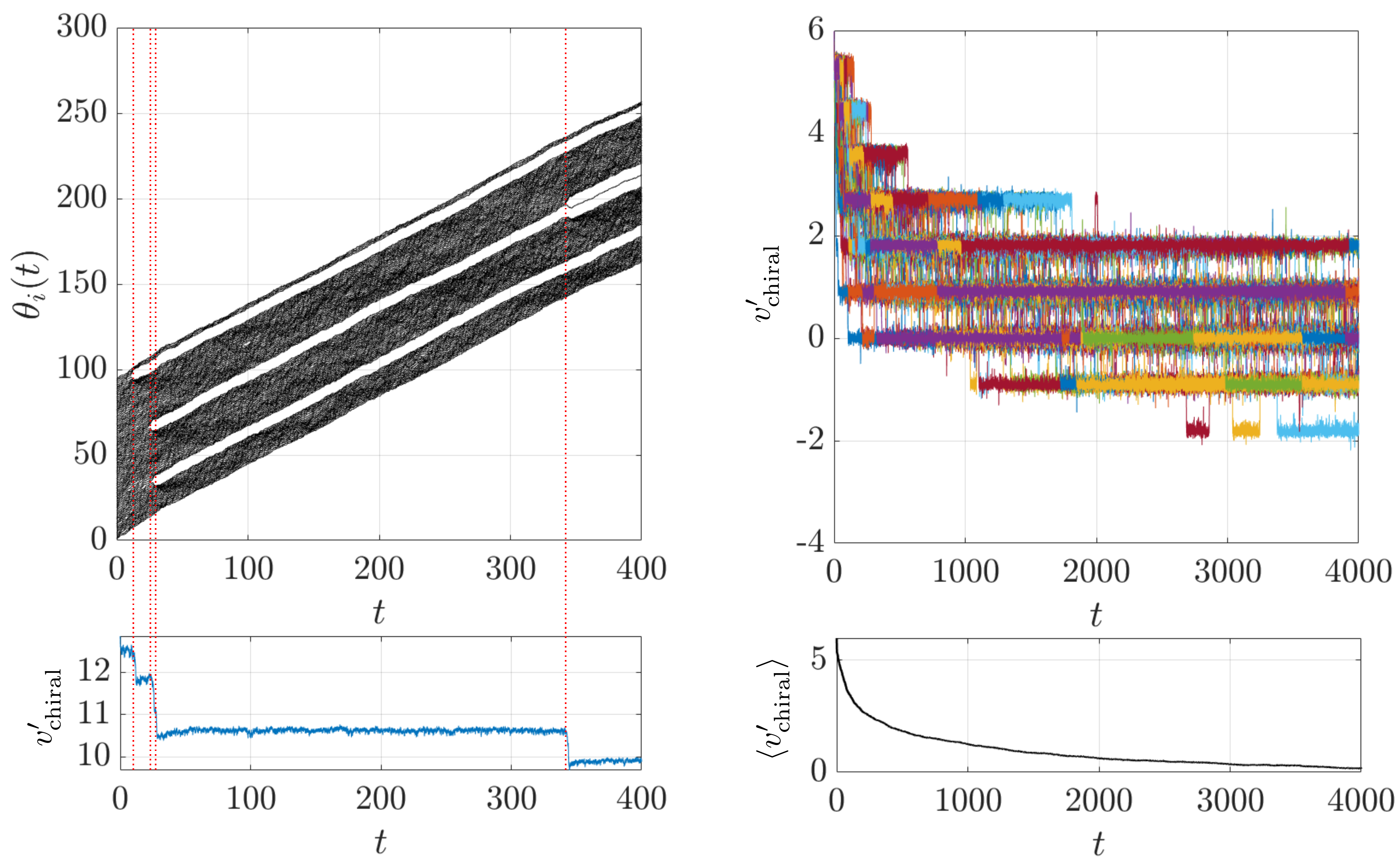}
	\caption{Stability of chiral phases: noise-induced phase transitions as a direct consequence of phase slips and non-reciprocity. Left panel: time evolution of the individual phases $\theta_i(t)$ (upper panel) and total drift velocity $v'_{\rm chiral} = v_{\rm chiral}/(2\pi (K_R-K_L)),~v_{\rm chiral} = \sum_{i=1}^{N}\frac{d\theta_{i}}{d t}$ (lower panel). The red dotted lines mark the non-equilibrium phase transitions due to the phase slips (See Sec.~\ref{sec:noise_chiral_phase_transition} for more details). Right panel: time evolution of total drift velocity $v'_{\rm chiral} $ for different noise realizations (upper panel) and average total drift velocity $\langle v'_{\rm chiral}\rangle $ (lower panel) (See Sec.~\ref{sec:noise_chiral_phase_transition} for more details). Parameters: Left panel: $K_R=2.5,~K_L=1.9,~D=0.15,~N=100.$ Right panel: $K_R=1.9,~K_L=1.0,~D=0.5,~N=100$, number of noise realizations ($N_d=200$).}
    \label{fig:phase_slip_noise}
\end{figure}

\begin{figure}[htbp]
	\centering
	\includegraphics[width=1\linewidth]{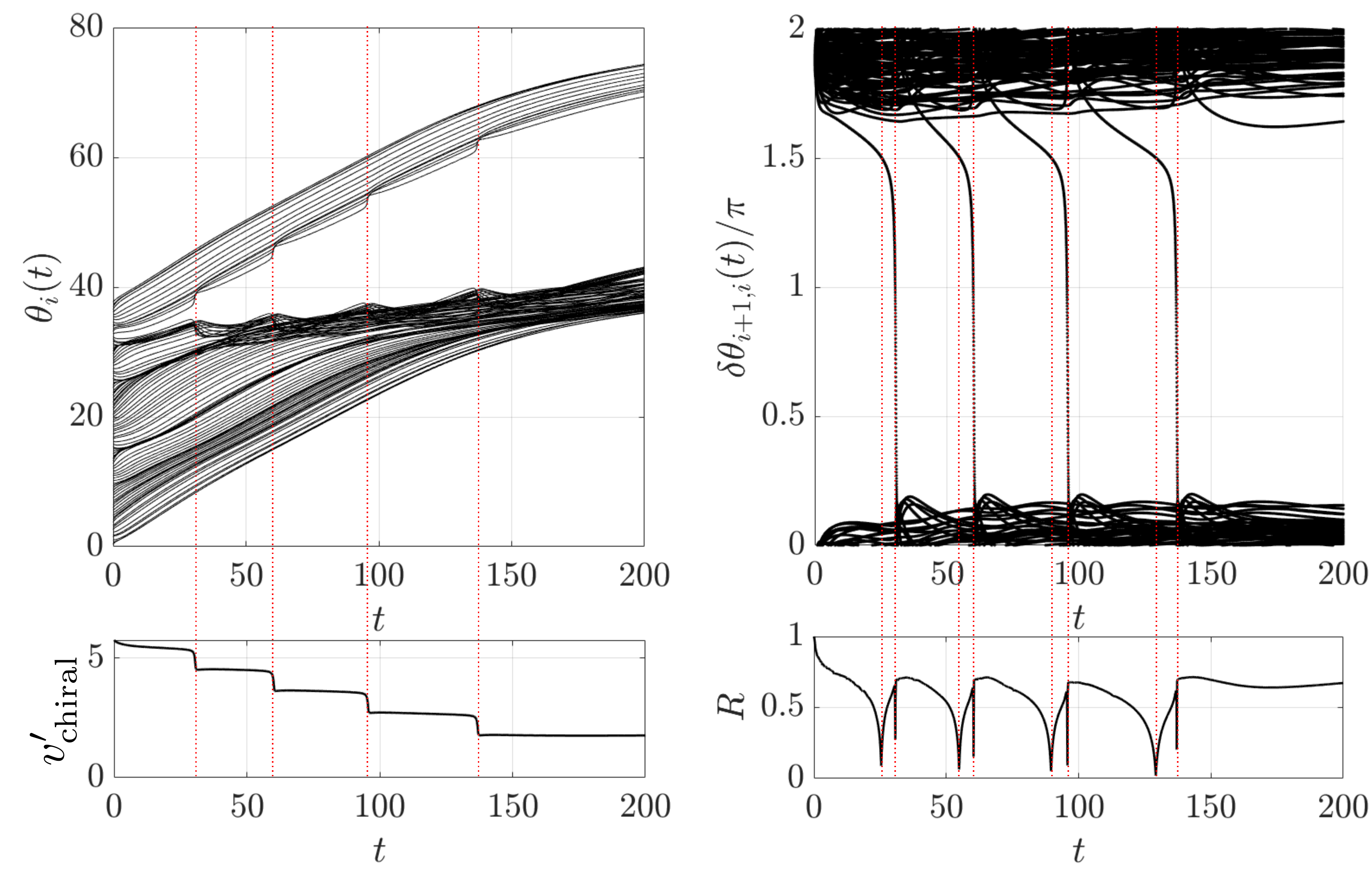}
	\caption{Stability of the chiral phases: disorder-induced non-equilibrium phase transitions as a direct consequence of phase slip and non-reciprocity. Left panel: Time evolution of the individual phases $\theta_i(t)$ (upper panel) and total drift velocity $v'_{\rm chiral} = v_{\rm chiral}/(2\pi (K_R-K_L))$ (lower panel). The red dotted lines mark the non-equilibrium phase transitions due to the phase slips (See Sec.~\ref{sec:noise_chiral_phase_transition} for more details). Right panel: Time evolution of nearest neighbor phase differences $\delta\theta_{i+1, i}(t)$ (upper panel) and total phase rigidity $R=\sum_{i=1}^{N}R_i/N$ (lower panel) (See Sec.~\ref{sec:noise_chiral_phase_transition} for more details). Parameters: $K_R=2.5,~K_L=1.9,~\Delta\omega=0.5,~N=100.$}
    \label{fig:phase_slip_frequency_disorder}
\end{figure}

\begin{figure}[htbp]
	\centering
	\includegraphics[width=1\linewidth]{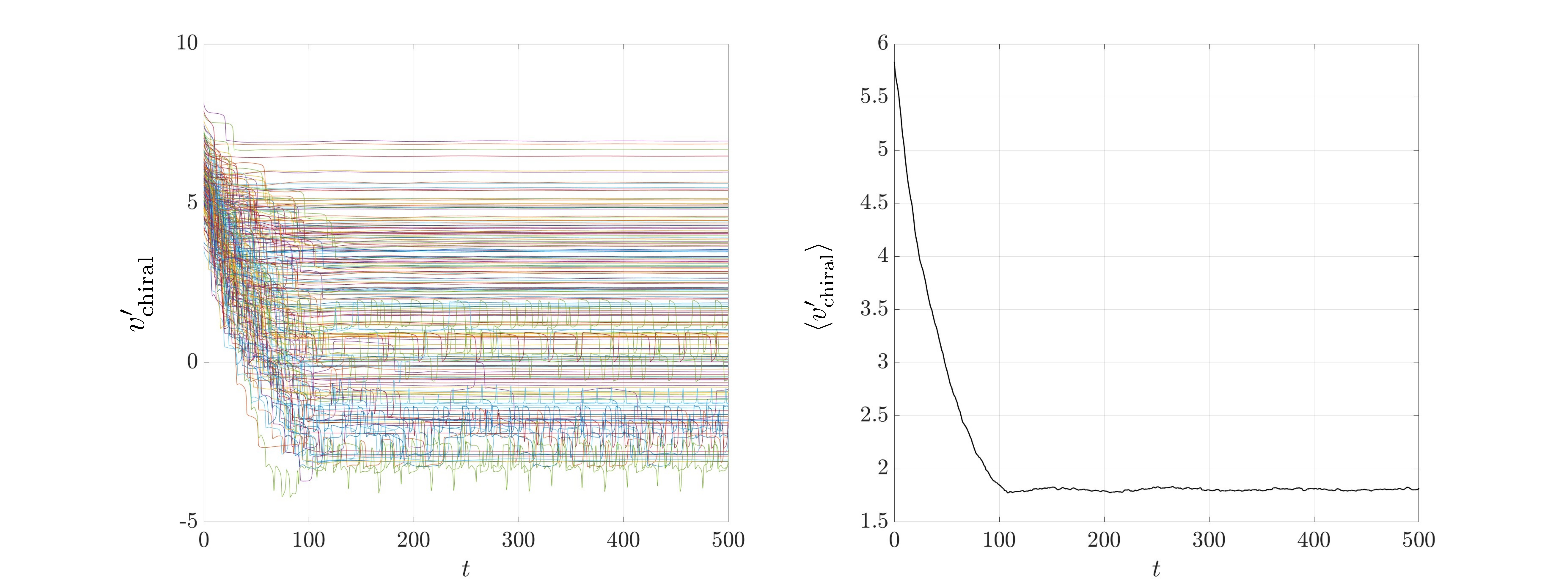}
	\caption{ Left panel: time evolution of total drift velocity $v'_{\rm chiral} $ for different noise realizations (upper panel) and average total drift velocity $\langle v'_{\rm chiral}\rangle $ (right panel) (See Sec.~\ref{sec:noise_chiral_phase_transition} for more details). Parameters: Left panel: $K_R=1.9,~K_L=1.0,~\Delta\omega=0.5,~N=100,$ number of disorder realizations $N_d=200$.}
	\label{fig:phase_slip_frequency_disorder2}
\end{figure}
\subsubsection{Stability of chiral phases: noise-induced non-equilibrium phase transitions}\label{sec:noise_chiral_phase_transition}
As discussed earlier that the different chiral phases are characterized by different winding numbers $W_p\in \mathbb{Z}$ which are only locally stable having finite basins of attraction. Therefore, random and unbounded noise will induce phase transitions between different winding number sectors due to the phase slips (See Sec.~\ref{sec:phase_slip_reciprocal} for the physical interpretation of phase slips) and eventually lead the system to the complete phase synchronized state. The main difference in comparison to the reciprocal case is that in the non-reciprocal case, the noise-induced phase transitions are characterized by the quantized jumps of the total common frequency $v_{\rm chiral} = \sum\limits_{i=1}^{N}\frac{d\theta_i}{d t}$ which is proportional to the topological winding number. In fact, for arbitrary small noise strengths, after averaging over a large number of noise realizations, we found that the system goes to a disordered phase in the thermodynamic limit after waiting a long enough time. Moreover, the finite common frequency or drift velocity decays rapidly initially and then approaches zero rather slowly verifying again that the chiral phases with smaller winding numbers are more stable and have larger basins of attraction (See Fig.~\ref{fig:phase_slip_noise}). 

We describe the numerical observations as follows:
if the initial conditions are such that the oscillators were in a basin of attraction of a chiral phase with finite drift velocity $v'_{\rm chiral}=v_{\rm chiral}/(2\pi \Delta K)  \in \mathbb{Z} $, then for a given ensemble of white noise phase slips occur and the drift velocity changes its value by some integer number. After a long enough time, the phase slip occurs around $V_{\rm tot} =0$ (See the time evolution of phases $\theta_i(t)$, total drift velocity $v'_{\rm chiral}(t)$, and its ensemble average $\langle v'_{\rm chiral}\rangle$ in Fig.~\ref{fig:phase_slip_noise}). This \textit{noise-induced non-equilibrium phase transitions} between different chiral phases are also named as \textit{non-reciprocal phase transitions}~\cite{Fruchart2021} which are marked by red dotted lines in the right panel of Fig.~\ref{fig:phase_slip_noise}. The average drift velocity of oscillators over noise realizations $\langle v_{\rm chiral} \rangle\neq 0$ decreases rapidly initially then rather slowly to $\langle v_{\rm chiral} \rangle = 0$ as time progresses (See Fig.~\ref{fig:phase_slip_noise}).

It is important to note here that because of the unbounded nature of the random white noise, there can be phase transitions from stable to unstable chiral phases \textit{i.e} the total drift velocity can increase as well for a given noise realization (See Fig.~\ref{fig:phase_slip_noise} (top right)). There are phase transitions even from $W_p=0$ phase to stable or unstable chiral phases for one noise realization. Moreover, irrespective of the initial conditions and noise strengths, the system will eventually approach a disordered phase after averaging over a large number of noise realizations in the thermodynamic limit.

We found similar results in the presence of frequency disorder (See Figs.~\ref{fig:phase_slip_frequency_disorder}, \ref{fig:phase_slip_frequency_disorder2}). In Fig.~\ref{fig:phase_slip_frequency_disorder}, we show the time evolution of the individual phases $\theta_i(t)$ (upper panel) and total drift velocity $v'_{\rm chiral} = v_{\rm chiral}/(2\pi (K_R-K_L))$ (lower panel). The red dotted lines mark the non-equilibrium phase transitions due to phase slips where the total drift velocity changes in a quantized manner. In Fig.~\ref{fig:phase_slip_frequency_disorder}, we also show the time evolution of nearest neighbor phase differences $\delta\theta_{i+1, i}(t)$ (upper panel) and total phase rigidity $R=\sum_{i=1}^{N}R_i/N$ (lower panel). The phase rigidity $R_i(t)$ (defined in Eq.~\eqref{def:phase_rigidity}) is calculated using the right and left eigenstates of the Jacobian as a function of time. The total phase rigidity vanishes two times corresponding to $\delta\theta_{i+1, i}(t)=\pi/2,3\pi/2$ in the time-interval of the phase slips when one of the phase differences undergoes a change of $2\pi$ (See the red dotted lines in Fig.~\ref{fig:phase_slip_frequency_disorder}). This is a remarkable result and it is the basis for our analytical analyses of noise-induced non-equilibrium phase transitions in the upcoming discussion. In Fig.~\ref{fig:phase_slip_frequency_disorder2}), we show the time evolution of the total drift velocity $v'_{\rm chiral}$ (left panel) and its disorder average (right panel) which eventually approaches the steady state value.

\textbf{Analytical modeling of noise-induced phase transitions:}
Similar to the LCKM with reciprocal couplings, the non-equilibrium phase transitions between different chiral phases can also be understood by modeling the adiabatic evolution of the phases as follows:
\begin{align}
    \theta_{i}(t)= \frac{2\pi W_p i}{N} (1-t) + \frac{2\pi W_{p^{\prime}} i}{N} t,~~(0\leq t\leq 1), 
\end{align}
here $W_p$ and $W_{p^{\prime}}$ are the winding numbers characterizing different chiral phases. At $t=0$, the system is in a chiral phase with winding number $W_p$ and evolves in an adiabatic manner to a chiral phase with winding number $W_{p^{\prime}}$ at $t=1$ as a consequence of the phase slip in the presence of noise. In the interval of the phase slip, one of the phase differences $\delta\theta_{1, N}$ undergoes a change from $\delta\theta_{1, N}$ to $\delta\theta_{1, N} + 2\pi$, and so it passes through the points $\delta\theta_{1, N}=\pi/2,3\pi/2$. 
Now we discuss how the Jacobian matrix evolves during the interval of the phase slip.
We want to study the Jacobian matrix for different chiral phases as well as for transitions between two chiral phases. For this purpose, it is convenient to re-write the above Jacobian matrix in the following general form with notations:
$J_{i,i} = -\tilde{K}_{R}-\tilde{K}_L
\equiv J_{0},~J_{i,i+1} = \tilde{K}_R,~ J_{i,i-1} = \tilde{K}_L,~\forall~i \in \lbrace 2,3,..., N-1\rbrace$ ,
$J_{1,1} = -\tilde{K}_R-\tilde{K}_{L}^{\prime}
\equiv J_{0}+J_{0L}$ with 
$J_{0L}=\tilde{K}_L - \tilde{K}_{L}^{\prime}$,
$J_{N,N} = -\tilde{K}_L-\tilde{K}_{R}^{\prime}
\equiv J_{0}+J_{0R}$ with 
$J_{0R}=\tilde{K}_R - \tilde{K}_{R}^{\prime}$, and 
$J_{N,1} = \tilde{K}_{R}^{\prime}$, $J_{1,N} = \tilde{K}_{L}^{\prime}$,
\begin{align} \label{general_jacobian}
\hat{\mathcal{J}}&=\hat{\mathcal{J}}_{\rm bulk} + \hat{\mathcal{J}}_{\rm boundary}\nonumber\\
&=\sum_{i=2}^{N-1}\tilde{K}_{R}| i+1\rangle\langle i|+\tilde{K}_{L}| i-1\rangle\langle i|+
J_{0}|i\rangle\langle i|\nonumber\\
&+\tilde{K}_{R}| 2\rangle\langle 1|+\tilde{K}_{L}| N-1\rangle\langle N|+\tilde{K}_{R}^{\prime}| 1\rangle\langle N|+\tilde{K}_{L}^{\prime}| N\rangle\langle 1|+(J_{0}+J_{0R})| N\rangle\langle N|+(J_{0}+J_{0L})| 1\rangle\langle 1|, 
\end{align}
here $\tilde{K}_{R} = K_R\cos\delta\theta_{i+1,i}$, $\tilde{K}_{L} = K_L\cos\delta\theta_{i-1,i}$, $\tilde{K}_{R}^{\prime} = K_{R}\cos\delta\theta_{1,N}$,
$\tilde{K}_{L}^{\prime} = K_{L}\cos\delta\theta_{N,1}$ with $\delta\theta_{i+1,i}=\theta_{i+1}-\theta_i$. Since the noise-induced phase transitions occur as a result of phase slip where one of the phase differences (say $\delta\theta_{1, N}\equiv \delta\theta$) changes by $2\pi$, the Jacobian matrix realizes the Hatano-Nelson model Hamiltonian with generalized boundary conditions in the time interval of the phase slip (see Sec.~\ref{sec:def_GBC}). In other words, in the time interval of the phase slip ($\delta\theta\rightarrow \delta\theta +2\pi$), the Jacobian realizes periodic boundary condition (PBC) at $\delta\theta=0,2\pi$, open boundary condition (OBC) at $\delta\theta =\pi/2,~3\pi/2$, and generalized boundary conditions (GBC) otherwise. We will focus on Eq.~\eqref{general_jacobian} for further discussion.

In the following section, we prove using this simple model that the noise-induced non-equilibrium phase transitions are characterized by the exceptional points in the spectrum of the Jacobian, which we call noise-induced \textit{exceptional phase transitions}.

\subsubsection{Non-Bloch band theory of GBZ for the Jacobian matrix}\label{sec:non_Bloch_theory_Jaco}

In this section, we discuss the non-Bloch band theory for the Jacobian matrix. Although the determination of the eigenvalues and eigenvalues are similar to the case of the HN model with GBC (See Sec.~\ref{sec:non_Bloch_theory}), we sketch the calculation here as well. We start with the eigenvalue equation,
\begin{align}
\hat{\mathcal{J}}|\Psi \rangle = E|\Psi\rangle~~{\rm with}~~|\Psi\rangle = \sum_{i=1}^{N}\psi_i|i\rangle.
\end{align}
The eigenstates $|\Psi\rangle = \sum_{i=1}^{N}\psi_i|i\rangle$ with eigenvalue $E$ need to simultaneously satisfy the bulk as well as boundary equations. We analyze these equations separately in the following: 

\textbf{Bulk equations:} The bulk equations which are given by,
\begin{align}\label{bulk_eqn}
\tilde{K}_L \psi_{i-1} + (J_{0} - E) \psi_i + \tilde{K}_R \psi_{i+1}  = 0 ~~{\rm for}~~i\in\lbrace 2,3,..., N-1\rbrace,
\end{align}
We solve the above equations with the ansatz for the wave function $\Psi_j = (z_j,z_j^2,z_j^3...,z_j^{N-1},z_j^N)^T$. Accordingly, the bulk equations transform as follows,
\begin{align}
    \tilde{K}_{L} z_{j}^{i-1} + (J_{0} - E) z_{j}^i + \tilde{K}_{R} z_{j}^{i+1} = 0,
    ~{\rm or}~~\tilde{K}_{L} z_{j}^{-1} + (J_{0} - E) + \tilde{K}_{R} z_{j} = 0,
\end{align}
which has following two non-trivial ($z_{j}\neq 0$) solutions for a given eigenvalue $E$,
\begin{align}
    z_{j} \equiv z_{1,2} = -\frac{(J_0 -E)}{2\tilde{K}_{R}} \pm \sqrt{\Bigg(\frac{J_0 -E}{2\tilde{K}_{R}}\Bigg)^2-\Bigg(
    \frac{\tilde{K}_{L}}
    {\tilde{K}_{R}}\Bigg)}, 
\end{align}
It is straightforward to note that for any eigenvalue $E$, the two solutions are related as follows
\begin{align}
    z_{1}z_{2} = \frac{\tilde{K}_{L}}
    {\tilde{K}_{R}}\equiv r^2, ~~{\rm with}~~ r=\sqrt{\frac{\tilde{K}_{L}}
    {\tilde{K}_{R}}}.
\end{align}
In general, the superposition of these two linearly independent solutions $\Psi_1 = (z_1,z_1^2,z_1^3,...,z_1^{N-1},z_1^N)^T$ and $\Psi_2 = (z_2,z_2^2,z_2^3,...,z_2^{N-1},z_2^N)^T$ \textit{i.e.} $\Psi=c_1 \psi_1 +c_2 \Psi_2 = (\psi_1,\psi_2,\psi_3,...,\psi_{N-1},\psi_{N})^T$ with wave functions
\begin{align}
    \psi_i = c_1 z_1^i + c_2 z_2^i,~~{\rm for}~~i\in\lbrace 2,3,..., N-1\rbrace,
\end{align} 
which forms the general solution of the bulk equations. 

\textbf{Boundary equations:} The boundary equations are given by,
\begin{align}\label{boundary_eqn_i}
\tilde{K}_{L}^{\prime} \psi_{N} + (J_0 + J_{0L} - E) \psi_1 + \tilde{K}_R \psi_{2}  &= 0 \nonumber\\
\tilde{K}_{R}^{\prime} \psi_{1} + (J_0 + J_{0R} - E) \psi_N + \tilde{K}_L \psi_{N-1}  &= 0,
\end{align}
which after further simplification using the bulk equations, reduce to,
\begin{align}\label{boundary_eqn_f}
 \tilde{K}_L \psi_{0}-J_{0L}\psi_1 -\tilde{K}_{L}^{\prime} \psi_{N}  &= 0 \nonumber\\
\tilde{K}_{R}^{\prime} \psi_{1} +  J_{0R}  \psi_N - \tilde{K}_{R} \psi_{N+1}  &= 0 .
\end{align}
The general solution must satisfy the generalized boundary conditions in order to be the correct physical solution for our system. With the help of the general solution, the boundary equations (\ref{boundary_eqn_f}) can be written in the following eigenvalue form,
\begin{align}
  \mathcal{J}_B \begin{pmatrix}
  c_1\\
  c_2\end{pmatrix} =0, ~~{\rm with}~~ \mathcal{J}_B = \begin{pmatrix}
  \tilde{K}_{L}-J_{0L}z_1 
  - \tilde{K}_{L}^{\prime} z_1^{N}& \tilde{K}_{L}-J_{0L}z_2 
  - \tilde{K}_{L}^{\prime} z_2^{N} \\
  \tilde{K}_{R}^{\prime}z_1 + J_{0R}z_1^N 
  - \tilde{K}_{R} z_1^{N+1}& \tilde{K}_{R}^{\prime}z_2 + J_{0R}z_2^N 
  - \tilde{K}_{R} z_2^{N+1}.
  \end{pmatrix}
\end{align}
Here $\mathcal{J}_B$ is the boundary matrix. 

\textbf{Eigenstate solutions and generalized Brillouin zone:}
In order to have non-trivial solutions for $c_1$ and $c_2$, ${\rm Det} [\mathcal{J}_B] =0$ need to be satisfied together with $z_1 z_2 =r^2$ which takes the following form,
\begin{align}\label{eqn_for_sol_z}
    (z_1^{N+1}-z_2^{N+1})-
    \frac{(J_{0L}+J_{0R})}
    {\tilde{K}_{R}}(z_1^N-z_2^N) + 
    \frac{(J_{0L}J_{0R}-
    \tilde{K}_{L}^{\prime}\tilde{K}_{R}^{\prime})}
    {\tilde{K}_{R}^2}(z_1^{N-1}-z_2^{N-1})-
    \Bigg(\frac{\tilde{K}_{R}^{\prime}}{\tilde{K}_{R}}+\frac{\tilde{K}_{L}^{\prime}}{\tilde{K}_{L}}r^{2N}\Bigg)(z_1-z_2)=0.
\end{align}
The solutions of the above equation with $z_1 z_2 = r^2$ give rise to generalized Brillouin zones for the Jacobian of finite-size Kuramoto lattices with generalized boundary conditions.

The paired generalized momenta ($z_1,~z_2$) can be expressed in the polar co-ordinates as follows
\begin{align}
    z_1 = r e^{i\phi}, ~~z_2 = r e^{-i\phi}, ~~{\rm here}~~\phi\in\mathbb{C},
\end{align}
and thus Eq.~\eqref{eqn_for_sol_z} becomes
\begin{align}\label{eqn_for_sol_phi}
    \sin[(N+1)\phi]-\eta_0\sin[N\phi]+\eta_1\sin[(N-1)\phi]-\eta_2\sin[\phi]=0,
\end{align}
with 
\begin{align}
\eta_0 = \frac{(J_{0L}+J_{0R})}
    {\sqrt{\tilde{K}_{R}\tilde{K}_{L}}},~~\eta_1 = \frac{(J_{0L}J_{0R}-
    \tilde{K}_{L}^{\prime}\tilde{K}_{R}^{\prime})}
    {\tilde{K}_{R}\tilde{K}_{L}},~~\eta_2 = \Bigg(\frac{\tilde{K}_{R}^{\prime}}{\tilde{K}_{R}}r^{-N}+\frac{\tilde{K}_{L}^{\prime}}{\tilde{K}_{L}}r^{N}\Bigg).
\end{align}
The eigenvalues are given by
\begin{align}
    E&=-\tilde{K}_{R}-\tilde{K}_{L} + \tilde{K}_{R} z_i + \tilde{K}_{L} z_i^{-1},\nonumber\\
    &=-\tilde{K}_{R}-\tilde{K}_{L}+2\sqrt{\tilde{K}_{R}\tilde{K}_{L}}\cos[\phi], ~~{\rm with}~~\phi\in\mathbb{C}~~{\rm such~that}~~z_1z_2 =r^2,
\end{align}
with corresponding wave functions $\Psi =(\psi_1,\psi_2,...,\psi_N)^T$
\begin{align}
    \psi_n &= c_1 z_1^n + c_2 z_2^n,\nonumber\\
    &=2r^n((c_1+c_2)\cos[n\phi]+(c_1-c_2)i\sin[n\phi])
\end{align}

\begin{figure}[htbp]
		\centering
		\includegraphics[width=1\linewidth]{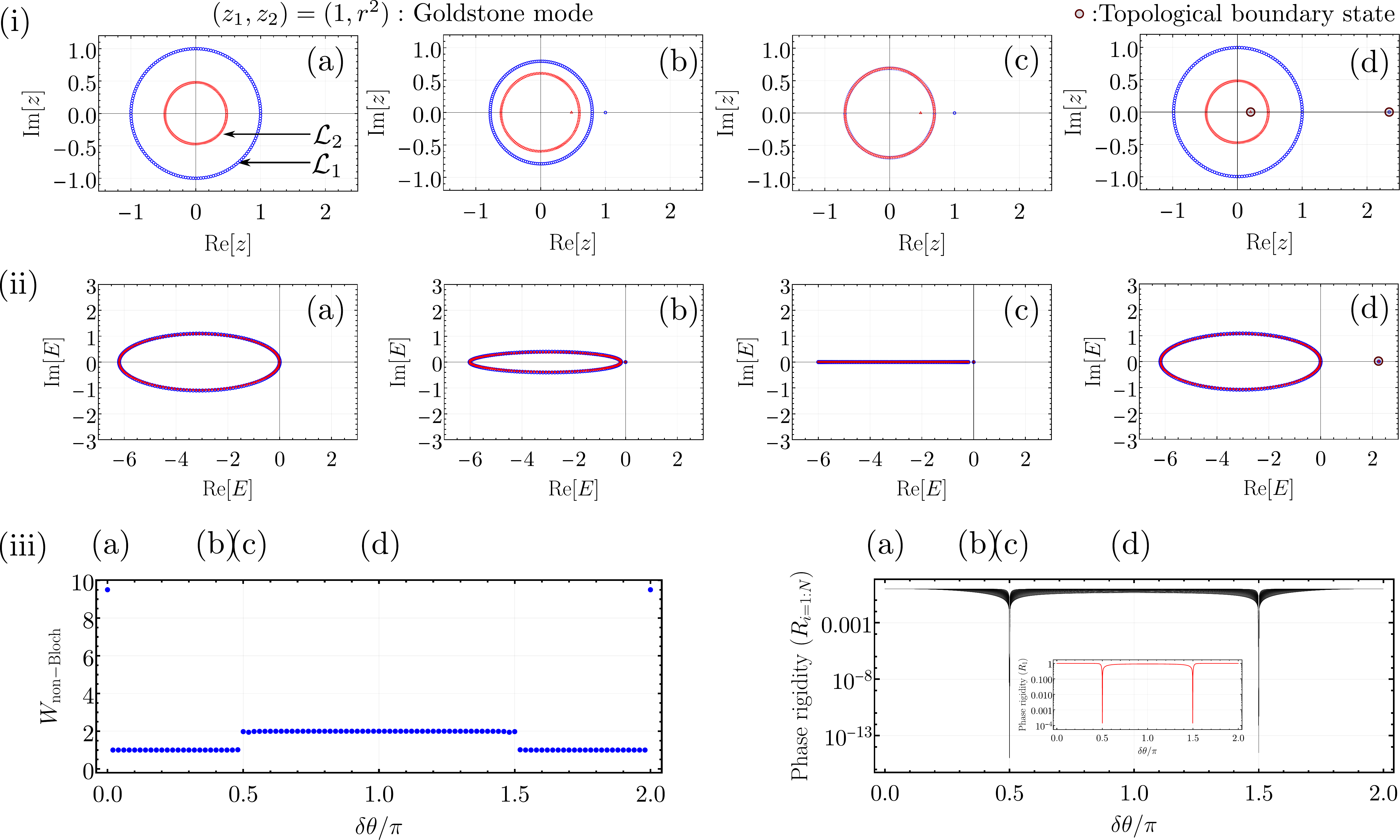}
		\caption{Exceptional and topological phase transition of GBZ in Kuramoto lattices. Panels (i) and (ii) show GBZs together with disjoint generalized momenta and eigenvalue spectra for different values of the phase differences $\delta\theta$ in the time interval of the phase slip ($\delta\theta_N\in[0,2\pi]$). Figures (left) of panels (iii) and (iv) show the behavior of non-Bloch topological invariant and phase rigidity in the time interval of the phase slip ($\delta\theta_N\in[0,2\pi]$) (See Secs.~\ref{sec:ET_GBZ_Jaco} and \ref{sec:TPT_GBZ_Jaco} for more details). Parameters: $N=160,~t'_L=t_L\cos(\delta\theta_N),~t_L=2.1,~t'_R=t_R\cos(\delta\theta_N),~t_R=1.0,~\epsilon_1=\epsilon_0+t_L-t'_L,~\epsilon_N=\epsilon_0+t_R-t'_R,~\epsilon_0=-t_R-t_L,~\delta\theta_N\in[0,2\pi].$}
		\label{fig:ET_TPT_GBC_5}
\end{figure}
\subsubsection{Noise induced phase transitions: exceptional transition of GBZ for the Jacobian matrix}\label{sec:ET_GBZ_Jaco}
In this section, we will prove that in the interval of the phase slip the Jacobian hosts exceptional points in its spectrum.

\textbf{Results from numerical diagonalization:}
In the interval of the phase slip, $\delta\theta$ passes through the points $\pi/2,3\pi/2$. At these points, the Jacobian realizes open boundary condition and we observe NHSE for the Jacobian. We also calculate the individual phase rigidity (defined in Eq.~\eqref{def:phase_rigidity}) for each of the eigenstates of the Jacobian matrix and find that the phase rigidity vanishes at $\delta\theta=\pi/2,~3\pi/2$ in the large system size limit (See Fig.~\ref{fig:ET_TPT_GBC_5}(iii) (left)). The vanishing phase rigidity can be understood as the consequence of the NHSE where the left eigenstates and the right eigenstates are localized at the opposite boundaries, causing the vanishing wavefunction overlap. At these points, the Goldstone mode ($\lambda_1=0$) coalesces with the normal mode which signifies the non-equilibrium phase transition and has been referred to as exceptional transition \cite{Fruchart2021}. 

\textbf{Results from non-Bloch band theory:}
In order to prove the existence of the exceptional transitions or exceptional points in the Jacobian spectra, we utilize the non-Bloch band theory and determine the GBZs in the full interval of the phase slip.  In the interval of the phase slip, for $t=0,~1$, the Jacobian realizes periodic boundary condition due to $\delta\theta=2\pi W_p/N,~2\pi W_{p'}/N$ and thus have two distinct GBZs (See Fig.~\ref{fig:ET_TPT_GBC_5} (i)(a)). Moreover, for $\delta\theta=\pi/2,~3\pi/2$, we indeed observe the merging of two GBZs which defines the exceptional points in the spectrum of the Jacobian (See Fig.~\ref{fig:ET_TPT_GBC_5} (i)(c). We call these phase transitions exceptional transitions of the GBZ.
We calculate the exact analytical solutions for eigenvalues and wave functions for open boundary conditions for $\delta\theta_{1, N}=\pi/2,~3\pi/2$ \textit{i.e.} $\tilde{K}_{L}^{\prime}=\tilde{K}_{R}^{\prime}=0$. In this case, Eq.~\eqref{eqn_for_sol_phi}
\begin{align}
    \sin[(N+1)\phi]-\Big(r+\frac{1}{r}\Big)\sin[N\phi]+\sin[(N-1)\phi]&=0,\nonumber\\
    \sin[N\phi]\Big(2\cos[\phi]-r-\frac{1}{r}\Big)&=0,
\end{align}
has following solutions
\begin{align}
    \phi_0=\pm i \ln r, ~~ \phi_k=k\pi/N, k\in \lbrace 1,...,N-1\rbrace. 
\end{align}
Here $\phi_0=\pm i \ln r$ corresponds to the Goldstone mode of the systems with $(z_1,z_2)=(1,r^2)$ which implies $E=0$. The corresponding wave function $\psi_0 = c_1 $ (const.) immediately gives the right eigenstate $\Psi = c_1(1,1,...,1,1)^T$. We also find that this Goldstone mode is always present as the system always respects the global rotation symmetry irrespective of system size or boundary conditions. For $\phi_k \neq \pm i\ln r$, the wave functions are proportional to $r^n\sin(n \phi_k),~\phi_k=k\pi/N$ which immediately implies localized eigenstates $\Psi (E\neq0)$ at one of the boundaries of the system. This proves the NHSE in our system. Therefore, in the time interval of the phase slip for $\delta\theta_{1, N}=\pi/2,~3\pi/2$, the Jacobian realizes the OBC and exhibits the NHSE.

In the time interval of the phase slip ($0\leq \delta\theta\leq 2\pi$), the Jacobian realizes the generalized boundary condition and the GBZs change accordingly. Figs.~\ref{fig:ET_TPT_GBC_5}(i)-(ii) shows the adiabatic evolution of the GBZs together with the eigenvalue spectra of the Jacobian matrix during half of the interval of the phase slip. The other half is symmetric. Therefore, during the phase transitions between different chiral phases, the Jacobian realizes the exceptional transition of the generalized Brillouin zone as a consequence of the NHSE.

\subsubsection{Noise-induced phase transitions: topological phase transition of GBZ for the Jacobian matrix: the emergence of positive eigenvalue bound state}\label{sec:TPT_GBZ_Jaco}
In the time interval of the phase slip for $\pi/2<\delta\theta<3\pi/2$, we observe a positive eigenvalue of the Jacobian which shows the onset of the phase instability during the non-equilibrium phase transition. This eigenmode is described by the disjoint generalized momenta and hence localized at the boundaries of the system (See Fig.~\ref{fig:ET_TPT_GBC_5}(i)(d)).

Now we discuss the intrinsic topology of the GBZ for the Jacobian characterizing the positive eigenvalue eigenstate localized at the boundary of the system. It is clear from the discussion in the previous sections that the boundary value problem governing the Jacobian in the time-interval of the phase slip emulates the boundary value problem of the HN model in generalized boundary conditions with the following boundary equations,
    \begin{align}
	H_B(c_1,~c_2)^\textrm{T}=0,~~\text{with}~H_\textrm{B}=
	\begin{pmatrix}
	A(z_1) & A(z_2) \\
	B(z_1) & B(z_2)
	\end{pmatrix},
	\end{align}
	where $A(z)=t_R-(t_R-t'_R) z - t'_R z^{N}$, 
	$B(z)=t_L' z + (t_L-t'_L) z^N -t_L z^{N+1}$. Here $t'_R=t_R\cos(\delta\theta),~t'_L=t_L\cos(\delta\theta)$ with $\delta\theta$ changing from $0$ to $2\pi$ in the time interval of the phase slip. The boundary equations can be rewritten as the eigenvalue problem,
	\begin{align}\label{eqn:Jaco_effective}
	\tilde{H}_B (c_1,c_2)^{\rm T}= (c_1,c_2)^{\rm T},~~{\rm with}~~\tilde{H}_B=&\begin{pmatrix}
	0& 	h_B^+(z_1,z_2) \\
	h_B^-(z_1,z_2)& 0
	\end{pmatrix},
 \end{align}
	where $ h_B^+(z_1,z_2)=A(z_2)/A(z_1)$, $h_B^-(z_1,z_2) = B(z_1)/B(z_2)$. The effective boundary matrix is generally a non-Hermitian matrix except for the open boundary condition when it becomes Hermitian. Moreover, it satisfies the chiral symmetry by construction (See Sec.~\ref{sec:sym_HB} for more details).

\end{widetext}

\end{document}